\newcommand{\redd}{$R_{\rm Edd}$}

\documentclass{aa}  
\usepackage{graphicx}
\usepackage{txfonts}
\usepackage{comment}
\usepackage{multirow}
\usepackage{xcolor}

\begin{document} 
\title{The universal power spectrum of Quasars in optical wavelengths: }
\subtitle{Break timescale scale directly with both black hole mass and accretion rate}
\titlerunning{Scaling of the quasar optical power spectra}
\authorrunning{P. Ar\'evalo et al.}
   \author{P. Ar\'evalo\inst{1,2}, E. Churazov\inst{3}, P. Lira\inst{4,2}, P. S\'anchez-S\'aez\inst{5,6}, S. Bernal\inst{1,2}, L. Hern\'andez-Garc\'ia\inst{1,6}, E. L\'opez-Navas\inst{1,2} \and P. Patel\inst{4,2}
          }
\institute{Instituto de F\'isica y Astronom\'ia, Universidad de Valpara\'iso, Gran Breta\~na 1111, Valpara\'iso, Chile\\
\email{patricia.arevalo@uv.cl}
\and Millennium Nucleus on Transversal Research and Technology to Explore Supermassive Black Holes (TITANS)
\and Max-Planck-Institut f\"ur Astrophysik, Karl-Schwarzschild-Str. 1, 85748, Garching, Germany
\and Departamento de Astronom\'ia, Universidad de Chile, Casilla 36D, Santiago, Chile
\and European Southern Observatory, Karl-Schwarzschild-Str. 2, 85748, Garching, Germany
\and Millennium Institute of Astrophysics (MAS),Monse\~nor S\'otero Sanz 100, Providencia, Santiago, Chile}



 
  \abstract
   {Optical variability of quasars is one of the few windows available to explore the behaviour of accretion discs around supermassive black holes. }
   {Establish the dependence of variability properties, such as characteristic timescales and variability amplitude, on basic quasar parameters such as black hole mass and accretion rate, controlling for the rest-frame wavelength of emission.}
   {Using large catalogues of quasars, we selected the g-band light curves for 4770 objects from the Zwicky Transient Facility archive. All selected objects fall into a narrow redshift bin, $0.6<z<0.7$, but cover a wide range of accretion rates in Eddington units (\redd) and black hole masses ($M$). We grouped these objects into 26 independent bins according to these parameters, calculated low-resolution $g$-band variability power spectra for each of these bins, and approximated the power spectra with a simple analytic model that features a break at a timescale $t_b$.}
   {We found a clear dependence of the break timescale $t_b$ on \redd , on top of the known dependence of $t_b$ on the black hole mass $M$. In our fits, $t_b\propto M^{0.65 - 0.55}$ \redd $^{0.35 - 0.3}$, where the ranges in the exponents correspond to the best-fitting parameters of different power spectrum models. This mass dependence is slightly steeper than found in other studies. Scaling $t_b$ to the orbital timescale of the innermost stable circular orbit (ISCO), $t_{\rm ISCO}$, results approximately in $t_{b}/t_{\rm ISCO} \propto ($\redd$/M)^{0.35}$. In the standard thin disk model, (\redd$/M)\propto T_{\rm max}^{4}$, where $T_{\rm max}$ is the maximum disk temperature, so that $t_{b}/t_{\rm ISCO}$ appears to scale approximately with the maximum temperature of the disc to a small power. The observed values of $t_b$ are $\sim 10$ longer than the orbital timescale at the light-weighted average radius of the disc region emitting in the (observer frame) $g$-band. The different scaling of the break frequency with $M$ and \redd\ shows that the shape of the variability power spectrum cannot be solely a function of the quasar luminosity, even for a single rest-frame wavelength. Finally, the best-fitting models have slopes above the break in the range between --2.5 and --3. A slope of --2, as in the damped random walk models, fits the data significantly worse.}
   {}


   \maketitle
%

\section{Introduction}
The flux variability of quasars was recognised early on as one of the few tools to study the structure and dynamics of their central engine and immediate environs, as reviewed by \citet{Cackett2021}. Different studies of the continuum optical variability have concluded that at least part of the variations must originate in the accretion disc, as opposed to representing only a reprocessed response to the highly variable X-ray emission, both on long and short timescales \citep[e.g.][]{Krolik91, Arevalo08,Ai2010,Edelson15,Lira15,Smith18}. Therefore, optical variations are a viable tool to constrain the behaviour of the accretion disc. Several authors have made significant progress in correlating variability properties with basic quasars parameters, with increasingly large samples and longer or more frequently sampled light curves and more complex methods \citep{Bauer09, Caplar17, Sanchez-Saez18, Li18, Luo20, Tachibana20,Arevalo23,Tang2023}. In particular, the search for a characteristic timescale of variations has been hampered by the long lightcurves that would be required to measure it \citep[e.g.][]{Kozlowski17}. However, \cite{Burke21} recently measured a strong, positive correlation between the damping timescale and black hole mass, as obtained from the fit of Damped Random Walk models to high-quality, light curves of quasars. In this work, we aim to establish whether the characteristic timescale also depends on another fundamental quasar parameter, namely the accretion rate, quantified through the Eddington ratio \redd .

The characteristic variability timescale, measured as a bend in the power spectrum of the lightcurves or as the de-correlation timescale of a Damped-Random-Walk (DRW) model \citep{Kelly09} can depend on the rest-frame wavelength of the lightcurves. In particular, longer characteristic timescales are predicted for longer wavelengths if the variations are produced on a local characteristic timescale (e.g. the radius-dependent thermal timescale of the accretion disc) and different wavelengths are preferentially emitted at different radii in the disc. Some observational results favour this scenario, for example, \citet{Sun2014} found that the amplitude of variability depends on the wavelength on short timescales but not on long timescales, which means that the power spectrum shape must depend on the emitted wavelength. \cite{Xin2020} find that optical and UV light curves are not perfectly correlated, using Galaxy Evolution Explorer and the Catalina Real-time Transient Survey light curves of over 1300 quasars, implying that the power spectra in different optical/UV bands are not simply scaled by amplitude. More recently, \citet{Stone2022} found that the characteristic timescale of variations depends on the emitted wavelength, from a sample of 190 quasars with decades-long light curves. Therefore, to refine the search for correlations between characteristic timescales and physical parameters, the light curves should necessarily probe the same rest-frame wavelength for all objects.

In this work we explore the optical power spectra of quasars in a single band and a narrow redshift range (see Sec. \ref{sec:data}), to avoid potential dependencies of the power spectrum shape on the rest-frame wavelength of emission.  We calculate the median power spectra for bins in mass, $M$, and \redd ,  to establish the dependence of the power spectrum break frequency and normalisation on both parameters. The variance is estimated using the Mexican Hat power spectrum \citep{Arevalo12}, which is well suited to light curves with gaps and uneven sampling (see Sec. \ref{sec:methods}) and the resulting power spectra are presented in Sec. \ref{sec:power spectrum}. Model fitting to a bending power law model, with different slopes, is done by simulating light curves with specific model parameters and folding them through the same sampling and variance estimation process as the real data, as described in Sec. \ref{sec:modeling}. Best-fitting models and scaling parameters were searched by minimising the difference between data and model power spectra. With this minimisation we show that the scaling relations of amplitude and characteristic timescale of the variability with mass and \redd\ are largely independent of the model and select the better-fitting power spectrum model parameters in Sec. \ref{sec:comparison}. A basic interpretation of the scaling relations in terms of accretion disc parameters is presented in Sec.\ref{sec:discussion} and the conclusions are summarised in Sec. \ref{sec:conclusion}. 

\section{Data}
\label{sec:data}
The selection of quasars included in this work and the light curves used are the same as those presented in \citet{Arevalo23}. In summary, we use g-band light curves obtained by the Zwicky Transient Facility (ZTF, \citealt{Masci19}) through their public data release DR14 of the 4770 quasars that: i) are in the redshift range 0.6<z<0.7; ii) have mass and \redd\ estimated through Sloan Digital Sky Survey (SDSS) spectra by \citet{Rakshit20} with reported statistical errors on mass less than 0.2 dex; iii) have a mean g-band ZTF magnitude less than 20, and that iv) their ZTF light curves, restricted to data obtained with a single CCD, excluding bad nights and multiple observations in the same night, have at least 90 data points and are at least 900 days long. Some objects have 2 light curves because they were observed with two different CCDs on different days and both light curves conform to the minimum standards. The work described below used the resulting 5433 valid light curves, which have a median length of 1547 days and a median number of points of 233. We note that since all galaxies are at z>0.6 they appear almost point-like in the ZTF images. Therefore the PSF photometry on the science images, which is used to produce the light curves of the data release, is sufficiently accurate.

\section{Estimation of the Variance}
\label{sec:methods}
The procedure to obtain the variance of the light curves follows the procedure described in \citet{Arevalo23}. Below we briefly summarise the method. 

We estimated the variability power density on 4 well-separated timescales with the Mexican hat filter \citep{Arevalo12} and multiplied them by the corresponding frequency to obtain the variance at the four timescales. Since the light curves are mean-subtracted and divided by the mean flux, these variances are dimensionless and represent the amplitude of variability relative to the mean flux, squared. 
The timescales chosen were 30, 75, 150 and 300 days in the rest-frame of the quasars. The additional variance produced by photometric errors of the light curve was subtracted from the measured variances of each light curve using simulated light curves of pure noise, constructed from the error bars of the real light curves. We further correct these noise estimates by carrying out the same procedure of variance estimation and noise subtraction for ZTF lightcurves of non-variable stars, which should result in net variances distributed around 0. We found small deviations from 0 for the median net variances, mostly towards negative values. We fitted the relation between median net variances and apparent magnitude with fourth-order polynomials and added this "missing variance" as a function of magnitude to the net variance of each quasar, for each timescale. 
These were always small corrections as can be seen in appendix B of \citet{Arevalo23}.  

We split the sample of quasars on a grid of $M$ and \redd\ in order to measure the power spectrum dependence on these two basic parameters. The ranges in $M$ and \redd\ used for the grid are displayed in the line and column headings, respectively, in Table \ref{tab:sample} and the table contains the number of light curves included in each bin. The analysis below will only use bins with 10 or more lightcurves, in order to obtain reliable estimates of the uncertainty of the variances, by bootstrapping the measured variances within each bin as described in \citet{Arevalo23}. 

\begin{table*}[]
    \centering
    \begin{tabular}{|cc|r r r r r r|}
\hline
   & &\multicolumn{6}{c}{range in log(\redd)} \\
 &  &-2 -- -1.7&-1.7 -- -1.3&-1.3 -- -1&-1 -- -0.7&-0.7 -- -0.3& -0.3 --0\\
\hline
\multirow{6}{40pt}{Range in $\log(M/M_\odot)$}
&7.5--7.8&     0&   1&   2&  49& 173&  58\\
&7.8--8.2&    2&    5&  102& 429& 376&  71\\
&8.2--8.5&    1&   99&  580& 711& 236&  45\\
&8.5--8.8&    40& 356&  634& 333&  80&   14\\
&8.8--9.2&    126& 304& 234&  65&  18&   0\\
&9.2--9.5&     74&  71&  35&  4&   1&   0\\
\hline

    \end{tabular}
    \caption{Number of light curves included in each $M-$\redd\  bin. The line and column headings display the ranges in $M$ and \redd , respectively, that limit each bin.}
\label{tab:sample}
\end{table*}

\section{Power Spectrum of the data}
\label{sec:power spectrum}
Light curves can be characterised through the power spectrum, which quantifies the amount of variance found on different timescales or equivalently, different frequencies $f=1/t$. Here we collect the variance estimates in four timescales to construct a low-resolution power spectrum for each object, covering one decade in frequency.  We note that whenever we plot power spectra we will multiply the power density by the corresponding frequency so that the quantity plotted is Power $\times$ frequency.

We split the quasar sample in bins of both \redd\ and mass as described in Sec.\ref{sec:methods} and plot the median power spectrum of each $M-$\redd\  bin in Figs. \ref{fig:PDS_M} and \ref{fig:PDS_REdd}. The error bars in the plots represent the standard deviation of the medians obtained for 1000 random re-samples of the data in each bin. We note that since all the quasars in the sample are at approximately the same redshift $(0.6\leq z\leq0.7)$ and all the light curves were observed with the same filter, they sample the same rest-frame wavelength $\lambda\sim 2900 $\AA . This allows us to determine the dependence of variability amplitude on $M$ and \redd\ without having to account for the dependence of amplitude on the emitted wavelength \citep[e.g.][]{Sun2014,Sanchez-Saez18, Stone2022}. 

\begin{figure*}[h]%
\centering{
\includegraphics[width=0.45\textwidth]{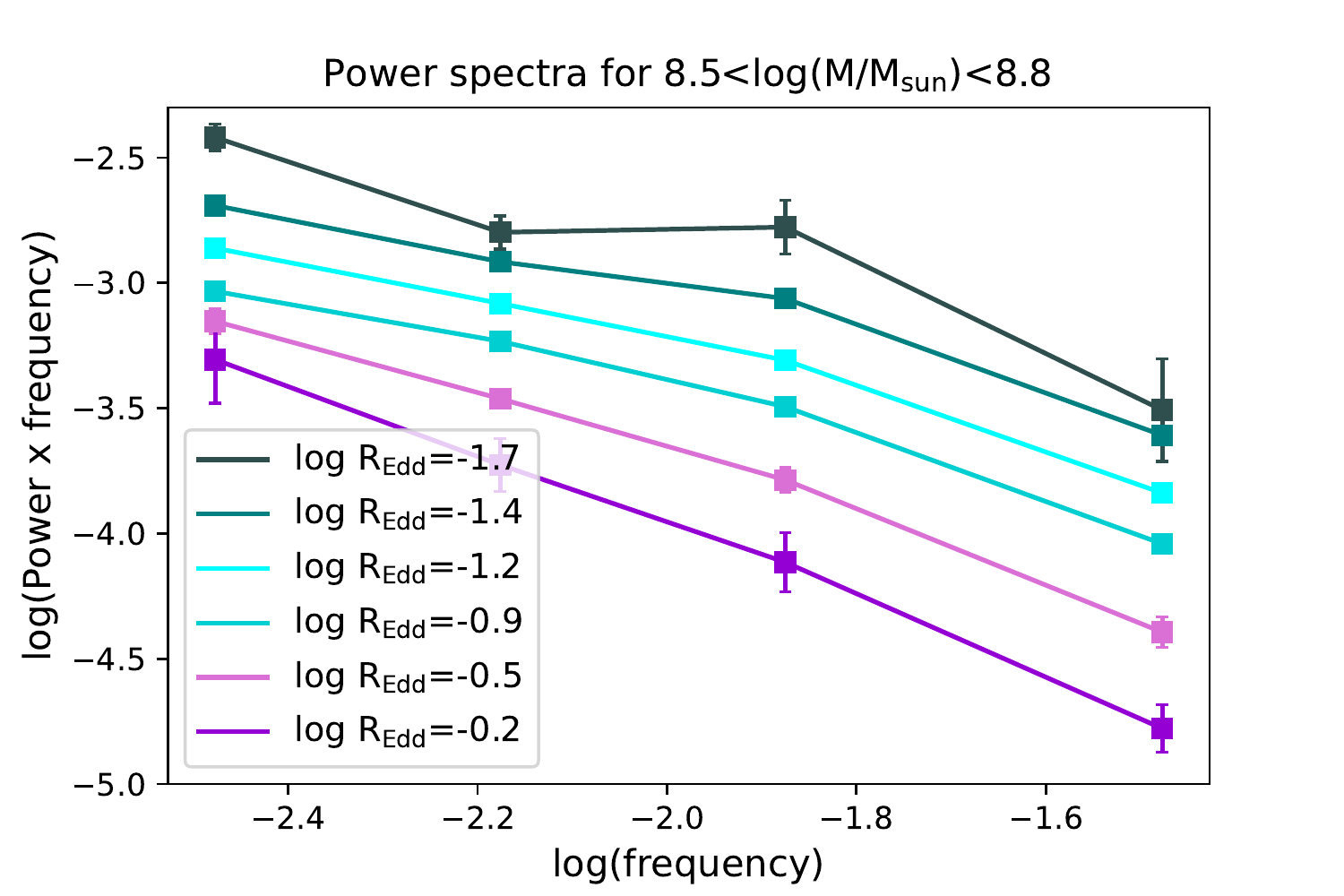}
\includegraphics[width=0.45\textwidth]{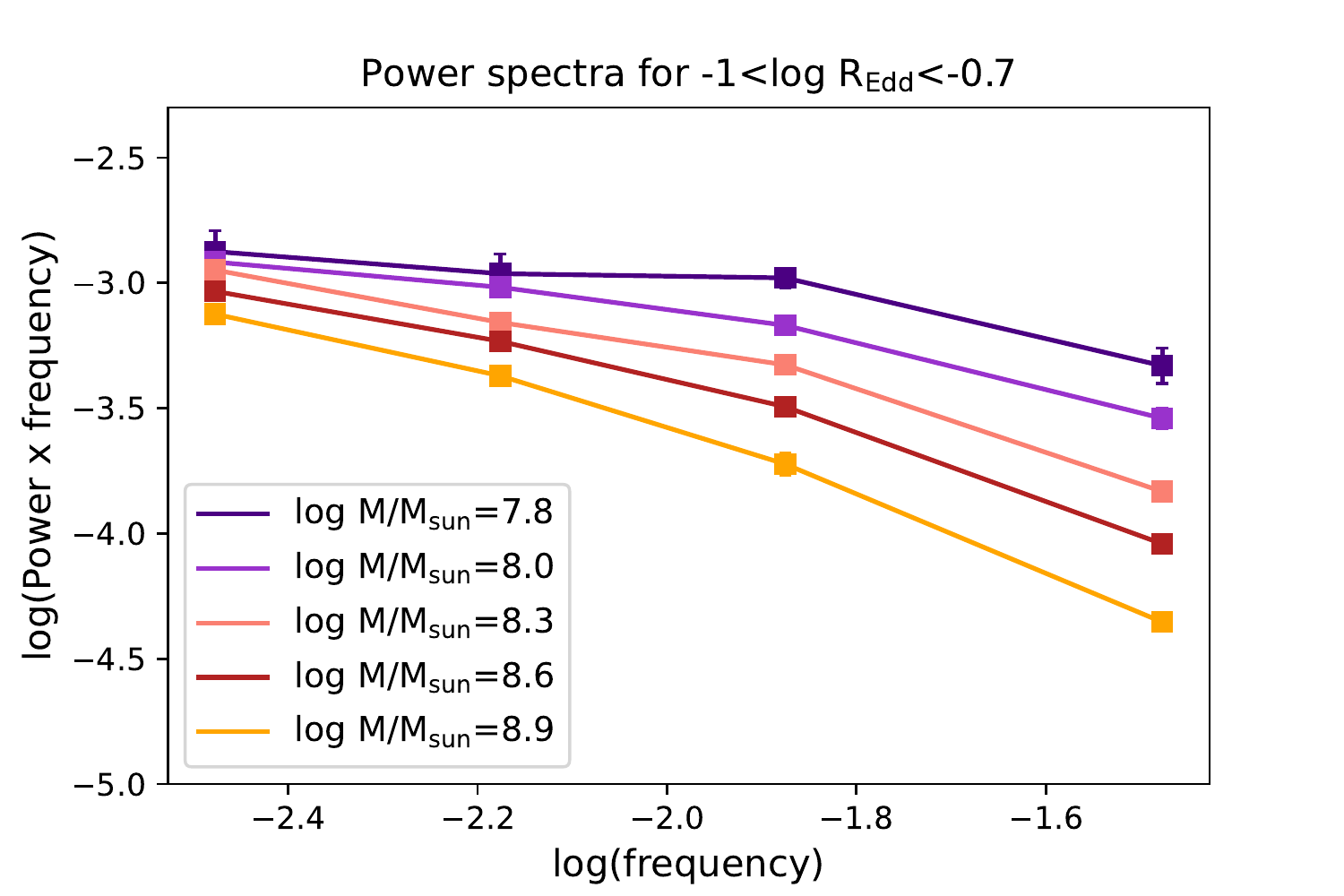}
}
\caption{Low-resolution power spectra for different values of \redd\ within a narrow range in black hole mass (left) and different black hole masses but similar values of \redd\ (right). The markers represent the median variance (i.e. Power$\times$ frequency) of all quasars falling in each $M-$\redd\  bin and temporal frequency, and the error bars represent the root-mean-squared (rms) of the median for each bin obtained by bootstrapping the samples. The power spectra depend on both parameters but in different ways, the mass changes the slope (or location of the bend in the power spectrum as described in Sec.\ref{sec:power spectrum}) and the accretion rate modifies mostly the normalisation, although some change in slope is apparent. Frequencies in are in units of day$^{-1}$.
}\label{fig:PDS_1M_1REdd}
\end{figure*}

\begin{figure*}[h]%
\centering{
\includegraphics[width=0.45\textwidth]{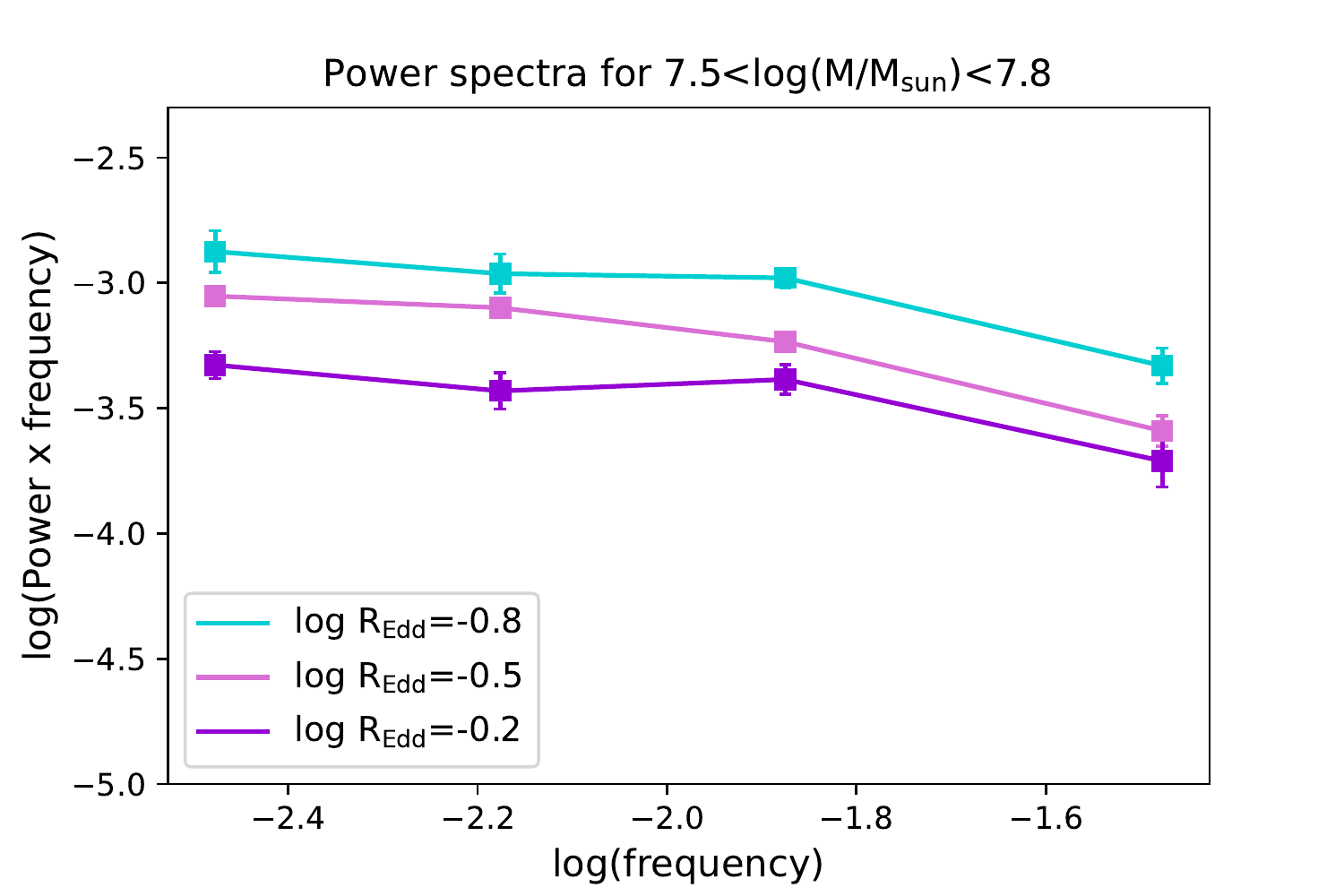}
\includegraphics[width=0.45\textwidth]{PDS_M4_correctedmask.pdf}
\includegraphics[width=0.45\textwidth]{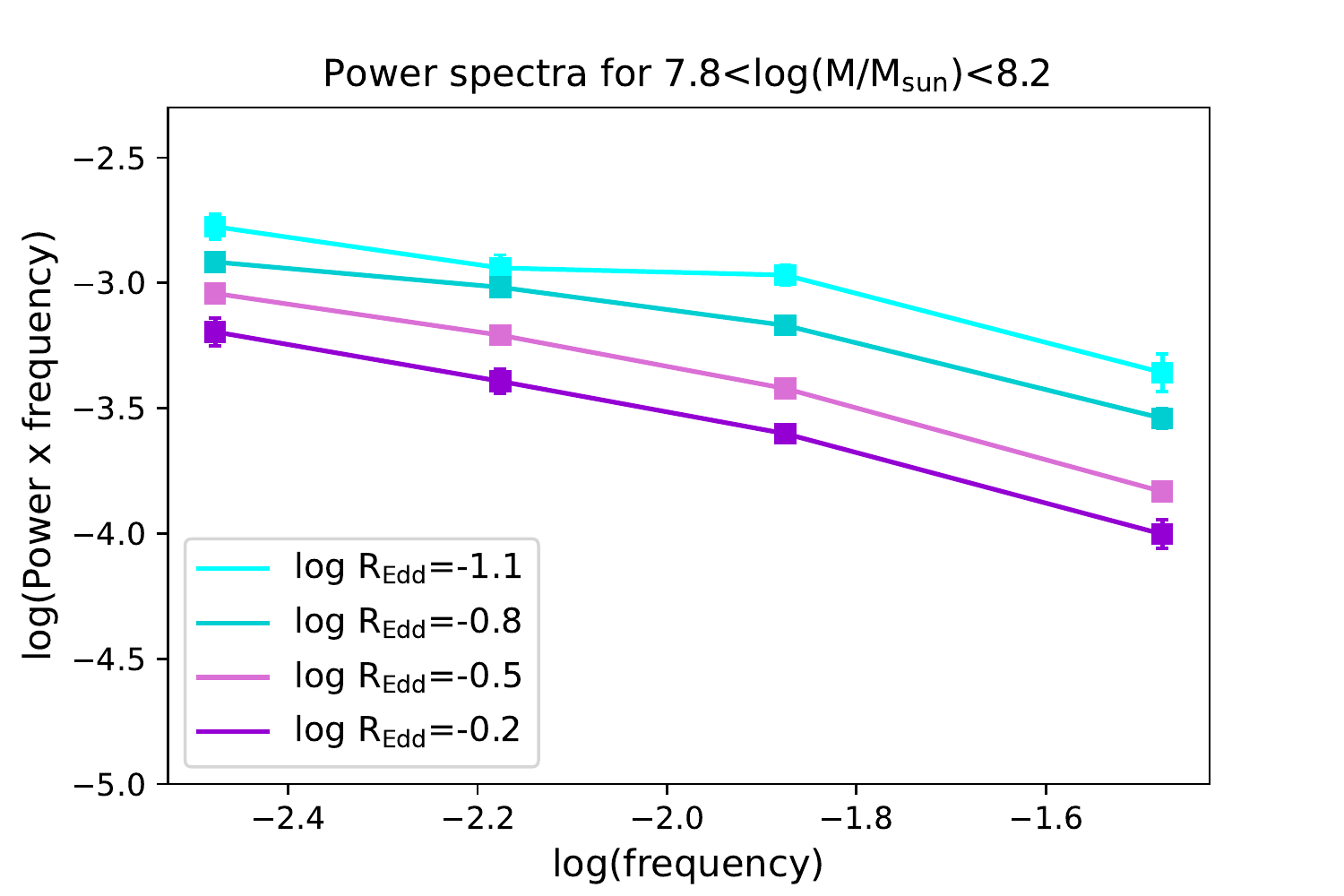}
\includegraphics[width=0.45\textwidth]{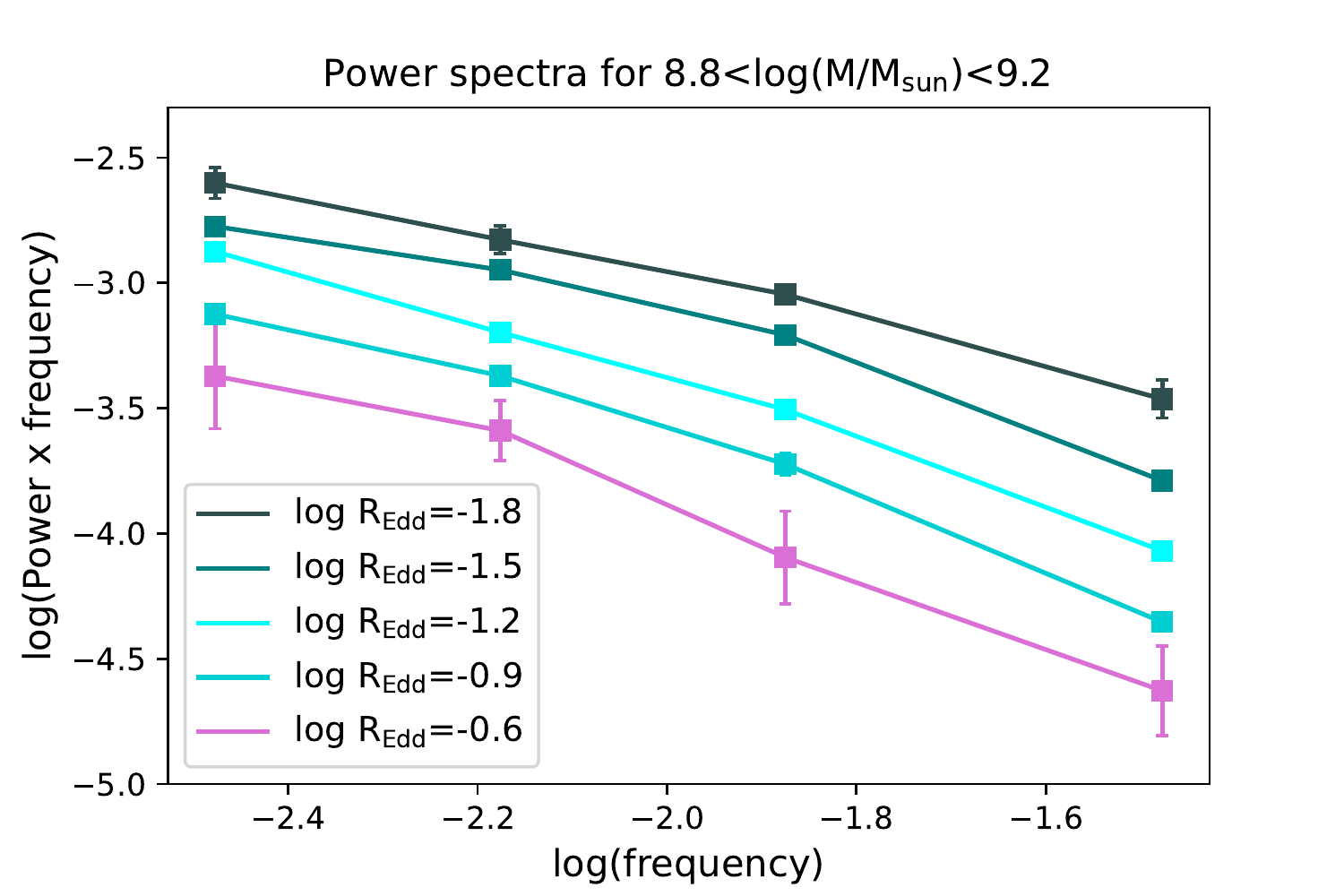}
\includegraphics[width=0.45\textwidth]{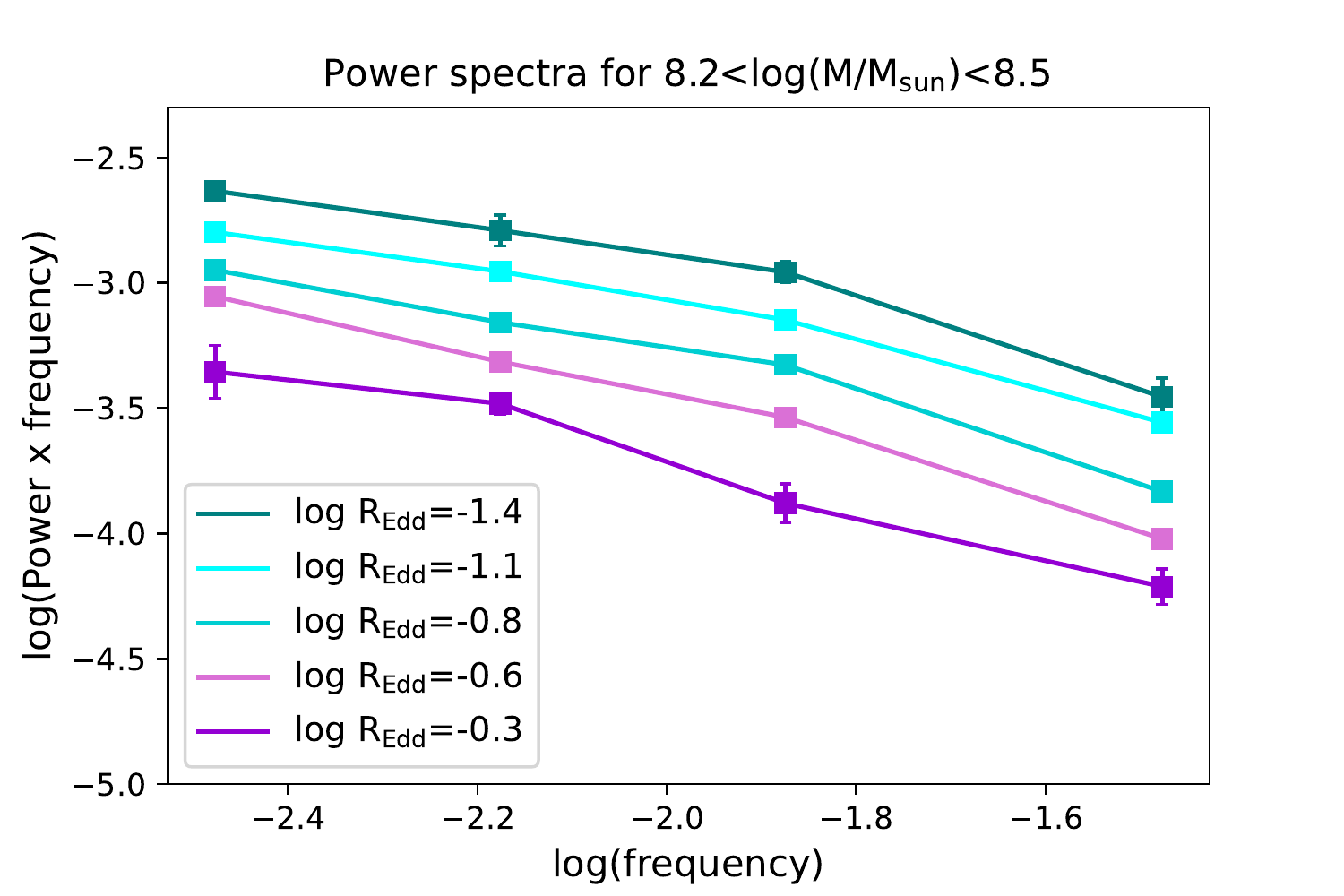}
\includegraphics[width=0.45\textwidth]{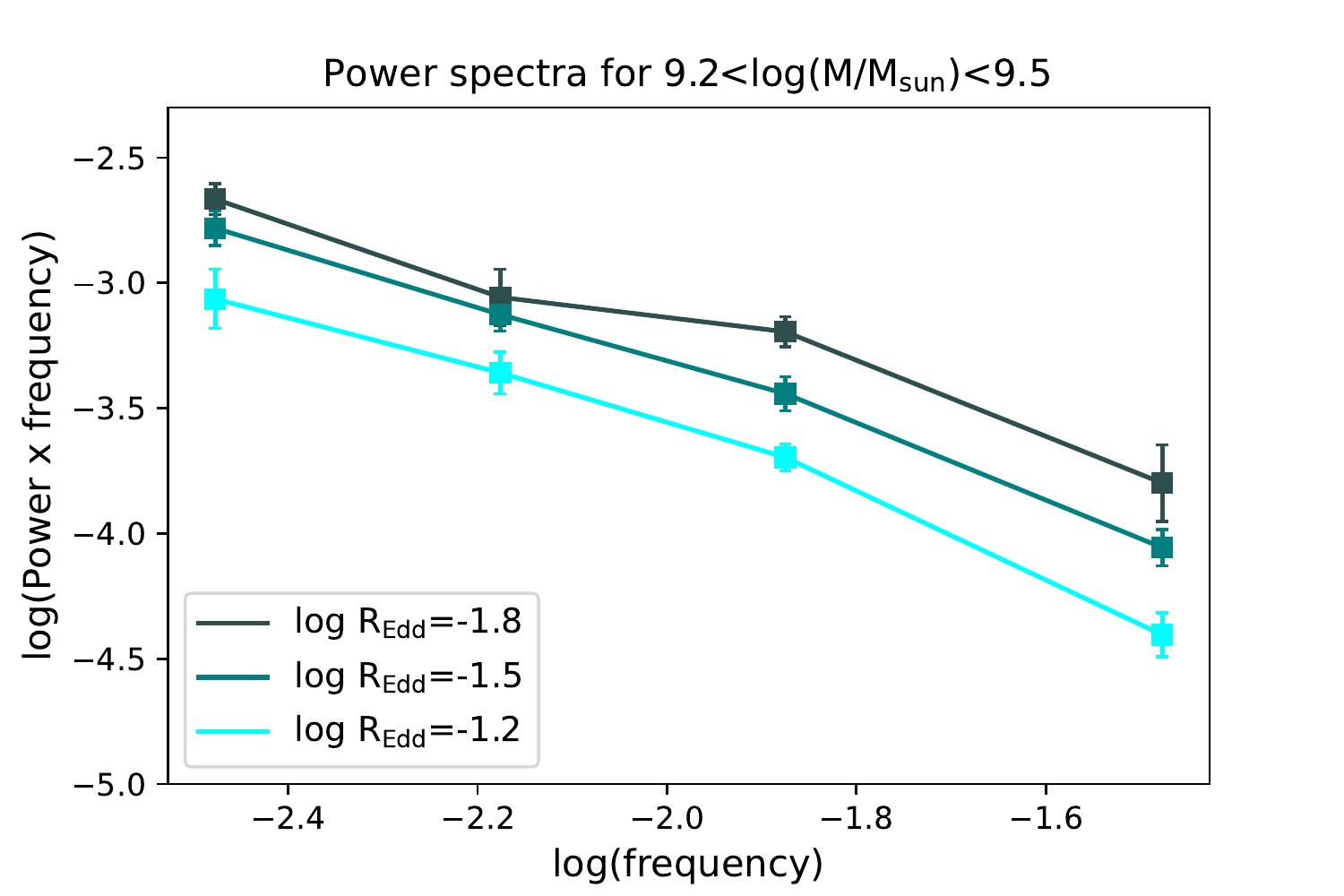}
}
\caption{Similar to Fig. \ref{fig:PDS_M} but arranged in separate panels for different mass bins, and different colours used to differentiate by accretion rate level.} \label{fig:PDS_M}
\end{figure*}

\begin{figure*}[h]%
\centering{
\includegraphics[width=0.45\textwidth]{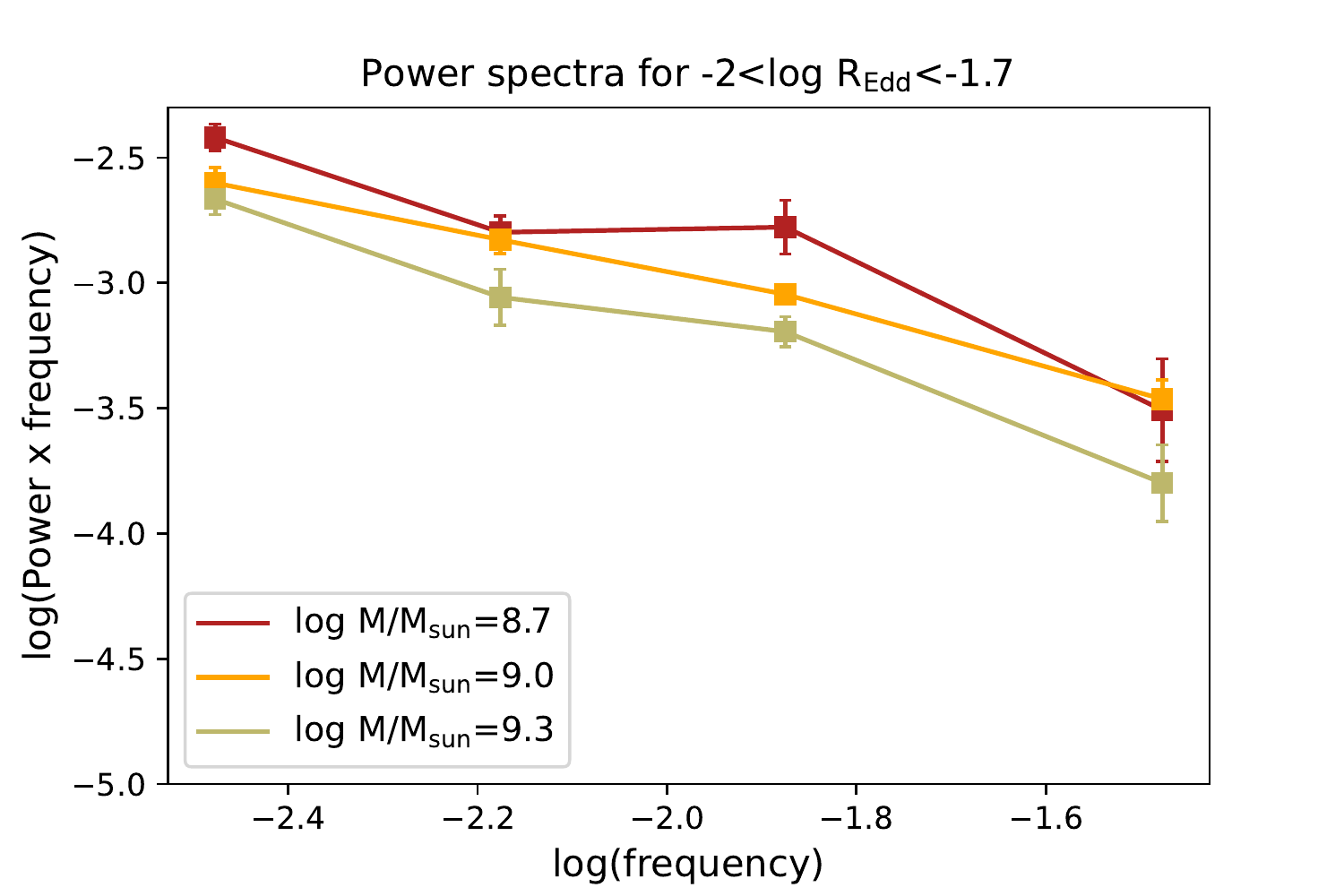}
\includegraphics[width=0.45\textwidth]{PDS_REdd4_correctedmask.pdf}
\includegraphics[width=0.45\textwidth]{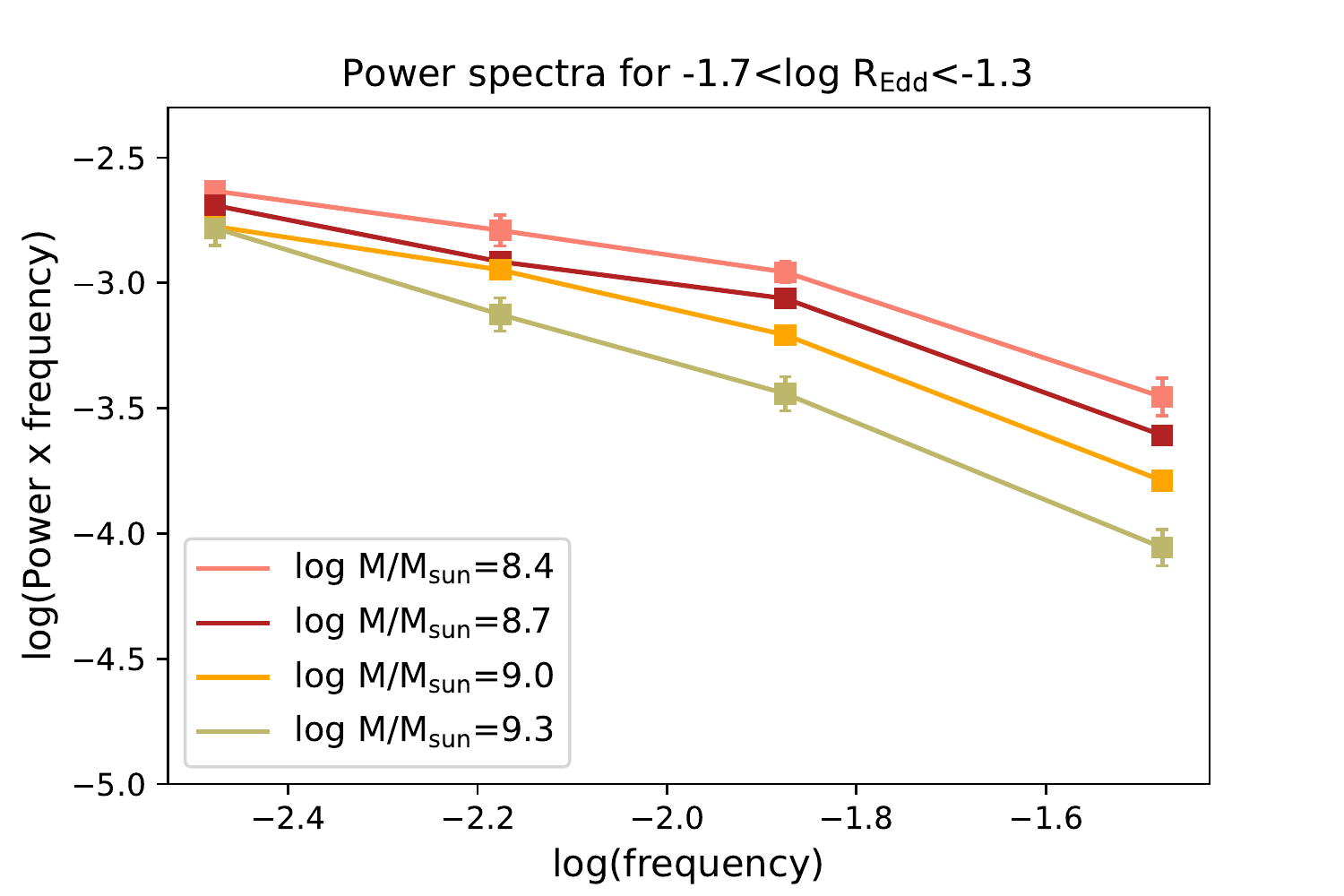}
\includegraphics[width=0.45\textwidth]{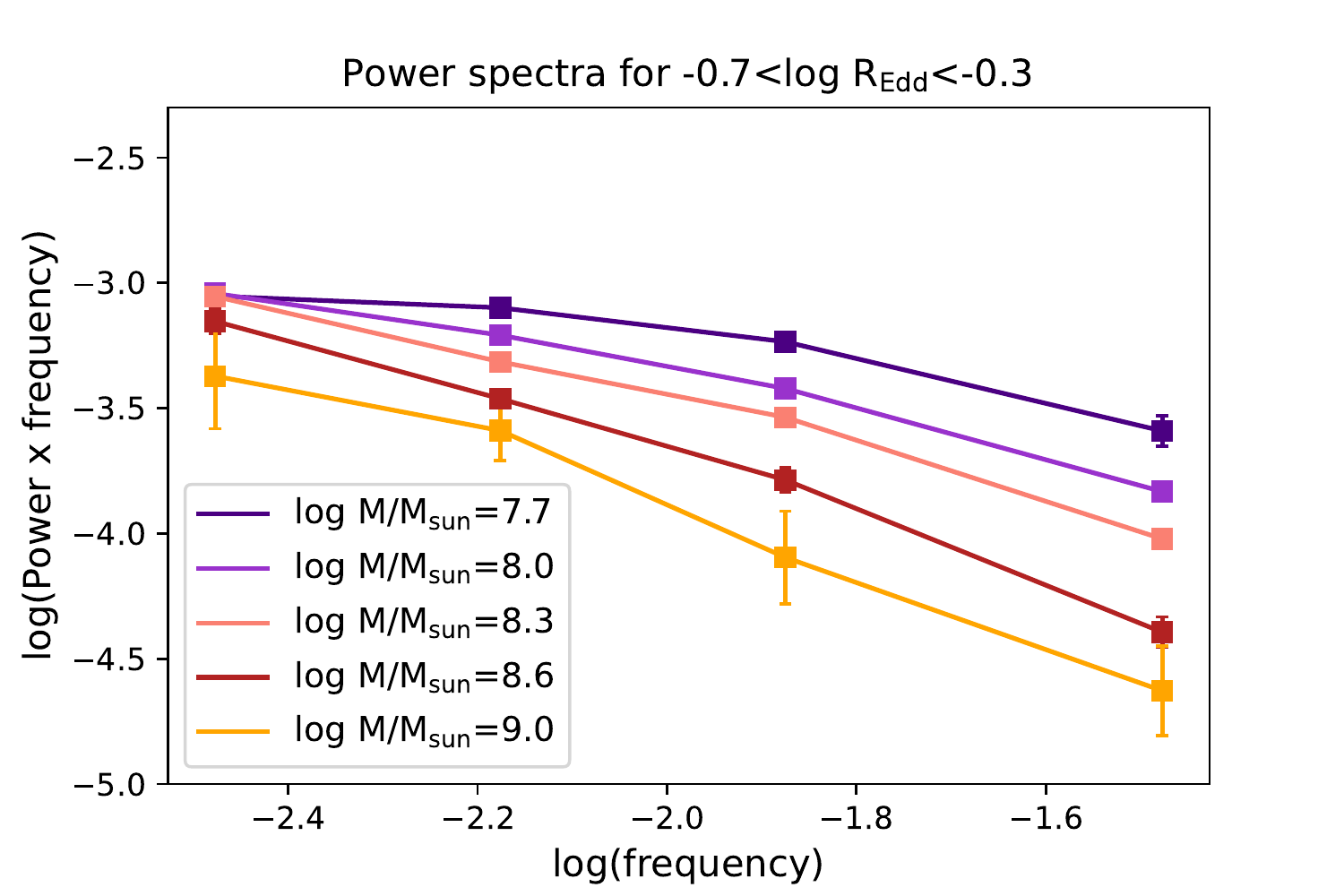}
\includegraphics[width=0.45\textwidth]{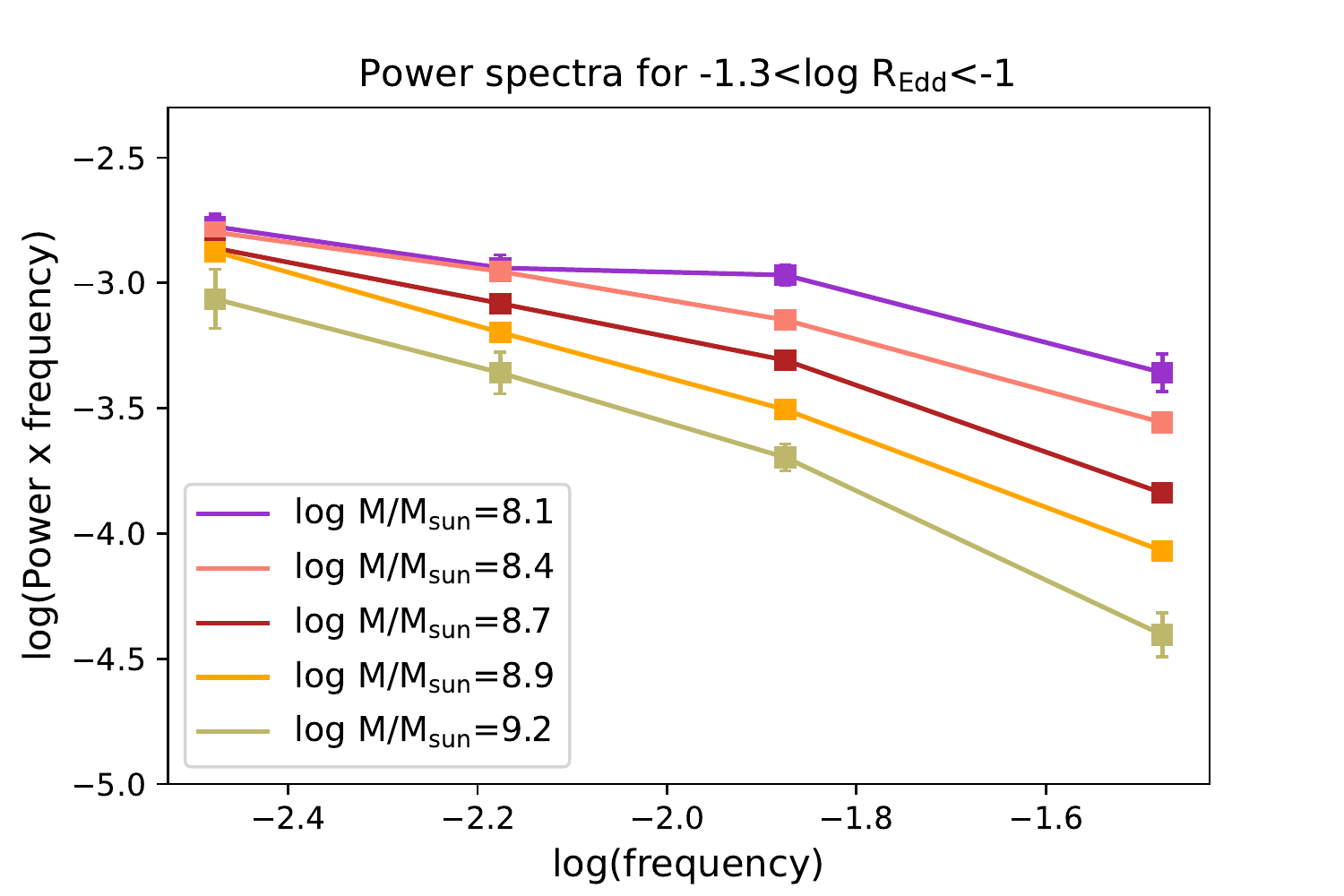}
\includegraphics[width=0.45\textwidth]{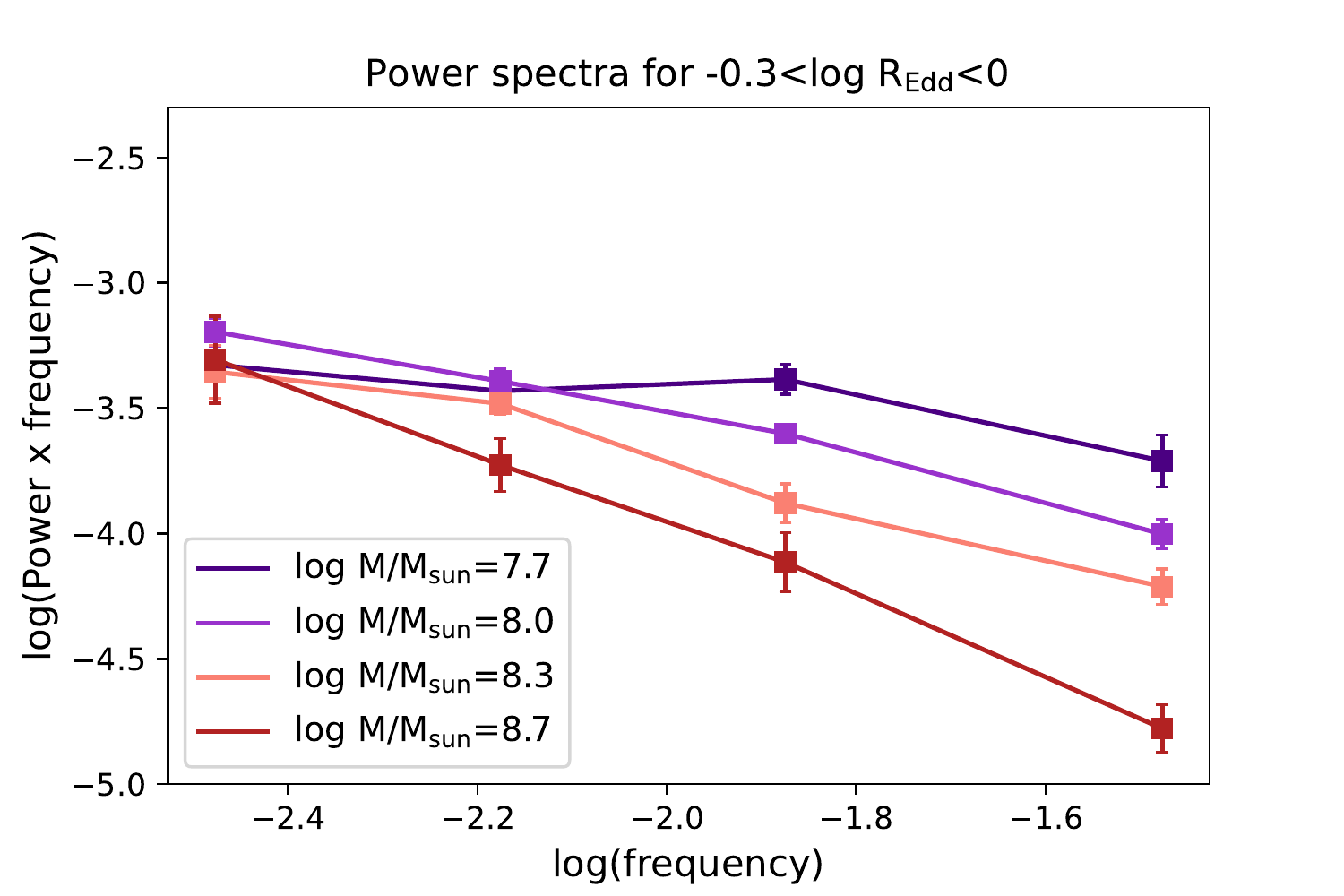}
}
\caption{Low-resolution power spectra  (plotted as Power $\times$ frequency) for different \redd\ bins in different panels and different black hole masses in different colours. The markers represent the median variance of all quasars falling in each $M-$\redd\  bin and temporal frequency, and the error bars represent the rms of the median for each bin obtained by bootstrapping the samples.  Frequencies in are in units of day$^{-1}$}. In most \redd\ bins, the power spectra steepen for higher masses, with the exception of the first panel, where all masses are high and all spectra are steep.\label{fig:PDS_REdd}
\end{figure*}

 Figure \ref{fig:PDS_1M_1REdd} compares the dependence of the power spectra on \redd, for objects within a narrow range in mass (left), to the dependence of the power spectra on mass, for quasars within a narrow range in \redd\ (right). Evidently, the power spectrum depends on both parameters but in different ways: the Eddington ratio affects mostly the normalisation and the black hole mass affects mostly the slope, in the frequency range sampled by the light curves.

Figure \ref{fig:PDS_M} highlights the dependence of the power spectra on \redd . On each panel, the power spectra at three or more accretion rate levels are shown in different colours, and the different panels correspond to different mass bins. The most noticeable difference between the power spectra is the normalisation, with higher accretion rates having increasingly lower normalisation.   In each panel, or equivalently, black hole mass bin, the power spectra of different \redd\ levels are nearly uniformly spaced. Noting that the bins are split uniformly in log(\redd ), this spacing implies that log(var) scales approximately linearly with log(\redd ). For each mass bin, the power spectra show lower amplitudes for higher accretion rates, and the decrease is almost independent of frequency (in each plot, the power spectra are almost parallel to each other). However, a small steepening of the power spectra at higher accretion rates is apparent. This implies that the shape of the power spectrum depends to some extent on \redd , either via slope or the bend frequency.

Figure \ref{fig:PDS_REdd} shows the dependence of the power spectra on black hole mass. Each panel contains one bin in \redd\ to compare the power spectra for different masses, coded by colour, and similar \redd . In all but the first panel the mass dependence becomes obvious, where higher masses correspond to increasingly steeper slopes. In the first panel, all slopes are similarly steep, but we note that the range of masses included is smaller and they are all in the high mass end of the sample. A similar steepening of the power spectra with increasing mass was observed by \citet{Caplar17}, for a large sample of about 28.000 quasars with a range of redshifts up to $z\sim 2.2$, observed in the $r$-band by the Palomar Transient Factory survey.

This dependence of the power spectrum slope on the black hole mass can be reconciled with a universal power spectrum shape for all quasars if this shape corresponds to a bending power law and the bend frequency scales with mass, with a higher bend frequency for lower black hole masses, as has been shown by \citet[e.g.][]{Burke21}. Therefore, if we plotted the frequency axes in Fig. \ref{fig:PDS_REdd} relative to the bend frequency of each mass bin (i.e. $f/f_b$),  then the spectra of lower mass objects would shift to the left and that of higher mass objects would shift to the right. In principle, these shifts could make the differently coloured spectra align onto a single power spectrum, with a bending power-law shape. 

\begin{figure}[h]%
\centering{
\includegraphics[width=0.47\textwidth]{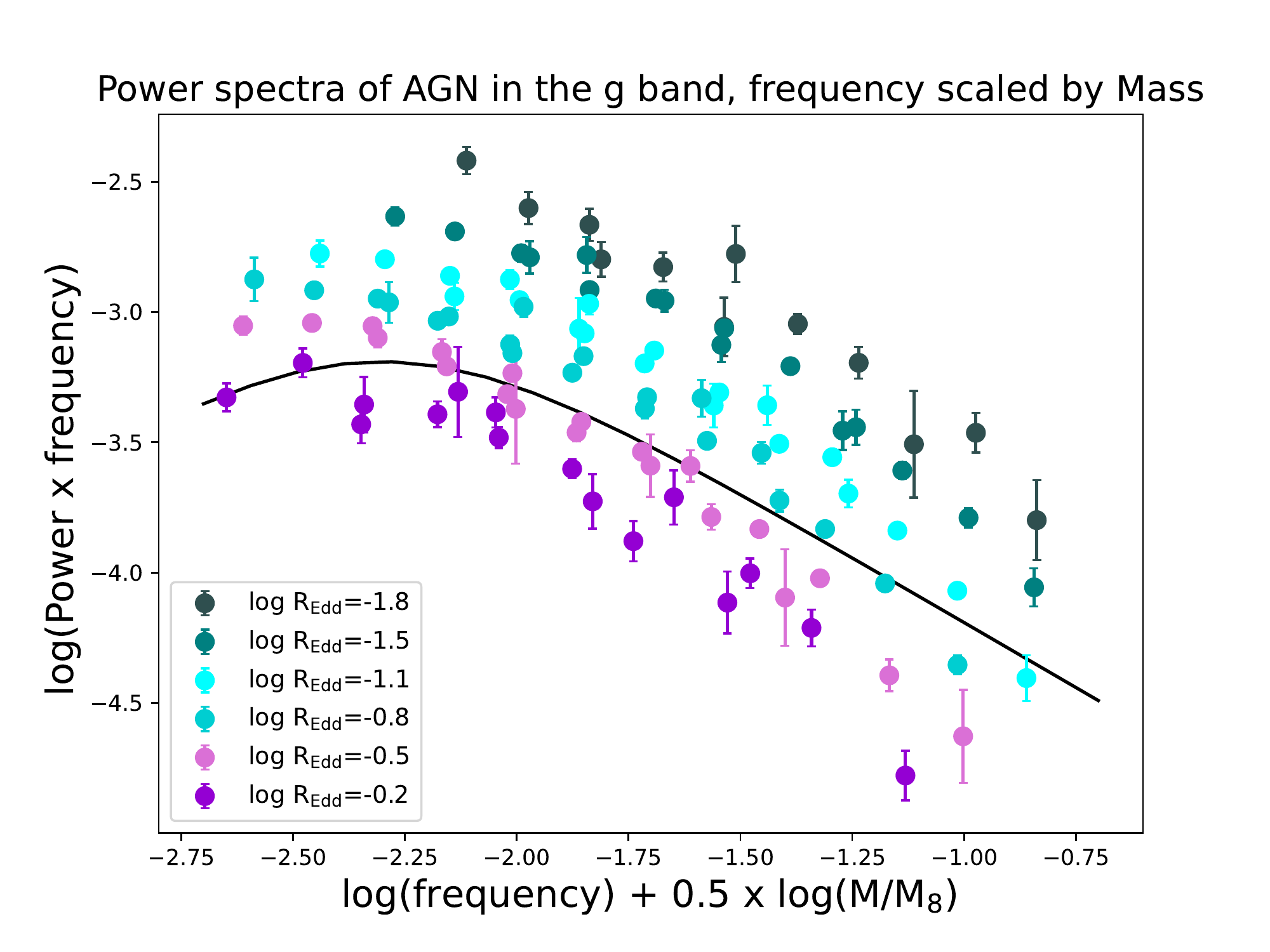}
}
\caption{Power spectra (plotted as Power $\times$ frequency) of all the sources, binned by mass and Eddington ratio. The variances have been calculated for timescales of 30, 75, 150 and 300 days, in this plot the corresponding frequencies $f=1/t$ have been multiplied by $(M/M_8)^{0.5}$ to put them on a common x-axis scaled to the bend timescale approximately as measured by \cite{Burke21}. The colour represents the different bins in \redd . The black solid line represents a bending power law model with $\alpha_L=0$ and $\alpha_H=-2$, similar to the DRW model, with a bend timescale of 200 days for a $10^8 M_\odot$ black hole. (We note that as power is multiplied by frequency in the plot, the part of the black line with $\alpha_L=0$ appears rising with frequency). Frequencies are in units of day$^{-1}$}.\label{fig:PDS_all}
\end{figure}

\citet{Burke21} found that the bend timescale of the power spectrum scales with mass approximately as $t_b\propto M^{0.5}$, and that the characteristic timescale for a $10^8 M\odot$ black hole is about 200 days. Following this prescription we re-scaled the frequencies of our power spectra by multiplying them by their corresponding value of $(M/M_8)^{0.5}$. The result of this scaling is shown in Fig. \ref{fig:PDS_all}, colour coded by \redd\ as in Fig. \ref{fig:PDS_M} but this time including all the mass bins in the same panel. There is a good alignment of the power spectra for different masses and equal \redd , consistent with a universal power spectrum shape where the bend frequency scales with mass and the normalisation scales with \redd , although with some scatter. The black solid line represents a bending power law model with $\alpha_L=0$ and $\alpha_H=-2$ with the bend at [200 days]$^{-1}$, comparable to the DRW model with the bend frequency at the location found by \citet{Burke21}, plotted only as a guideline. The comparison between spectral shapes cannot be done directly in this plot, however, since the light curve sampling and variance estimation methods distort the spectra to some extent. Therefore, we will model the spectral shape using a "folded" model as described below. 

\section{Modelling of the Power Spectrum}
\label{sec:modeling}
Power spectra of AGN, in the optical as well as the X-ray regime, resemble bending or broken power laws, with at least one characteristic break frequency \citep[e.g.][and references therein]{McHardy06,Burke21,Stone2022}. This shape includes the power spectrum of a Damped-Random-Walk (DRW) model, often used for fitting optical power spectra of quasars \citep[e.g.][]{Kelly09,Kozlowski10,MacLeod10,MacLeod2012,Sanchez-Saez18,Suberlak2021,Burke21,Stone2022}, when the low-frequency power-law slope is 0 and the high-frequency slope is -2. This particular model is shown by the solid black line in Fig. \ref{fig:PDS_all}, where the bend frequency was set at 1/(200 days) following the findings of \citet{Burke21}, and with an arbitrary normalisation. We note that although this model is close to the shape and apparent bend frequency of our power spectra, it does not reproduce perfectly the measured slopes. Other authors have also found deviations from the DRW model for other samples of quasars and other light curve sets \citep[e.g.][]{Mushotzky2011, Zu13, Kasliwal15,Caplar17,Sanchez-Saez18, Stone2022}. We will therefore  use a more general bending power law model that allows for different power law slopes and is often used to model quasar power spectra in the X-ray regime \citep[e.g.][]{McHardy2004}:
\begin{equation}
    P(f)= \frac{A f^{\alpha_L}}{1+(f/f_b)^{\alpha_L-\alpha_H}},
    \label{eq:PDS}
\end{equation}

In this model, frequencies below the break approximate a power law of slope $\alpha_L$ and above the break a power law of slope $\alpha_H$.

As discussed in Sec. \ref{sec:power spectrum}, for a given mass, the logarithm of the variance scales approximately linearly with the logarithm of \redd . The black hole mass in turn appears to modify the break frequency, so that the frequency range probed, i.e. 1/300 to 1/30 days$^{-1}$ covers part of the power spectra below the break for the lowest mass bins and above the break for the highest mass bins, again showing a uniform separation of log(variance) for a uniform separation of log(M). Additionally, there is a weaker dependence of the power spectrum shape on \redd . Therefore, we will attempt to fit all 104 power spectral points shown in Fig. \ref{fig:PDS_all} together, with a bending power law model where the bend frequency and normalisation have the following dependencies:
 \begin{eqnarray}
\log{f_b} &=& \log{B} + C \times\log{M/M_8} +D \times\log{R_{Edd}/0.1} \label{eq:fb}\\
\log{A}&=& \log{A1} + F \times\log{R_{Edd}/0.1},\label{eq:A}
\end{eqnarray}
where $f_b$ and $B$ will be in units of days$^{-1}$. We note that $A1$ and $B$ will be the normalisation of the power spectrum and the bend frequency, respectively, for quasars of $10^8M_\odot$ and \redd = 0.1, which are the most typical in our sample. 

It is unclear whether a universal power spectrum would have equal variances (i.e. $P\times f$) at the bend frequency or at zero frequency, for a given \redd\ and different masses. If the low-frequency slope is $\alpha_L =-1$ then both cases are equivalent, but if the low-frequency slope is flatter then the choices are to align the low-frequency spectra or to match the variance at the break. We will explore the case where the universal power spectrum implies equal variance at the bend frequency. In this case, the normalisation A1 has an additional dependence on the break frequency:
\begin{equation}
    A1= A2\times B/f_b \times f_b^{-a_L}\label{eq:A2}
\end{equation}
With this scaling, the variance, $P\times f$ will be equal to $A2\times B$ at the bend frequency for all mass bins when \redd = 0.1 and for the other \redd\ ranges it will scale with \redd\ only, i.e. $P(f_b)\times f_b = A2\times B \times$ (\redd/0.1)$^{F}$.

We will fit different power spectrum models to obtain the values of A2, B, C, D and F that best describe the data. 

\subsection{PDS model fitting with simulated lightcurves}

The sampling pattern of the lightcurves and the method used to obtain the variance inevitably affect the measurements to some degree. Therefore, we resort to simulated light curves using the model defined in Eq.\ref{eq:PDS} and time samplings taken from the real observations, to compare the measured variances with a `folded' model. The light curves were simulated following the prescription of \citet{Timmer95}, with a total length of 1200 days and a generation sampling of 0.1 days, both in the quasar rest-frame. We allowed the model parameters to depend on the physical properties $M$ and \redd\ as described in Eqs.\ \ref{eq:fb}, \ref{eq:A} and \ref{eq:A2}. We searched a grid in the parameters $B,C,D$ and $F$. 

We note that the dependence of the normalisation on \redd\ is set by the parameter $F$, but the overall normalisation is arbitrary. We set it by minimising the difference between the sum of the data variances and the sum of the model variances, to obtain the best-fitting value of $A2$ for each set of the other parameters. The steps in $C,D$ and $F$ were linear and the steps on $B$ were multiplicative. Different step sizes were used to refine the search for the best fit, with a final step size of 1.2 in $B$, 0.05 in $C$, $D$ and $F$.

For each set of model parameters, we calculated the break frequency $f_b$ and normalisation $A$ for each $M-$\redd\  bin, and created 100--300 simulated light curves with a fixed set of 30 real sampling patterns re-scaled to the quasars rest-frame times. These light curves were then used to measure the variance using the Mexican Hat method at the same four timescales as was done with the real data. We then averaged the logarithm of the resulting variances to obtain the `folded model' variance for each $M-$\redd\  bin and timescale. 

The figure of merit, $fom$, recorded for each set of parameters was 
\begin{equation}
    fom=\sum_{bins,timescales}\frac{(\tilde{var}-var)^2}{err^2},
    \label{eq:fom}
\end{equation} where  $\tilde{var}$ is the folded model for one $M-$\redd\  bin and timescale and var is the variance for the corresponding bin and timescale of the real data. The errors considered, $err$, are the errors on $var$ of the real data, obtained through bootstrapping as described above.

\section{PDS Model comparison}\label{sec:comparison}
Optical light curves of quasars are often modelled as a DRW \citep[e.g][]{Kelly09,Kozlowski10,MacLeod10,Burke21}, which produces a power spectrum slope of 0 below a break frequency and a slope of --2 above. To compare with previous works we include a  model with a similar shape, i.e. $\alpha_L=0$ and $\alpha_H=-2$.  \citet{Guo2017} modelled the scatter in observed DRW parameters of SDSS Stripe 82 quasars by allowing a low-frequency slope that is steeper than the DRW model. They conclude that this effect indeed can explain the observed scatter and that the steepest the low-frequency slope can be is $\alpha_L=-1.3$, so we will include also steeper values of $\alpha_L=-1$ and $\alpha_L=-0.5$, i.e. within the acceptable range according to that study. Additionally, as in Fig.\ref{fig:PDS_all} the measured high-frequency power spectra appear steeper than the DRW model plotted in black, we will also use steeper high-frequency slopes $\alpha_H=-2.5$ and $\alpha_H=-3$. 

We initially explored a grid in the parameters with $B=0.002-0.02$ in multiplicative steps of 1.5, $C=-0.8- -0.3$ in steps of 0.1, $D=-0.6- 0.0$ in steps of 0.1 and $F=-1.0- 0.0$ in steps of 0.05, creating 100 simulated light curves per $M-$\redd\  bin to produce the folded model. If the minimum $fom$ fell on a limiting value of one or more parameters, the fit was repeated with a wider parameter range. Table \ref{tab:first_run} shows the best $fom$ obtained for each pair of slopes, together with the best-fitting values of all the parameters.

\begin{table}[]
    \centering
    \begin{tabular}{|c c |c c c c c|c|}
    \hline
$\alpha_L$ & $\alpha_H$ & $A2$ & $B$ & $C$ &$D$ & $F$ & $fom/N$  \\
\hline      
     \hline
    0     &-2  & 15 & 0.0011 & -0.7  & 0.0 & -0.8  & 7.7\\ 
    -0.5  &-2  & 28 & 0.0008 & -0.6 & -0.1 & -0.65 & 7.8\\ 
    -1    &-2  & 160 & 0.0003 & -0.7 & -0.1& -0.7  & 7.8 \\ 
    \hline
    0     &-2.5& 1.9  & 0.0045  &  -0.6 & -0.3 & -0.45 &2.3\\ 
    -0.5  &-2.5& 1.9 & 0.0045 & -0.6  & -0.2 & -0.55 & 2.0\\ 
    -1    &-2.5& 0.83 & 0.0068 &  -0.6 & -0.4 & -0.35 &2.7 \\ 
    \hline
    0     & -3 & 1.1  & 0.0068 &  -0.5 &-0.2 & -0.5 & 4.3 \\ 
    -0.5  & -3 & 1.0 & 0.0068 & -0.5  & -0.2 & -0.5 & 3.3 \\ 
    -1    & -3 & 0.48 & 0.0100 &  -0.6 & -0.2 & -0.55 & 2.1 \\ 
         \hline
    \end{tabular}
    \caption{Best-fitting parameters and corresponding figure of merit for the 9 power spectrum models tested, with low and high-frequency slopes given in the first two columns. The parameters are explained in Eqs. Eqs. \ref{eq:PDS},\ref{eq:fb},\ref{eq:A},\ref{eq:A2} and $N$ is the number of power spectrum points, i.e. the number of timescales times the number of $M-$\redd\  bins. First, coarser grid run.}
    \label{tab:first_run}
\end{table}

Considering the low resolution of the grid, the best-fitting parameters are fairly consistent among the different models, except for the bend frequency $B$, which is higher for the models with steeper high-frequency slopes, and consequently, the normalisation at the bend $A2$ differs.  Although there is not much difference between the $fom$ values between models with equal $\alpha_L$, there is a significant difference between models with different $\alpha_H$, where $\alpha_H=-2.5$ and $\alpha_H=-3$ fit consistently better than $\alpha_H=-2$, so we will not consider the latter models further. 

A second round of simulations was done for the better-fitting models: all three models with $\alpha_H=-2.5$ and $\alpha_L=-0.5, 0$ for $\alpha_H=-3$. These fits consisted of significantly more iterations, with 300 simulations per $M-$\redd\  bin and per set of parameters, with steps of 1.2 in $B$ and 0.05 in $C,D$ and $F$. The best-fitting values for all parameters and the minimum $fom$ are presented in Table \ref{tab:second_run}. Although models with $fom$ values around 2 do not appear to be very good representations of the data, they are sufficiently good if the errors on $M$ and \redd\ are considered, as discussed below in Sec. \ref{sec:fom}. Therefore, we did not introduce further parameters to the model.

\begin{table}[]
    \centering
    \begin{tabular}{|c c |c c c c c|c|}
    \hline
$\alpha_L$ & $\alpha_H$ & $A2$ & $B$ & $C$ &$D$ & $F$ & $fom/N$  \\
\hline      
    \hline
    0     &-2.5& 1.5  & 0.0052  &  -0.65 & -0.30 & -0.50 & 2.0\\ 
    -0.5  &-2.5& 1.4 & 0.0052 & -0.60  & -0.30 & -0.45 & 1.9\\ 
    -1    &-2.5& 1.1 & 0.0058 &  -0.60 & -0.35 & -0.40 &2.3 \\ 
    \hline
    -0.5  & -3 & 0.71 & 0.0086 & -0.60  & -0.35 & -0.40 & 2.4 \\ 
    -1    & -3 & 0.42 & 0.0104 &  -0.55 & -0.35 & -0.40 & 1.7 \\ 
         \hline
    \end{tabular}
    \caption{Best-fitting parameters and corresponding figure of merit for the 5 better-fitting power spectrum models, with low and high-frequency slopes given in the first two columns. The parameters are explained in Eqs. \ref{eq:PDS},\ref{eq:fb},\ref{eq:A},\ref{eq:A2} and $N$ is the number of power spectral points, i.e. the number of timescales times the number of $M-$\redd\  bins. Second, finer grid run.}
    \label{tab:second_run}
\end{table}

The minimum $fom/N$ as a function of the bend frequency $B$ is shown in Fig. \ref{fig:chi_B}, i.e. for each value of $B$, the $fom/N$ was marginalised over all the other parameters. The different panels show the results for the 5 different power spectrum models shown in Table \ref{tab:second_run}. 
Evidently, the bend frequency shows a well-defined minimum in $fom/N$, within the accessible frequency range. The best-fitting bend frequencies are in the range $f_b=0.005-0.01$ days$^{-1}$, for $M=10^8M_\odot$ and \redd = 0.1, being higher for steeper power law models. Similar plots for the other parameters are shown in Figs. \ref{fig:chi_C}, \ref{fig:chi_D}, \ref{fig:chi_F}. Although these show shallower minima, they do show a constrained value different from 0 for all parameters. 

The best-fitting values for the mass dependence of the bend frequency, $f_b\propto M^{C}$ range from $C=-0.65$ for $\alpha_L=0$, $\alpha_H=-2.5$ to $C=-0.55$ for $\alpha_L=-1$, $\alpha_H=-3$, with $C=-0.6$ for the power spectrum models with intermediate slopes. The best-fitting value for the dependence of the bend frequency on \redd , $f_b\propto $ \redd $^{D}$ are either $D=-0.3$ or $D=-0.35$ for all models. Upper and lower limits for all the parameters, calculated as explained below in Sec. \ref{sec:errors}, are listed in Table \ref{tab:error_bounds}. Notably, all models produce consistent values of $C$, in the sense that all their uncertainty ranges overlap, and also of $D$ while, for each model, $C$ and $D$ are significantly different, i.e. the scaling of the bend frequency with mass is different to its scaling on \redd . Finally, the scaling of the normalisation on \redd , $A\propto$ \redd $^F$, has best-fitting values of $F=-0.4$ to $F=-0.5$ and the uncertainty ranges for this parameter overlap for all models.

Figure \ref{fig:chi_DF} shows the minimum $fom/N$ for fixed values of $D$ on differently coloured lines, as a function of $F$. We recall that $D$ determines the dependence of the bend frequency $f_b$ on \redd\ as $f_b\propto$\redd $^D$ and F determines the dependence of the normalisation of the power spectrum on \redd\ as $A\propto$ \redd $^F$. Since both parameters relate to the dependence of the power spectrum on \redd\ they are partially degenerate, which produces the flat profiles of $fom/N$ vs $D$ and also $fom$ vs $F$ to a lesser extent. It is clear, however, that extreme values of $D$ produce overall worse fits, for any value of F, and $D=0$ (i.e. no dependence of $f_b$ on \redd ) produces significantly worse fits. 

\begin{figure*}
    \centering
    \includegraphics[width=0.7\textwidth]{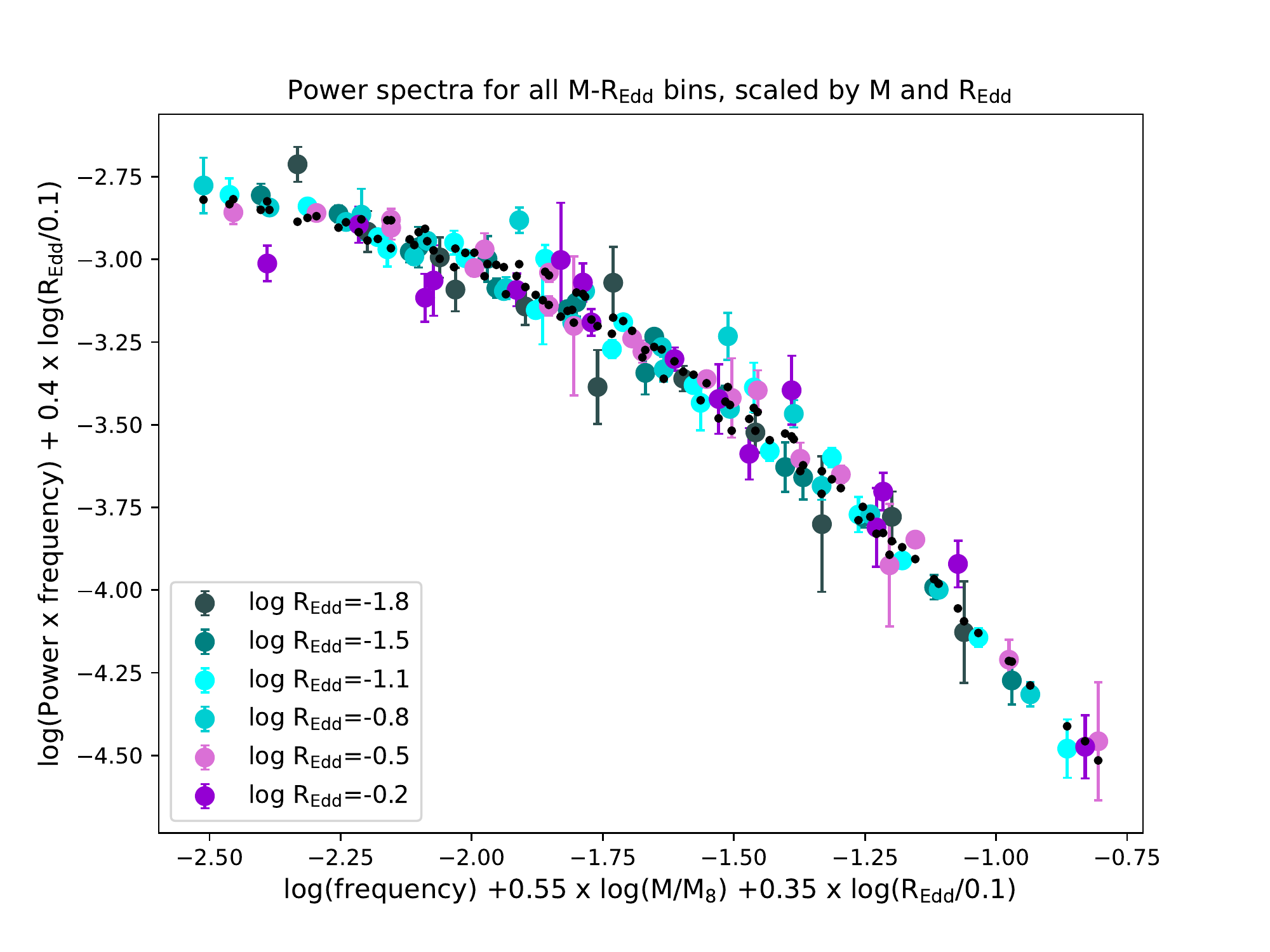}
     \caption{The large dots, colour-coded by \redd\ represent the power spectra of the data (plotted as Power $\times$ frequency), after the frequencies and amplitudes have been re-scaled according to the mass and \redd\ of each bin and the best-fitting scaling parameters C=-0.55, D=-0.35,  F=-0.4. The black dots represent a 'folded' model constructed from a bending power law model with $\alpha_L=-1$ and $\alpha_H=-3$.  Frequencies are in units of day$^{-1}$.  }
    \label{fig:PDS_bestfit}
\end{figure*}

\begin{figure}
    \centering
    \includegraphics[width=0.55\textwidth]{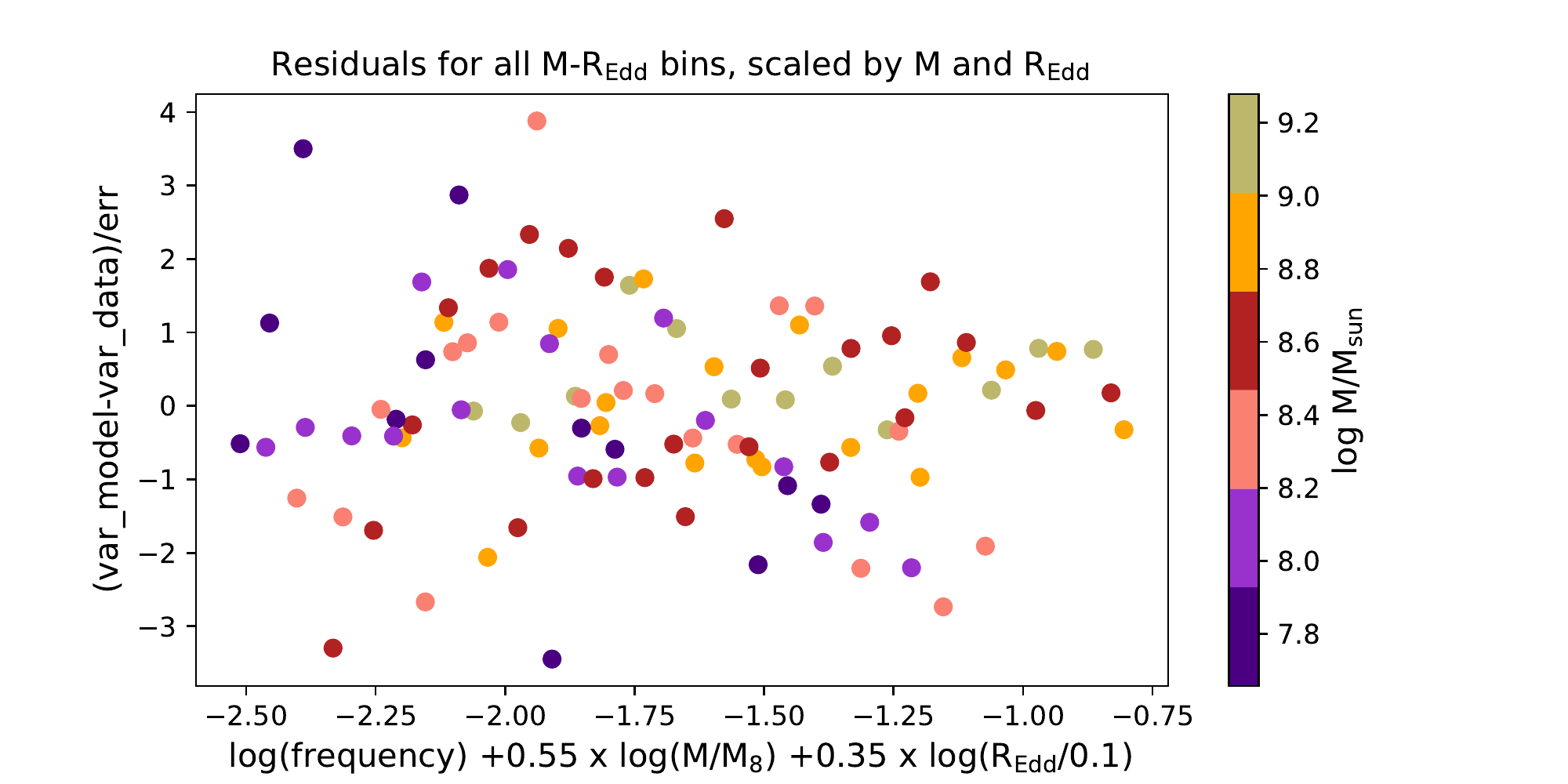}
    \caption{Residuals of the model and data plotted in Fig.\ref{fig:PDS_bestfit}, color coded by mass.  Frequencies are in units of day$^{-1}$.}
    \label{fig:residuals}
\end{figure}

The best-fitting values of parameters $C,D$ and $F$ can be used to scale the frequency and variance of the power spectra of all $M-$\redd\  bins to that of the fiducial bin, i.e. $M=10^8 M_\odot$, \redd =$0.1$. This scaling results in the power spectrum shown in large dots in Fig. \ref{fig:PDS_bestfit}. One of the best-fitting `folded' models, with $\alpha_L=-1$ and $\alpha_L=-3$, is over-plotted in small black dots. The model follows the data quite closely and shows some scatter, produced by the sampling of the light curves and the stochastic nature of the `folding' process. Figure \ref{fig:residuals} shows the difference between the model and data points, divided by the error on the data points, colour-coded by mass. Although some bins have a large deviation from the model, they are not preferentially of one \redd\ or mass, as can be seen in the colour codes of both figures.  

\subsection{Error bounds on the fitted parameters}
\label{sec:errors}
To estimate the confidence intervals of the parameters we start by considering the parameter values where $fom = fom_{\rm min}+\Delta$, where $\Delta$ is set to produce a given confidence interval assuming a $\chi^2$ distribution with 4 degrees of freedom since 4 parameters are being searched simultaneously over a grid.
For 90\% confidence intervals, $\Delta=7.8$ 
which equates to an increase in $fom/N$ of 7.8/104=0.075. 

\begin{figure}
    \centering
    \includegraphics[width=0.5\textwidth]{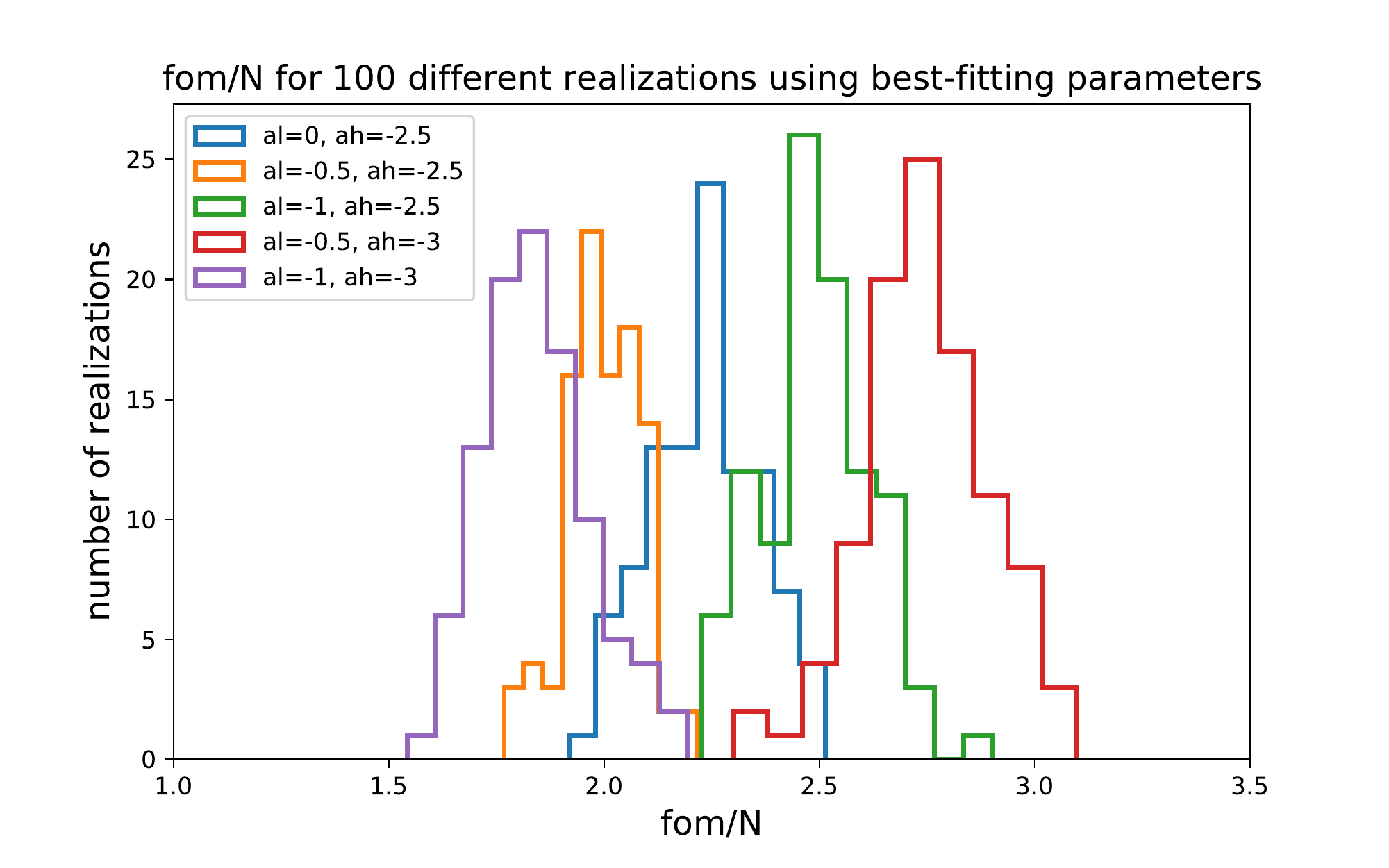}
    \caption{Distributions of values of $fom/N$ obtained for 100 different realisations of each model, using only the best-fitting parameters in each case. The different models are characterised by the low and high-frequency slopes of their power spectra and these are written in the legend. }
    \label{fig:chi2s_realizations}
\end{figure}

This value of $\Delta$ is too small however, given that the folded model is calculated through a stochastic process that produces a larger uncertainty on the $fom_{\rm min}$ of each realisation of the fitting routine. In other words, different realisations of the same model calculation, with the same model parameters result in slightly different values of the $fom/N$, with a scatter larger than this $\Delta=0.075$. In Fig.\ref{fig:chi2s_realizations} we show the distribution of values of $fom/N$ obtained for the best-fitting parameters of each model in Table \ref{tab:second_run} for 100 realisations of each. For each realisation, 300 lightcurves were generated for each $M-$\redd\ bin, to make the $fom$ values comparable to the second round of simulations. Evidently, each model has a distribution of values of $fom/N$ for the best-fitting parameters, with a standard deviation of 0.1--0.2. Therefore we will consider as acceptable values all parameter combinations that return a $fom/N$ equal or lower than the 90\% upper bound of this distribution of $fom/N$, for each model. The upper 90\% bounds of $fom/N$ are listed in the fourth column in Table \ref{tab:fom_other} and represent an addition of 0.2 to 0.6 to $fom_{min}/N$. These limit values of $fom/N$ are marked by the grey horizontal lines in Figs. \ref{fig:chi_B}, \ref{fig:chi_C}, \ref{fig:chi_D} and \ref{fig:chi_F}.

\begin{table*}[]
    \centering
    \begin{tabular}{|c c|c c c c c c c c|}
    \hline  
    $\alpha_L$ & $\alpha_H$ & $B_l$ &$B_h$ & $C_l$ & $C_h$ & $D_l$ & $D_h$ & $F_l$ & $F_h$\\
    \hline
    
    \hline
    0     &-2.5& 0.0035 & 0.0055 & -0.7 & -0.53 & -0.46 & -0.12 & -0.68 & -0.3 \\ 
    -0.5  &-2.5& 0.0042 & 0.0058 & -0.68 & -0.54 & -0.36 & -0.16 & -0.62& -0.42 \\
    -1    &-2.5& 0.0044 &0.0081 &-0.71 &-0.51 & -0.47 & -0.02 & -0.76 &-0.31 \\ 
    \hline
    -0.5  & -3 & 0.007 & 0.0097 & -0.63 & -0.47 & -0.4 & -0.19 & -0.56 & -0.32 \\ 
    -1    & -3 & 0.0095 & 0.0131 & -0.67 & -0.5 & -0.41 & -0.12 & -0.64 & -0.32 \\ 
        \hline
    \end{tabular}
    \caption{Lower and upper limits for the power spectrum model parameters, for each combination of low and high-frequency slopes written in the first two columns. The break frequencies $B$ are in units of days$^{-1}$.}
    \label{tab:error_bounds}
\end{table*}

\begin{table}[]
    \centering
    \begin{tabular}{|c c|c c c c |}
    \hline  
    $\alpha_L$ & $\alpha_H$ &$fom/N$& $fom_{90}/N$ &$fom_{\rm scatter}/N$&$fom_{\rm ind}/N$\\
    \hline
    
    \hline
    0     &-2.5& 2.0&2.4& 1.8& 1.4\\
    -0.5  &-2.5& 1.9&2.1&1.6 & 1.1\\
    -1    &-2.5& 2.3&2.7&1.7 & 0.95\\
    \hline
    -0.5  & -3 & 2.4&3.0& 2.1& 1.4\\
    -1    & -3 & 1.7&2.0&1.3 &0.8 \\
        \hline
    \end{tabular}
    \caption{Figure of merit for each model in Table \ref{tab:second_run}, $fom/N$ corresponds to the best fitting model, where errors in the denominator only consider the scatter in variances in the data bins;  $fom_{90}/N$ are the upper 90\% bounds of the distribution of $fom/N$ values for different realisations of each model with its best-fitting parameters, described in Sec. \ref{sec:errors};  $fom_{\rm scatter}/N$ and $fom_{\rm ind}/N$ are re-scaled versions of $fom/N$, incorporating also the errors on mass and \redd, first using only the scatter of these parameters within each bin ($fom_{\rm scatter}/N$) and then propagating the errors on individual measurements of both values ($fom_{\rm ind}/N$) as described in Sec. \ref{sec:fom}.}
    \label{tab:fom_other}
\end{table}

\subsection{Note on the figure of merit}
\label{sec:fom}
The $fom$ has been calculated using only the uncertainty in the variance of the binned data, not the uncertainty on the $M$ or \redd\ of the data bins. The latter errors should not be included for fitting purposes because their propagation to the error on the variance is model-dependent, so including them could bias the fit towards parameter values where the dependence of the variance on mass and \redd\ is larger and therefore the errors are also larger and the $fom$ becomes smaller. We note that all the fitting described above, as well as all the errors used for the figures, represent only the scatter on the measured variances in each data bin. 

The error on the mass and \redd\ of the bins does affect the figure of merit since they will add to the difference between the model and data points but are not considered in the errors in the denominator in Eq. \ref{eq:fom}. To estimate how much the $fom$ could improve if these errors were considered, we took the best-fitting parameters for each model, propagated the error on mass and \redd\ measured from the scatter of these values in the bin, to an error on the model variance and combined these errors quadratically with the error on the variance from the data. Replacing these combined errors in the denominator of equation \ref{eq:fom} reduced the figures of merit from the values in Table \ref{tab:second_run} to the values listed in the fifth column in Table \ref{tab:fom_other} ($fom_{\rm scatter}/N$).

Alternatively, we can use the uncertainties in the values of $M$ and \redd\ for individual objects to estimate the errors on $M$ and \redd\ of each bin, not just the scatter in reported values as above. We assumed a systematic error of 0.3 dex in the single epoch mass estimate and an error of 50\% in \redd\ as noted by \citet{Rakshit20} to be relevant errors for their catalogue.
We calculated $dM=M/\sqrt{N_{lc}}$ and $dR_{Redd}=0.5 R_{Redd}/\sqrt{N_{lc}}$, where $N_{lc}$ is the number of light curves in the relevant $M-$\redd\  bin and $dM$ and $dR_{Redd}$ will be the errors in the mass and \redd\ of the bin, respectively. 
Propagating these errors and then combining them with the uncertainty in the variance of each bin produced the values listed in the sixth column in Table \ref{tab:fom_other} ($fom_{\rm ind}/N$).
These errors on mass and \redd\ of the individual measurements in the catalogue can therefore explain the remaining difference between the model and data points. 

\section{Discussion}
\label{sec:discussion}
The data presented above show that the optical power spectra, at around 2900 \AA\ in the rest-frame, conform with a universal bending power-law spectrum, where the bend frequency scales inversely with both $M$ and \redd\ and the normalisation scales inversely with \redd . Comparing the dependence of the power spectra on \redd\ (Fig.\ref{fig:PDS_M}) to the dependence of the power spectra on mass (Fig.\ref{fig:PDS_REdd}) we see their very distinct effects: the \redd\ mostly changes the normalisation while the mass has the largest effect on the location of the break in the power spectrum. As a result, the mass determines which part of the power spectrum is being sampled by the light curves, resulting in different slopes visible in the frequency range probed. An immediate consequence of this is that the power spectrum cannot be simply a function of luminosity, which is the product of the two. For example, a low-mass, high-accretion rate object will have a flatter power spectrum with lower normalisation than a high-mass, low-accretion rate object of the same luminosity. Below we discuss the implications of these results in terms of the accretion disc structure and other considerations for comparing these results to other data sets. 

\subsection{Interpretation of the scaling relations}
The orbital timescale at a given location in the disc, for example at the ISCO, scales linearly with black hole mass. Re-scaling the break timescales by mass, so that they can be interpreted in terms of the orbital timescales at a given radius in units of $R_g$, results in $t_b/t_{ISCO}\propto M^{0.55}$\redd$^{0.35}/M = ($\redd$/M)^{0.35} M^{-0.1}$, where we have used the best fitting parameters for the model with $\alpha_L=-1$ and $\alpha_H=-3$. Alternatively, a value of $C=-0.65$ is allowed within the uncertainty range for the same model. With this value, the relation becomes $t_b/t_{ISCO}\propto M^{0.65}$\redd$^{0.35}/M= ($\redd$/M)^{0.35}$, i.e. a bend timescale that scales linearly with the orbital timescale of the innermost orbit times a function of the combination \redd$/M$. In the standard thin disc model \citep{Shakura73} the maximal surface temperature $T$ is proportional to (\redd$/M)^{0.25}$, suggesting a connection between the characteristic timescale and the disc temperature. 

The size of the emitting region of a given wavelength $\lambda^*$ can be estimated by the largest radius in the disc where the temperature is above $T^*=0.29/\lambda^*$ with $\lambda^*$ in cm and $T$ in K. The orbital timescale at this radius is $t_{orb}^*=2 \pi r^{*3/2} R_g/c$ or equivalently $t_{orb}^*=r^{*3/2} t_{ISCO}$, where the dimensionless radius $r=R/R_g$ and $R_g=GM/c^2$. Assuming a radial temperature dependence of $T(r)=T_{max}r^{-3/4}$ gives  
\begin{equation}
    t_{orb}^*=(T^*/T_{max})^{3/2\times -4/3} t_{ISCO}= (T^*/T_{max})^{-2} t_{ISCO}.
\end{equation}
If $T_{max} \propto ($\redd$/M)^{0.25}$ then 
\begin{equation}
\label{eq:t_orb}
t_{orb}^* \propto T^{*-2} ({\rm R_{Edd}} /M)^{0.5} t_{ISCO} \propto \lambda^{*2} ({\rm R_{Edd}} /M)^{0.5} t_{ISCO}.
\end{equation} 
This relation between the orbital timescale at the outer edge of the optical emitting region, $t_{orb}^*$, and $M$ and \redd\ is slightly steeper than the relation we find between these quantities and the power spectrum bend timescale $t_b\propto $\redd$^{-D}M^{-C-1}t_{ISCO}=($\redd$/M)^{0.35} M^{-0.1} t_{ISCO}$. 

We note that re-replacing $t_{ISCO}$ by $M$ in Eq. \ref{eq:t_orb} results in $t_{orb}^* \propto \lambda^{*2} ({\rm R_{Edd}} M)^{0.5}\propto \lambda^{*2} L_{\rm bol}^{0.5}$, as calculated for example by \citet{MacLeod10}. Our fits to the data can be expressed in terms of $L_{\rm bol} \propto M \times $\redd\ as $t_b\propto $ \redd $^{-D}M^{-C}\propto L_{\rm bol}^{-D} M^{D-C}=L_{\rm bol}^{0.35} M^{-0.2}$ or $t_b\propto L_{\rm bol}^{0.55}$ \redd $^{0.2}$, i.e. the difference between $C$ and $D$ implies that the bend timescale cannot depend on $L_{bol}$ alone. We note that although both parameters $C$ and $D$ have associated errors, the allowed ranges for $C$ and $D$ values do not overlap, for any of the fitted models, as shown in Table \ref{tab:error_bounds}.

Returning to the standard thin disc model, if we characterise the optical emitting region by a light-weighted radius, rather than the outermost radius, the relation between orbital timescale and mass and \redd\ changes somewhat. We used thin disc model prescriptions for the emitted flux as a function of mass, \redd\ and radius to obtain the black body temperature as a function of radius, $T(r)$.  We start from the flux emitted by a thin accretion disc \citep[e.g. eq. 23 in][]{Spruit1995}:
\begin{equation}
\sigma T^4 =F_{\rm acc}= \frac{3GM\dot M f}{8\pi R^3}
\label{eq:Facc}
\end{equation}
and assume that the mass accretion rate $\dot M$ in cgs units corresponds to: 
\begin{equation}
\dot M=\frac{L}{\eta c^2}= \frac{L_{\rm Edd} R_{\rm Edd}}{\eta c^2}=\frac{1.26\times10^{38} m R_{\rm Edd}}{\eta c^2}. 
\end{equation}
Replacing this value in equation \ref{eq:Facc} to obtain a relation in terms of the mass in solar units $m=M/M_\odot$ and the Eddington ratio \redd=$L/L_{\rm Edd}$ gives:
\begin{eqnarray}
\sigma T^4&=&\frac{3GM f}{8 \pi R_g^3 r^3  } \frac{1.26\times10^{38} m R_{\rm Edd}}{\eta  c^2}\\
\sigma T^4&=&\frac{3 f 1.26\times10^{38} m R_{\rm Edd}}{8 \pi R_g^2 r^3  \eta } =\frac{3 f 5.6\times 10^{27} R_{\rm Edd}}{8 \pi r^3  \eta m}\\
T(r)&=&\left(\frac{1.2\times 10^{31} f}{\eta r^3  }\frac{R_{\rm Edd}}{m }\right)^{0.25} \rm K
\end{eqnarray}
where $r=R/R_g$, $R_g=GM/c^2=1.5\times 10^5 m$ cm and $\sigma$ is the Stefan Boltzmann constant. Then the energy flux per unit frequency interval emitted by (both sides of) a ring of area $dA=2\pi  r dr R_g^2$ at a given frequency $\nu$ is 
\begin{equation}
dL = dA \times \frac{ 4\pi h \nu^3}{c^2(e^{h\nu/kT(r)}-1)}.
\end{equation}
These annular fluxes were used to calculate a light-weighted average radius as
\begin{equation}
r_{lw}= \frac{\int_{r_{min}}^{r_{out}} rdL}{\int_{r_{min}}^{r_{out}}dL}=\frac{\int_{r_{min}}^{r_{out}} (e^{h\nu/kT(r)}-1)^{-1}r^2dr}{\int_{r_{min}}^{r_{out}} (e^{h\nu/kT(r)}-1)^{-1}rdr}.
\end{equation}
For these calculations we assumed $f=1-\sqrt{r_{min}/r}$, $r_{min}=6$ and $\eta=0.1$. The outer radius of the emission region, in units of the gravitational radius $R_g$, was set at $r_{out}=3000$.

\begin{figure}
    \centering
    \includegraphics[width=0.5\textwidth]{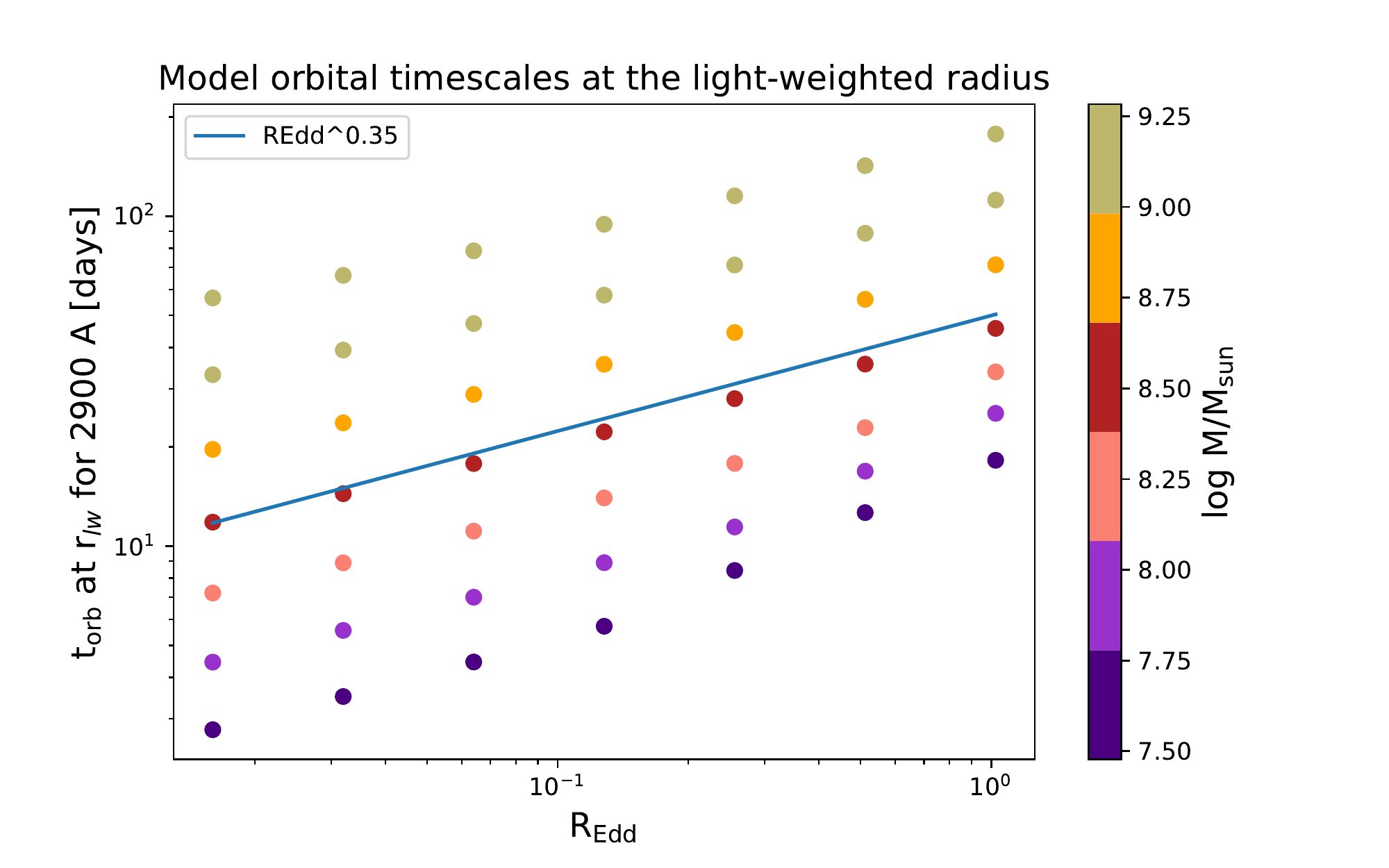}
    \includegraphics[width=0.5\textwidth]{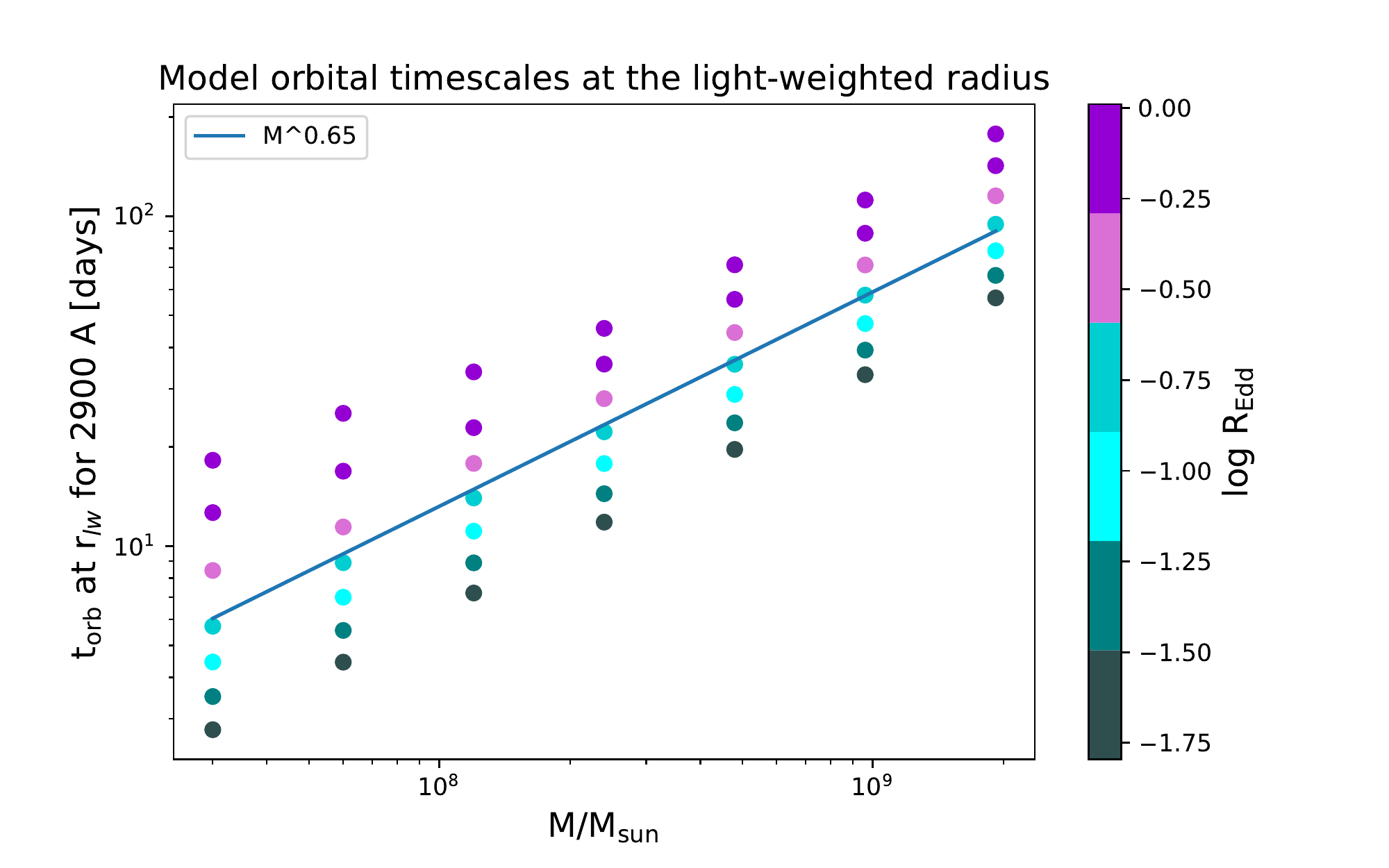}
    \caption{Standard thin disc model predictions for the orbital timescale at the light-weighted radius for emission at 2900\AA , as a function of black hole mass and \redd . The top panel shows the timescales as a function of $M$, color coded by \redd , together with a solid line displaying a $t_{orb}\propto M^{0.65}$ relation. The bottom panel shows the same model points as a function of \redd , color coded by mass and the solid line displays a $t_{orb}\propto$\redd$^{0.35}$ relation.}
    \label{fig:model}
\end{figure}

For a given mass, the relation between the light-weighted radius $r_{lw}$ and \redd\ is slightly flatter than the relation of the outermost radius and \redd . In Fig.~\ref{fig:model} we show the orbital timescale, i.e. $t_{orb}=2\pi r_{lw}^{3/2} R_g/c$, with $r_{lw}$ in units of $R_g$, as a function of both mass and \redd , calculated for parameter ranges similar to those of our data.  For a given \redd , or colour in the top panel in Fig.~\ref{fig:model}, the relation between orbital timescale and mass follows closely a $t_{orb}\propto M^{0.65}$ relation, plotted as a blue solid line, and different masses show approximately parallel relations. The bottom panel displays the relation between $t_{orb}$ and \redd\ which, for a given mass, closely follows a $t_{orb} \propto $ \redd$^{0.35}$ relation, plotted in the solid line, with approximately parallel relations for other masses. Therefore, the orbital timescale at a light-weighted radius for the optical emitting region scales with $M$ and \redd\ in approximately the same way as the bend timescale. 

As can be seen in Fig. \ref{fig:model} the orbital timescale at the light-weighted radius of 2900 \AA , for the standard thin disc, and for $M$ $=10^8$M$_{\odot}$ and \redd = 0.1 is about 10 days. The bend timescale $t_b=1/f_b$ that we find for these black hole parameters is between 90 and 200 days, depending on the power spectrum model. The ratio between the bend and orbital timescales is therefore 9 to 20. As pointed out before by several authors \citep[e.g.][]{Kelly09,MacLeod10,Burke21}, the bend timescale could correspond to the thermal timescale of a thin disc, which is expected to be around 10 times the orbital timescale, independently of the accretion rate and most disc parameters \citep[e.g][and references therein]{Spruit1995}. 

We note however that this modelling with a standard thin disc produces a bolometric correction factor $k_{5100}=L_{bol}/L_{5100}$ that depends on the disc temperature, and therefore also on the ratio \redd $/M$. The values of \redd\ used for the fitting have been calculated with a constant bolometric correction factor $k_{5100}=9.26$ \citep{Rakshit20}. In Appendix \ref{sec:thindisc} we explore the effect of correcting the values of \redd\ to the values predicted by the thin disc model, given the $L_{5100}$ and mass of each quasar, to produce a more self-consistent model comparison. The result is that with this correction, the relation between $f_b$ and mass is stronger ($f_b\propto M^{-0.66}$ instead of $f_b\propto M^{-0.55}$) and the relation of $f_b$ to the "true" \redd\ is weaker, but still of the same sign, i.e. $f_b\propto$ \redd$ _{\rm model}^{-0.25}$), for the model with $\alpha_L=-1$ and $\alpha_H=-3$, still fairly consistent with the model scalings described above. 

Apart from the scaling of the bend frequency, we also find that the normalisation of the power spectrum at the break scales with \redd\  approximately as $A \propto $ \redd $^{-0.45}$. In the interpretation described above this would mean that smaller optical emitting regions characteristic of lower \redd\ sources and therefore also lower temperature discs, are not only characterised by shorter timescales but also by larger amplitude of fluctuations. This scaling might help in constraining accretion disc models that incorporate prescriptions for its variations in luminosity such as the inhomogeneous disc model with thermal perturbations in clumps proposed by \citet{Dexter2011} and expanded by \citet{Cai2016} or the Corona Heated Accretion Disc model of \citet{Sun2020}, perhaps by having fewer independently-varying regions that emit in the observed wavelength for cooler discs.   

Incidentally, relating the bend frequency to the size of the emitting region implies that the bend frequency should be wavelength-dependent, where shorter wavelengths, being produced by a smaller region, would have higher bend frequencies.  
\citet{Stone2023_corr} recently found a relation between rest-frame wavelength and the de-correlation timescale of the DRW model, $\tau_{DRW}$, but with a weaker dependence of $\tau_{DRW}\propto\lambda^{0.4+/-0.1} $, from very long (20 years) light curves and DRW modelling of 190 quasars. As they argue, this weaker dependence of timescale with $\lambda$ can be expected, for example, for a radially-averaged thermal timescale, as described for the model of \citet{Sun2020}.  

Finally, we note that, as has been shown in every power spectrum analysis before \citep[e.g][]{deVries05, Caplar17,Burke21,Stone2022}, there is significant variability power on timescales much shorter than the bend timescale. In other words, the variability is not simply concentrated on one characteristic timescale but, instead, there is variability on a continuum of timescales, with smaller amplitudes for increasingly shorter timescales. In the interpretation described above, this could be achieved if all the optical emitting regions produced variations on their local thermal timescales. In this case, the smaller radii in the disc would produce shorter timescale fluctuations and would modulate a smaller fraction of the 2900 \AA\ emission, both by having a smaller area and by having a higher temperature, so that a larger fraction of their luminosity is emitted at shorter wavelengths than observed.\footnote{For radii $r$ where the observed wavelength is already in the Rayleigh-Jeans tail of the black body spectrum, the emitted flux per unit area scales as $F_\lambda \propto T$. If $T\propto r^{-3/4}$ and the annulus area scales as $r^2$ then the luminosity emitted by each annulus in this wavelength is $L_\lambda \propto  r^{-3/4} r^2=r^{5/4}$, i.e. it is smaller for more central annuli.}

\subsection{Other implications}
We note that, in our modelling, the bend frequency of the power spectrum scales both with mass and with \redd\ but the dependence on \redd\ is about half as strong as the dependence on mass. Therefore, for smaller or more heterogeneous samples, the dependence on \redd\ might not be noticeable. One effect of not accounting for this dependence is that it will `leak' into the dependence of $f_b$ on mass, for example, if mass and \redd\ are anti-correlated in the sample. In the most extreme case, where objects in a sample have $M\propto$\redd $^{-1}$ then, for example, $f_b\propto M^{-0.6}$ \redd $^{-0.3}$ would result in $f_b\propto M^{-0.6}M^{+0.3}=M^{-0.3}$, flattening the dependence on mass. In less well anti-correlated samples, such as the one used in this paper, the effect would be weaker but still present. 

The best fitting parameters and $fom$ obtained by fixing $D=0$ (i.e. neglecting any dependence of the bend frequency on \redd ) are summarised in Table \ref{tab:second_run_D0}.  Comparing Table \ref{tab:second_run_D0} to Table \ref{tab:second_run}, where $D$ was a free parameter exemplifies the effect described in the previous paragraph, where neglecting the dependence of $f_b$ on \redd\ results in a shallower dependence of $f_b$ on mass, i.e. a smaller value of $|C|$.  We note also that all these models fit the data significantly worse than models with free $D$. However, these fits are useful to show the overall dependence of variance as a function of accretion rate. In all models, this relation results in either $P\times f \propto$\redd $^{-0.75}$ or $P\times f \propto$\redd $^{-0.7}$. Although part of this dependence results from the scaling of power spectrum normalisation with \redd\ and the rest from the dependence of the bend frequency on \redd, this relation is useful to predict the variance expected for equal mass and different accretion rates, at least at timescales shorter than the bend in the power spectrum.

\begin{table}[]
    \centering
    \begin{tabular}{|c c |c c c c c|c|}
    \hline
$\alpha_L$ & $\alpha_H$ & $A2$ & $B$ & $C$ & $D$ & $F$ & $fom$  \\
\hline      
    \hline
    0     &-2.5& 2.2  & 0.0041  &  -0.60 & 0.0 & -0.75 & 4.5 \\ %
    -0.5  &-2.5& 2.8 &0.0036 &-0.55 &0.0 &-0.75& 2.7\\ 
    -1    &-2.5& 1.2 &0.0058 &-0.60 &0.0 &-0.75 &2.8 \\ 
    \hline
    -0.5  & -3 & 0.92 &0.0072 &-0.50 &0.0 &-0.7 &4.5\\ 
    -1    & -3 & 0.57 &0.0086 &-0.50 &0.0 &-0.7 &3.1 \\ 
         \hline
    \end{tabular}
    \caption{Best-fitting parameters and corresponding figure of merit for the 5 better-fitting power spectrum models, with low and high-frequency slopes given in the first two columns, when the bend frequency is allowed to depend only on mass and not on \redd\ (i.e. when parameter $D$ is fixed to $D=0$).}
    \label{tab:second_run_D0}
\end{table}

\subsection{The shape of the Power Spectrum}
The universal power spectrum model fits our data well, particularly for a high-frequency slope of $\alpha_H=-2.5$, which allows low-frequency slopes of $\alpha_L=-0.5$ or 0, or for $\alpha_H=-3$ and $\alpha_L=-1$, all with similar goodness of fit. Models with a flatter high-frequency slope of $\alpha_H=-2$, similar to the DRW model, resulted in significantly worse fits.  This last result is consistent with previous findings by \citet{Caplar17} who used lightcurves from the Palomar Transient Factory (PTF/iPTF) surveys as well as \citet{Stone2022} using light curves from SDSS, PanSTARRS-1 and the Dark Energy Survey and \citet{Mushotzky2011} using Kepler light curves, among others, all of which showed high-frequency power spectra that are steeper than $\alpha_H=-2$. Regarding the low-frequency slope, a value of $\alpha_L=-1$ was preferred in our analysis only for the model with $\alpha_H=-3$, while for $\alpha_H=-2.5$ any low-frequency slope (i.e. $\alpha_L=0, -0.5, -1$) was acceptable. This result probably relates to the location of the model bend, which is at a higher frequency in the case of $\alpha_H=-3$ and therefore the low frequency slope is better constrained in this model. We note that a similarly steep low-frequency slope of $\alpha_L=-1$ was found by \citet{Simm16} from Pan-STARRS-1 data of 90 quasars using CARMA \citep{Kelly14} modelling. 

All our models with $\alpha_H=-2.5$ gave bend frequencies with overlapping confidence ranges, and the models with $\alpha_H=-3$ resulted in significantly higher bend frequencies. This result might indicate that a continuously bending power-law is a better description of the true power spectrum than a single-bend power-law model since steeper slopes fit better with bends at higher frequencies.  

\section{Conclusion}
\label{sec:conclusion}
We present a power spectral analysis of optical light curves taken in the $g$-band by the ZTF project using 5433 light curves for 4770 individual quasars at 0.6<z<0.7. This narrow redshift range allows us to probe the dependence of the power spectral parameters on mass and \redd\ without having to correct for uncertain dependencies of the power spectrum on the rest-frame wavelength of emission. The light curves have a median length of 1547 days and a median of 233 good data points per light curve. The masses and \redd\ of all the quasars used in this study were obtained by \citet{Rakshit20} by modelling SDSS spectra homogeneously.

We used the Mexican hat method \citep{Arevalo12} to estimate the variance of the light curves at 4 different timescales, in the range of 30 to 300 days in the rest-frame of the quasars (already presented in \citealt{Arevalo23}). With these variance estimates, we construct low-resolution power spectra for 26 bins in a grid of mass and \redd , with $7.5 <\log{M/M_\odot}<9.5$ and $-2 <$ \redd\ $<0$. Each bin had a minimum of 14 lightcurves. The power spectra can be approximately described by a power-law shape with a negative slope. When separated in groups of similar \redd\ the power spectra show a clear dependence on mass, where higher mass objects show steeper power spectra and the spacing between high-frequency log(power) grows approximately linearly with the separation in log(M). When split in groups of similar mass, the power spectra of different \redd\ appear almost parallel, with higher normalisation for lower \redd . In this case, too the spacing between the logarithm of the power spectra scales approximately linearly with the separation in log(\redd). There is still a small dependence of the slope on \redd\ but this is less marked than the dependence on mass.

The power spectrum models tested were smoothly bending power-law models, characterised by a bend frequency $f_b$, low- and high-frequency power-law slopes $\alpha_L$ and $\alpha_H$ and a normalisation. The $f_b$ was allowed to depend on the mass and \redd\ and the normalisation on \redd, based on the findings described in the previous paragraph. We note that the power spectrum model used had the same variance at the bend frequency for objects of different mass and equal \redd . For low-frequency slopes $\alpha_L\neq -1$ this normalisation implies that the power at 0 frequency is different for objects of different masses. The best-fitting bend timescales, for $M=10^8M_\odot$ and \redd = 0.1, range from $96^{+9}_{-20}$ days for the model with $\alpha_H=-3$ and $\alpha_L=-1$ to $192^{+46}_{-10}$ days for the model with $\alpha_H=-2.5$ and $\alpha_L=-0.5$. The latter model is the closest to the DRW used in \cite{Burke21} and for this model, our bend timescales are consistent with their results. The best fitting slope pairs were $\alpha_H=-3$, $\alpha_L=-1$ and $\alpha_H=-2.5$ with either $\alpha_L=-0$ or $\alpha_L=-0.5$.  DRW-type high-frequency slopes $\alpha_H=-2$ gave significantly worse fits. 

We find, for the first time, a dependence of the break frequency on \redd\ in the optical power spectra of quasars, with the bend frequency scaling as $f_b \propto M^{-0.67- -0.5}\times $\redd$^{-0.41- -0.12}$, where the ranges in the exponent values correspond to a power-spectral model with $\alpha_L=-1$ and $\alpha_H=-3$, but they are similar to the ranges of the other well-fitting models. As the exponents for the dependence on mass and on \redd\  are different, the bend timescale cannot be simply a function of bolometric luminosity, which is the product of both parameters. Also as the exponent of the dependence on \redd\ is significantly different from 0, the bend timescale is not a function of mass alone. Interestingly, the higher \redd\ objects show less high-frequency power, this sense is opposite to the behaviour of the X-ray power spectra of quasars, where higher accretion rate objects have more high-frequency variability \citep[e.g.][and references therein]{McHardy06}. This can be explained if the bend timescale corresponds to the thermal timescale of the light-weighted radius of the emitting region of the observed wavelength, in the context of the standard thin disc model. In this case, for a given mass and different accretion rates, the characteristic radius increases with temperature and therefore with accretion rate and so the relevant timescale becomes longer. A simple calculation with thin disc prescriptions shows that this light-weighted radius for emission at 2900 \AA\ has an orbital timescale that scales with $M^{0.65}$ and \redd$^{0.35}$, similar to the scaling we find for $t_b=1/f_b$, at least for masses and accretion rates in the same range as the data. 

The dependence of $f_b$ on mass goes from $f_b\propto M^{-0.65}$ to $ f_b\propto M^{-0.55}$ depending on the power spectrum model and is slightly steeper than that previously published by \citet{Burke21}. This might be partly due to the dependence of $f_b$ on \redd\ and the anti-correlation between mass and \redd\ that usually exists in flux-limited quasar samples which, combined, results in a flatter mass dependence if the dependence of $f_b$ on \redd\ is not included. Alternatively, a flatter mass dependence could result from a positive relation between bend timescale and wavelength. High-mass objects are preferentially found at higher redshifts, so their rest-frame wavelength is shorter than for lower-mass objects in the same sample observed with the same filter. If the bend timescale indeed scales with wavelength as predicted by the thin disc model, then the high mass, high redshift objects would appear to have shorter bend timescales than if they were observed in the same rest-frame wavelength, reducing the difference with the low mass, low redshift objects. 

Measuring the wavelength dependence of the bend timescale is challenging since the effect is subtle, the uncertainties on the bend timescale are large, and the dependence on the other black hole parameters must be controlled in a sufficiently accurate manner. Such a correlation, however, has been found using the characteristic timescale of variability measured with the Structure Function, by \citet[][and references therein]{MacLeod10} where the reported dependence is very weak. A similar analysis but estimating the characteristic timescale with the DRW model ($\tau_{\rm DRW}$) and limiting the study to objects whose $\tau_{\rm DRW}$  is much shorter than the length of the light curves was presented recently by \citet{Stone2022}. This later work found a stronger correlation with rest-frame wavelength but still weaker than predicted by the thin disc model. Long-term, well-sampled light curves of many more objects, in multiple optical bands and to high depths, as will be provided in the future by the Vera Rubin LSST \citep{Ivezic2019} will likely shed more light on this issue. 

It is also possible that not only the characteristic timescale of variability depends on wavelength but that the slope of the power spectrum might depend on this parameter as well. Results in this direction will be presented in the future by Patel et al. (in prep.) who calculate the ratio of variances on different timescales (i.e. the power spectrum slope) as a function of rest-frame wavelength for objects of equal mass and \redd . In addition, a replication of the work presented in this paper but using the other available ZTF band ($r$) will show whether the fitted power spectrum parameters depend strongly on wavelength.  

{\bf Acknowledgements:} The authors would like to thank Prof. H. Netzer for providing his model results. We also thank the anonymous referee for their useful comments and suggestions. The authors acknowledge financial support from the National Agency for Research and Development (ANID) grants: Millennium Science Initiative Program ICN12\_12009 (PSS,LHG), and NCN$19\_058$ (PA, PL, EL, SB); FONDECYT Regular 1201748 (PL); FONDECYT Postdoctorado 3200250 (PSS); Programa de Becas/Doctorado Nacional 21212344 (SB), 21200718 (EL) and 21222298 (PP), and from the Max-Planck Society through a Partner Group grant (PA, SB).

Based on observations obtained with the Samuel Oschin Telescope 48-inch and the 60-inch Telescope at the Palomar
Observatory as part of the Zwicky Transient Facility project. ZTF is supported by the National Science Foundation under Grant
No. AST-2034437 and a collaboration including Caltech, IPAC, the Weizmann Institute for Science, the Oskar Klein Center at
Stockholm University, the University of Maryland, Deutsches Elektronen-Synchrotron and Humboldt University, the TANGO
Consortium of Taiwan, the University of Wisconsin at Milwaukee, Trinity College Dublin, Lawrence Livermore National
Laboratories, and IN2P3, France. Operations are conducted by COO, IPAC, and UW.
\\

%
\bibliographystyle{aa} 
\bibliography{bibliography.bib} 

\begin{thebibliography}{46}
\expandafter\ifx\csname natexlab\endcsname\relax\def\natexlab#1{#1}\fi

\bibitem[{{Ai} {et~al.}(2010){Ai}, {Yuan}, {Zhou}, {Wang}, {Dong}, {Wang}, \& {Lu}}]{Ai2010}
{Ai}, Y.~L., {Yuan}, W., {Zhou}, H.~Y., {et~al.} 2010, \apjl, 716, L31

\bibitem[{{Ar{\'e}valo} {et~al.}(2012){Ar{\'e}valo}, {Churazov}, {Zhuravleva}, {Hern{\'a}ndez-Monteagudo}, \& {Revnivtsev}}]{Arevalo12}
{Ar{\'e}valo}, P., {Churazov}, E., {Zhuravleva}, I., {Hern{\'a}ndez-Monteagudo}, C., \& {Revnivtsev}, M. 2012, \mnras, 426, 1793

\bibitem[{{Ar{\'e}valo} {et~al.}(2023){Ar{\'e}valo}, {Lira}, {S{\'a}nchez-S{\'a}ez}, {Patel}, {L{\'o}pez-Navas}, {Churazov}, \& {Hern{\'a}ndez-Garc{\'\i}a}}]{Arevalo23}
{Ar{\'e}valo}, P., {Lira}, P., {S{\'a}nchez-S{\'a}ez}, P., {et~al.} 2023, arXiv e-prints, arXiv:2304.14228

\bibitem[{{Ar{\'e}valo} {et~al.}(2008){Ar{\'e}valo}, {Uttley}, {Kaspi}, {Breedt}, {Lira}, \& {McHardy}}]{Arevalo08}
{Ar{\'e}valo}, P., {Uttley}, P., {Kaspi}, S., {et~al.} 2008, \mnras, 389, 1479

\bibitem[{{Bauer} {et~al.}(2009){Bauer}, {Baltay}, {Coppi}, {Ellman}, {Jerke}, {Rabinowitz}, \& {Scalzo}}]{Bauer09}
{Bauer}, A., {Baltay}, C., {Coppi}, P., {et~al.} 2009, \apj, 696, 1241

\bibitem[{{Burke} {et~al.}(2021){Burke}, {Shen}, {Blaes}, {Gammie}, {Horne}, {Jiang}, {Liu}, {McHardy}, {Morgan}, {Scaringi}, \& {Yang}}]{Burke21}
{Burke}, C.~J., {Shen}, Y., {Blaes}, O., {et~al.} 2021, Science, 373, 789

\bibitem[{{Cackett} {et~al.}(2021){Cackett}, {Bentz}, \& {Kara}}]{Cackett2021}
{Cackett}, E.~M., {Bentz}, M.~C., \& {Kara}, E. 2021, iScience, 24, 102557

\bibitem[{{Cai} {et~al.}(2016){Cai}, {Wang}, {Gu}, {Sun}, {Wu}, {Huang}, \& {Chen}}]{Cai2016}
{Cai}, Z.-Y., {Wang}, J.-X., {Gu}, W.-M., {et~al.} 2016, \apj, 826, 7

\bibitem[{{Caplar} {et~al.}(2017){Caplar}, {Lilly}, \& {Trakhtenbrot}}]{Caplar17}
{Caplar}, N., {Lilly}, S.~J., \& {Trakhtenbrot}, B. 2017, \apj, 834, 111

\bibitem[{{de Vries} {et~al.}(2005){de Vries}, {Becker}, {White}, \& {Loomis}}]{deVries05}
{de Vries}, W.~H., {Becker}, R.~H., {White}, R.~L., \& {Loomis}, C. 2005, \aj, 129, 615

\bibitem[{{Dexter} \& {Agol}(2011)}]{Dexter2011}
{Dexter}, J. \& {Agol}, E. 2011, \apjl, 727, L24

\bibitem[{{Edelson} {et~al.}(2015){Edelson}, {Gelbord}, {Horne}, {McHardy}, {Peterson}, {Ar{\'e}valo}, {Breeveld}, {De Rosa}, {Evans}, {Goad}, {Kriss}, {Brandt}, {Gehrels}, {Grupe}, {Kennea}, {Kochanek}, {Nousek}, {Papadakis}, {Siegel}, {Starkey}, {Uttley}, {Vaughan}, {Young}, {Barth}, {Bentz}, {Brewer}, {Crenshaw}, {Dalla Bont{\`a}}, {De Lorenzo-C{\'a}ceres}, {Denney}, {Dietrich}, {Ely}, {Fausnaugh}, {Grier}, {Hall}, {Kaastra}, {Kelly}, {Korista}, {Lira}, {Mathur}, {Netzer}, {Pancoast}, {Pei}, {Pogge}, {Schimoia}, {Treu}, {Vestergaard}, {Villforth}, {Yan}, \& {Zu}}]{Edelson15}
{Edelson}, R., {Gelbord}, J.~M., {Horne}, K., {et~al.} 2015, \apj, 806, 129

\bibitem[{{Guo} {et~al.}(2017){Guo}, {Wang}, {Cai}, \& {Sun}}]{Guo2017}
{Guo}, H., {Wang}, J., {Cai}, Z., \& {Sun}, M. 2017, \apj, 847, 132

\bibitem[{{Ivezi{\'c}} {et~al.}(2019){Ivezi{\'c}}, {Kahn}, {Tyson}, {Abel}, {Acosta}, {Allsman}, {Alonso}, {AlSayyad}, {Anderson}, {Andrew}, {Angel}, {Angeli}, {Ansari}, {Antilogus}, {Araujo}, {Armstrong}, {Arndt}, {Astier}, {Aubourg}, {Auza}, {Axelrod}, {Bard}, {Barr}, {Barrau}, {Bartlett}, {Bauer}, {Bauman}, {Baumont}, {Bechtol}, {Bechtol}, {Becker}, {Becla}, {Beldica}, {Bellavia}, {Bianco}, {Biswas}, {Blanc}, {Blazek}, {Blandford}, {Bloom}, {Bogart}, {Bond}, {Booth}, {Borgland}, {Borne}, {Bosch}, {Boutigny}, {Brackett}, {Bradshaw}, {Brandt}, {Brown}, {Bullock}, {Burchat}, {Burke}, {Cagnoli}, {Calabrese}, {Callahan}, {Callen}, {Carlin}, {Carlson}, {Chandrasekharan}, {Charles-Emerson}, {Chesley}, {Cheu}, {Chiang}, {Chiang}, {Chirino}, {Chow}, {Ciardi}, {Claver}, {Cohen-Tanugi}, {Cockrum}, {Coles}, {Connolly}, {Cook}, {Cooray}, {Covey}, {Cribbs}, {Cui}, {Cutri}, {Daly}, {Daniel}, {Daruich}, {Daubard}, {Daues}, {Dawson}, {Delgado}, {Dellapenna}, {de Peyster}, {de Val-Borro}, {Digel}, {Doherty}, {Dubois},
  {Dubois-Felsmann}, {Durech}, {Economou}, {Eifler}, {Eracleous}, {Emmons}, {Fausti Neto}, {Ferguson}, {Figueroa}, {Fisher-Levine}, {Focke}, {Foss}, {Frank}, {Freemon}, {Gangler}, {Gawiser}, {Geary}, {Gee}, {Geha}, {Gessner}, {Gibson}, {Gilmore}, {Glanzman}, {Glick}, {Goldina}, {Goldstein}, {Goodenow}, {Graham}, {Gressler}, {Gris}, {Guy}, {Guyonnet}, {Haller}, {Harris}, {Hascall}, {Haupt}, {Hernandez}, {Herrmann}, {Hileman}, {Hoblitt}, {Hodgson}, {Hogan}, {Howard}, {Huang}, {Huffer}, {Ingraham}, {Innes}, {Jacoby}, {Jain}, {Jammes}, {Jee}, {Jenness}, {Jernigan}, {Jevremovi{\'c}}, {Johns}, {Johnson}, {Johnson}, {Jones}, {Juramy-Gilles}, {Juri{\'c}}, {Kalirai}, {Kallivayalil}, {Kalmbach}, {Kantor}, {Karst}, {Kasliwal}, {Kelly}, {Kessler}, {Kinnison}, {Kirkby}, {Knox}, {Kotov}, {Krabbendam}, {Krughoff}, {Kub{\'a}nek}, {Kuczewski}, {Kulkarni}, {Ku}, {Kurita}, {Lage}, {Lambert}, {Lange}, {Langton}, {Le Guillou}, {Levine}, {Liang}, {Lim}, {Lintott}, {Long}, {Lopez}, {Lotz}, {Lupton}, {Lust}, {MacArthur}, {Mahabal},
  {Mandelbaum}, {Markiewicz}, {Marsh}, {Marshall}, {Marshall}, {May}, {McKercher}, {McQueen}, {Meyers}, {Migliore}, {Miller}, {Mills}, {Miraval}, {Moeyens}, {Moolekamp}, {Monet}, {Moniez}, {Monkewitz}, {Montgomery}, {Morrison}, {Mueller}, {Muller}, {Mu{\~n}oz Arancibia}, {Neill}, {Newbry}, {Nief}, {Nomerotski}, {Nordby}, {O'Connor}, {Oliver}, {Olivier}, {Olsen}, {O'Mullane}, {Ortiz}, {Osier}, {Owen}, {Pain}, {Palecek}, {Parejko}, {Parsons}, {Pease}, {Peterson}, {Peterson}, {Petravick}, {Libby Petrick}, {Petry}, {Pierfederici}, {Pietrowicz}, {Pike}, {Pinto}, {Plante}, {Plate}, {Plutchak}, {Price}, {Prouza}, {Radeka}, {Rajagopal}, {Rasmussen}, {Regnault}, {Reil}, {Reiss}, {Reuter}, {Ridgway}, {Riot}, {Ritz}, {Robinson}, {Roby}, {Roodman}, {Rosing}, {Roucelle}, {Rumore}, {Russo}, {Saha}, {Sassolas}, {Schalk}, {Schellart}, {Schindler}, {Schmidt}, {Schneider}, {Schneider}, {Schoening}, {Schumacher}, {Schwamb}, {Sebag}, {Selvy}, {Sembroski}, {Seppala}, {Serio}, {Serrano}, {Shaw}, {Shipsey}, {Sick}, {Silvestri},
  {Slater}, {Smith}, {Smith}, {Sobhani}, {Soldahl}, {Storrie-Lombardi}, {Stover}, {Strauss}, {Street}, {Stubbs}, {Sullivan}, {Sweeney}, {Swinbank}, {Szalay}, {Takacs}, {Tether}, {Thaler}, {Thayer}, {Thomas}, {Thornton}, {Thukral}, {Tice}, {Trilling}, {Turri}, {Van Berg}, {Vanden Berk}, {Vetter}, {Virieux}, {Vucina}, {Wahl}, {Walkowicz}, {Walsh}, {Walter}, {Wang}, {Wang}, {Warner}, {Wiecha}, {Willman}, {Winters}, {Wittman}, {Wolff}, {Wood-Vasey}, {Wu}, {Xin}, {Yoachim}, \& {Zhan}}]{Ivezic2019}
{Ivezi{\'c}}, {\v{Z}}., {Kahn}, S.~M., {Tyson}, J.~A., {et~al.} 2019, \apj, 873, 111

\bibitem[{{Kasliwal} {et~al.}(2015){Kasliwal}, {Vogeley}, \& {Richards}}]{Kasliwal15}
{Kasliwal}, V.~P., {Vogeley}, M.~S., \& {Richards}, G.~T. 2015, \mnras, 451, 4328

\bibitem[{{Kelly} {et~al.}(2009){Kelly}, {Bechtold}, \& {Siemiginowska}}]{Kelly09}
{Kelly}, B.~C., {Bechtold}, J., \& {Siemiginowska}, A. 2009, \apj, 698, 895

\bibitem[{{Kelly} {et~al.}(2014){Kelly}, {Becker}, {Sobolewska}, {Siemiginowska}, \& {Uttley}}]{Kelly14}
{Kelly}, B.~C., {Becker}, A.~C., {Sobolewska}, M., {Siemiginowska}, A., \& {Uttley}, P. 2014, \apj, 788, 33

\bibitem[{{Koz{\l}owski}(2017)}]{Kozlowski17}
{Koz{\l}owski}, S. 2017, \aap, 597, A128

\bibitem[{{Koz{\l}owski} {et~al.}(2010){Koz{\l}owski}, {Kochanek}, {Udalski}, {Wyrzykowski}, {Soszy{\'n}ski}, {Szyma{\'n}ski}, {Kubiak}, {Pietrzy{\'n}ski}, {Szewczyk}, {Ulaczyk}, {Poleski}, \& {OGLE Collaboration}}]{Kozlowski10}
{Koz{\l}owski}, S., {Kochanek}, C.~S., {Udalski}, A., {et~al.} 2010, \apj, 708, 927

\bibitem[{{Krolik} {et~al.}(1991){Krolik}, {Horne}, {Kallman}, {Malkan}, {Edelson}, \& {Kriss}}]{Krolik91}
{Krolik}, J.~H., {Horne}, K., {Kallman}, T.~R., {et~al.} 1991, \apj, 371, 541

\bibitem[{{Li} {et~al.}(2018){Li}, {McGreer}, {Wu}, {Fan}, \& {Yang}}]{Li18}
{Li}, Z., {McGreer}, I.~D., {Wu}, X.-B., {Fan}, X., \& {Yang}, Q. 2018, \apj, 861, 6

\bibitem[{{Lira} {et~al.}(2015){Lira}, {Ar{\'e}valo}, {Uttley}, {McHardy}, \& {Videla}}]{Lira15}
{Lira}, P., {Ar{\'e}valo}, P., {Uttley}, P., {McHardy}, I.~M.~M., \& {Videla}, L. 2015, \mnras, 454, 368

\bibitem[{{Luo} {et~al.}(2020){Luo}, {Shen}, \& {Yang}}]{Luo20}
{Luo}, Y., {Shen}, Y., \& {Yang}, Q. 2020, \mnras, 494, 3686

\bibitem[{{MacLeod} {et~al.}(2010){MacLeod}, {Ivezi{\'c}}, {Kochanek}, {Koz{\l}owski}, {Kelly}, {Bullock}, {Kimball}, {Sesar}, {Westman}, {Brooks}, {Gibson}, {Becker}, \& {de Vries}}]{MacLeod10}
{MacLeod}, C.~L., {Ivezi{\'c}}, {\v Z}., {Kochanek}, C.~S., {et~al.} 2010, \apj, 721, 1014

\bibitem[{{MacLeod} {et~al.}(2012){MacLeod}, {Ivezi{\'c}}, {Sesar}, {de Vries}, {Kochanek}, {Kelly}, {Becker}, {Lupton}, {Hall}, {Richards}, {Anderson}, \& {Schneider}}]{MacLeod2012}
{MacLeod}, C.~L., {Ivezi{\'c}}, {\v{Z}}., {Sesar}, B., {et~al.} 2012, \apj, 753, 106

\bibitem[{{Masci} {et~al.}(2019){Masci}, {Laher}, {Rusholme}, {Shupe}, {Groom}, {Surace}, {Jackson}, {Monkewitz}, {Beck}, {Flynn}, {Terek}, {Landry}, {Hacopians}, {Desai}, {Howell}, {Brooke}, {Imel}, {Wachter}, {Ye}, {Lin}, {Cenko}, {Cunningham}, {Rebbapragada}, {Bue}, {Miller}, {Mahabal}, {Bellm}, {Patterson}, {Juri{\'c}}, {Golkhou}, {Ofek}, {Walters}, {Graham}, {Kasliwal}, {Dekany}, {Kupfer}, {Burdge}, {Cannella}, {Barlow}, {Van Sistine}, {Giomi}, {Fremling}, {Blagorodnova}, {Levitan}, {Riddle}, {Smith}, {Helou}, {Prince}, \& {Kulkarni}}]{Masci19}
{Masci}, F.~J., {Laher}, R.~R., {Rusholme}, B., {et~al.} 2019, \pasp, 131, 018003

\bibitem[{{McHardy} {et~al.}(2006){McHardy}, {Koerding}, {Knigge}, {Uttley}, \& {Fender}}]{McHardy06}
{McHardy}, I.~M., {Koerding}, E., {Knigge}, C., {Uttley}, P., \& {Fender}, R.~P. 2006, \nat, 444, 730

\bibitem[{{McHardy} {et~al.}(2004){McHardy}, {Papadakis}, {Uttley}, {Page}, \& {Mason}}]{McHardy2004}
{McHardy}, I.~M., {Papadakis}, I.~E., {Uttley}, P., {Page}, M.~J., \& {Mason}, K.~O. 2004, \mnras, 348, 783

\bibitem[{{Mushotzky} {et~al.}(2011){Mushotzky}, {Edelson}, {Baumgartner}, \& {Gandhi}}]{Mushotzky2011}
{Mushotzky}, R.~F., {Edelson}, R., {Baumgartner}, W., \& {Gandhi}, P. 2011, \apjl, 743, L12

\bibitem[{{Netzer}(2019)}]{Netzer2019}
{Netzer}, H. 2019, \mnras, 488, 5185

\bibitem[{{Rakshit} {et~al.}(2020){Rakshit}, {Stalin}, \& {Kotilainen}}]{Rakshit20}
{Rakshit}, S., {Stalin}, C.~S., \& {Kotilainen}, J. 2020, \apjs, 249, 17

\bibitem[{{S{\'a}nchez-S{\'a}ez} {et~al.}(2018){S{\'a}nchez-S{\'a}ez}, {Lira}, {Mej{\'{\i}}a-Restrepo}, {Ho}, {Ar{\'e}valo}, {Kim}, {Cartier}, \& {Coppi}}]{Sanchez-Saez18}
{S{\'a}nchez-S{\'a}ez}, P., {Lira}, P., {Mej{\'{\i}}a-Restrepo}, J., {et~al.} 2018, \apj, 864, 87

\bibitem[{{Shakura} \& {Sunyaev}(1973)}]{Shakura73}
{Shakura}, N.~I. \& {Sunyaev}, R.~A. 1973, \aap, 24, 337

\bibitem[{{Simm} {et~al.}(2016){Simm}, {Salvato}, {Saglia}, {Ponti}, {Lanzuisi}, {Trakhtenbrot}, {Nandra}, \& {Bender}}]{Simm16}
{Simm}, T., {Salvato}, M., {Saglia}, R., {et~al.} 2016, \aap, 585, A129

\bibitem[{{Smith} {et~al.}(2018){Smith}, {Mushotzky}, {Boyd}, {Malkan}, {Howell}, \& {Gelino}}]{Smith18}
{Smith}, K.~L., {Mushotzky}, R.~F., {Boyd}, P.~T., {et~al.} 2018, \apj, 857, 141

\bibitem[{{Spruit}(1995)}]{Spruit1995}
{Spruit}, H.~C. 1995, in NATO Advanced Study Institute (ASI) Series C, Vol. 450, The Lives of the Neutron Stars, ed. M.~A. {Alpar}, U.~{Kiziloglu}, \& J.~{van Paradijs}, 355--376

\bibitem[{{Stone} {et~al.}(2022){Stone}, {Shen}, {Burke}, {Chen}, {Yang}, {Liu}, {Gruendl}, {Adam{\'o}w}, {Andrade-Oliveira}, {Annis}, {Bacon}, {Bertin}, {Bocquet}, {Brooks}, {Burke}, {Carnero Rosell}, {Carrasco Kind}, {Carretero}, {da Costa}, {Pereira}, {De Vicente}, {Desai}, {Diehl}, {Doel}, {Ferrero}, {Friedel}, {Frieman}, {Garc{\'\i}a-Bellido}, {Gaztanaga}, {Gruen}, {Gutierrez}, {Hinton}, {Hollowood}, {Honscheid}, {James}, {Kuehn}, {Kuropatkin}, {Lidman}, {Maia}, {Menanteau}, {Miquel}, {Morgan}, {Paz-Chinch{\'o}n}, {Pieres}, {Plazas Malag{\'o}n}, {Rodriguez-Monroy}, {Sanchez}, {Scarpine}, {Serrano}, {Sevilla-Noarbe}, {Smith}, {Suchyta}, {Swanson}, {Tarl{\'e}}, {To}, \& {DES Collaboration}}]{Stone2022}
{Stone}, Z., {Shen}, Y., {Burke}, C.~J., {et~al.} 2022, \mnras, 514, 164

\bibitem[{{Stone} {et~al.}(2023){Stone}, {Shen}, {Burke}, {Chen}, {Yang}, {Liu}, {Gruendl}, {Adam{\'o}w}, {Andrade-Oliveira}, {Annis}, {Bacon}, {Bertin}, {Bocquet}, {Brooks}, {Burke}, {Carnero Rosell}, {Carrasco Kind}, {Carretero}, {da Costa}, {Pereira}, {De Vicente}, {Desai}, {Diehl}, {Doel}, {Ferrero}, {Friedel}, {Frieman}, {Garc{\'\i}a-Bellido}, {Gaztanaga}, {Gruen}, {Gutierrez}, {Hinton}, {Hollowood}, {Honscheid}, {James}, {Kuehn}, {Kuropatkin}, {Lidman}, {Maia}, {Menanteau}, {Miquel}, {Morgan}, {Paz-Chinch{\'o}n}, {Pieres}, {Plazas Malag{\'o}n}, {Rodriguez-Monroy}, {Sanchez}, {Scarpine}, {Serrano}, {Sevilla-Noarbe}, {Smith}, {Suchyta}, {Swanson}, {Tarl{\'e}}, {To}, \& {DES Collaboration}}]{Stone2023_corr}
{Stone}, Z., {Shen}, Y., {Burke}, C.~J., {et~al.} 2023, \mnras, 521, 836

\bibitem[{{Suberlak} {et~al.}(2021){Suberlak}, {Ivezi{\'c}}, \& {MacLeod}}]{Suberlak2021}
{Suberlak}, K.~L., {Ivezi{\'c}}, {\v{Z}}., \& {MacLeod}, C. 2021, \apj, 907, 96

\bibitem[{{Sun} {et~al.}(2020){Sun}, {Xue}, {Brandt}, {Gu}, {Trump}, {Cai}, {He}, {Lin}, {Liu}, \& {Wang}}]{Sun2020}
{Sun}, M., {Xue}, Y., {Brandt}, W.~N., {et~al.} 2020, \apj, 891, 178

\bibitem[{{Sun} {et~al.}(2014){Sun}, {Wang}, {Chen}, \& {Zheng}}]{Sun2014}
{Sun}, Y.-H., {Wang}, J.-X., {Chen}, X.-Y., \& {Zheng}, Z.-Y. 2014, \apj, 792, 54

\bibitem[{{Tachibana} {et~al.}(2020){Tachibana}, {Graham}, {Kawai}, {Djorgovski}, {Drake}, {Mahabal}, \& {Stern}}]{Tachibana20}
{Tachibana}, Y., {Graham}, M.~J., {Kawai}, N., {et~al.} 2020, \apj, 903, 54

\bibitem[{{Tang} {et~al.}(2023){Tang}, {Wolf}, \& {Tonry}}]{Tang2023}
{Tang}, J.-J., {Wolf}, C., \& {Tonry}, J. 2023, Nature Astronomy, 7, 473

\bibitem[{{Timmer} \& {Koenig}(1995)}]{Timmer95}
{Timmer}, J. \& {Koenig}, M. 1995, \aap, 300, 707

\bibitem[{{Xin} {et~al.}(2020){Xin}, {Charisi}, {Haiman}, \& {Schiminovich}}]{Xin2020}
{Xin}, C., {Charisi}, M., {Haiman}, Z., \& {Schiminovich}, D. 2020, \mnras, 495, 1403

\bibitem[{{Zu} {et~al.}(2013){Zu}, {Kochanek}, {Koz{\l}owski}, \& {Udalski}}]{Zu13}
{Zu}, Y., {Kochanek}, C.~S., {Koz{\l}owski}, S., \& {Udalski}, A. 2013, \apj, 765, 106

\end{thebibliography}
%

\onecolumn

\appendix
\section{Re-calculating \redd\ according to the standard thin disc model}
\label{sec:thindisc}
According to the standard thin disc model, the surface temperature of the disc is a function of mass and \redd\ so that the luminosity at a fixed wavelength represents a different fraction of the total disc luminosity depending on these parameters. The bolometric correction used by \citet{Rakshit20} to calculate the \redd\ in their catalogue has a fixed value and therefore will over or underestimate the \redd\ for colder and hotter discs, respectively, according to this model. In particular, for the redshift range used in our work, the bolometric luminosity is estimated from the monochromatic luminosity at 5100 \AA\ and a bolometric correction factor $k_{5100}=9.26$, i.e $R_{\rm Edd,cat} =L_{5100}\times k_{5100}/L_{\rm Edd}$. In this section we will use the subscript cat to refer to the \redd\ obtained from the catalogue, which is the one used throughout this paper, to distinguish it from $R_{Edd,model}$, i.e. the Eddington ratio used as input in the disc model.

The dependence of the bolometric correction factor $k_{5100}$ on mass, accretion rate and black hole spin is described in detail by \citet{Netzer2019}. For a simple analysis, we fit their results for a spin parameter $a=0.7$,  kindly provided by Dr. Netzer. This value of $a$ produces values of $k_{5100, {\rm Netzer}}$ approximately half way between $a=0.998$ and $a=0, -1$. With these we can convert the \redd\ values as they would appear in the catalogue of \citet{Rakshit20} to the \redd\ values of the model as:

\begin{equation}
R_{\rm Edd,model}= \frac{L_{5100} \times k_{5100, {\rm Netzer}}}{L_{\rm Edd}} = R_{\rm Edd,cat} \frac{k_{5100, {\rm Netzer}}(M,R_{\rm Edd,cat})}{k_{5100}}.
\end{equation}

we note that for the fits we will evaluate the model values of $k_{5100, {\rm Netzer}}$ as a function of $R_{\rm Edd,cat}$, not of $R_{\rm Edd,model}$, in order to replace the values of $R_{\rm Edd,cat}$ in the power spectral fitting functions. To do this, we multiplied the luminosity at 5100 \AA\ (i.e. the x-axis in the third panel in figure 2 of \citealt{Netzer2019}) by $k_{5100}/(1.26\times10^{38}\times M)$ and fitted $k_{5100, {\rm Netzer}}$ as a function of this new variable. The best-fitting linear models to $\log(R_{\rm Edd,cat})$ vs $\log(k_{5100, {\rm Netzer}})$, for a spin parameter $a=0.7$ were:
\begin{eqnarray}
    \log(k_{5100, {\rm Netzer}})=0.39\log(R_{\rm Edd,cat})+1.3;& M =10^9M_\odot \\
    \log(k_{5100, {\rm Netzer}})=0.43\log(R_{\rm Edd,cat})+1.5;& M =10^8M_\odot \\
    \log(k_{5100, {\rm Netzer}})=0.49\log(R_{\rm Edd,cat})+2.1;& M =10^7M_\odot 
\end{eqnarray}
To simplify the analysis, we will take the relation for $M=10^8M\odot$ for the slope and scale the normalisation by mass. Taking the values of $k_{5100, {\rm Netzer}}$ at a single value of $R_{\rm Edd,cat}=0.1$ we obtain the relation between the bolometric correction and mass: $\log(k_{5100, {\rm Netzer}})= -0.38\log(M/M_8)+1.1$. Putting both together we obtain the approximate relation 
\begin{equation}
   \log(k_{5100, {\rm Netzer}})=0.4 \log(R_{\rm Edd,cat})-0.4\log(M/M_8)+1.5 
\end{equation}
and therefore the relation between  $R_{\rm Edd,model}$ and $R_{\rm Edd,cat}$ is 
\begin{eqnarray*}
    \log(R_{\rm Edd,cat})&=&\frac{\log(R_{\rm Edd,model})+0.4 \log(M/M_8)-1.5+\log(k_{5100})}{1.4}\\    
    \log(R_{\rm Edd,cat})&=&0.7\log(R_{\rm Edd,model})+0.3\log(M/M_8)-0.38
\end{eqnarray*}

Recalling the dependence of $f_b$ on these parameters and replacing $R_{\rm Edd,cat}$ gives:
 \begin{eqnarray*}
\log{f_b} &=& \log{B} + C \times\log(M/M_8)+D \times\log(R_{\rm Edd,cat}/0.1) \\
\log{f_b} &=& \log{B} + C \times\log(M/M_8) +D \times 0.7\log(R_{\rm Edd,model}) +D \times (0.3\log(M/M_8)-0.38-1) \\
\log{f_b} &=& \log{B} + (C+0.3D) \times\log(M/M_8) +0.7D \times\log R_{\rm Edd,model} -1.38D \\
\log{f_b} &=& \log{B} -0.655 \times\log(M/M_8) -0.245 \times\log R_{\rm Edd,model} +0.48 
\end{eqnarray*}
where in the last equation we have replaced $C$ and $D$ by their best fitting values, -0.55 and -0.35, respectively. Therefore, including this correction of the accretion rate, from the constant bolometric correction used in the catalogue to the value predicted for the measured $L_{5100}$ and mass by the thin disc model, results in an even stronger dependence of $f_b$ on $M$ and a weaker, but still negative dependence of $f_b$ on \redd . 

\section{Figures of merit as a function of the power spectral model parameters}
Below we show the dependence on the figure of merit $fom/N$ as a function of the 4 fitted parameters, for the 5 better-fitting models. 
\begin{figure*}[h]
    \centering
     \includegraphics[width=0.33\textwidth]{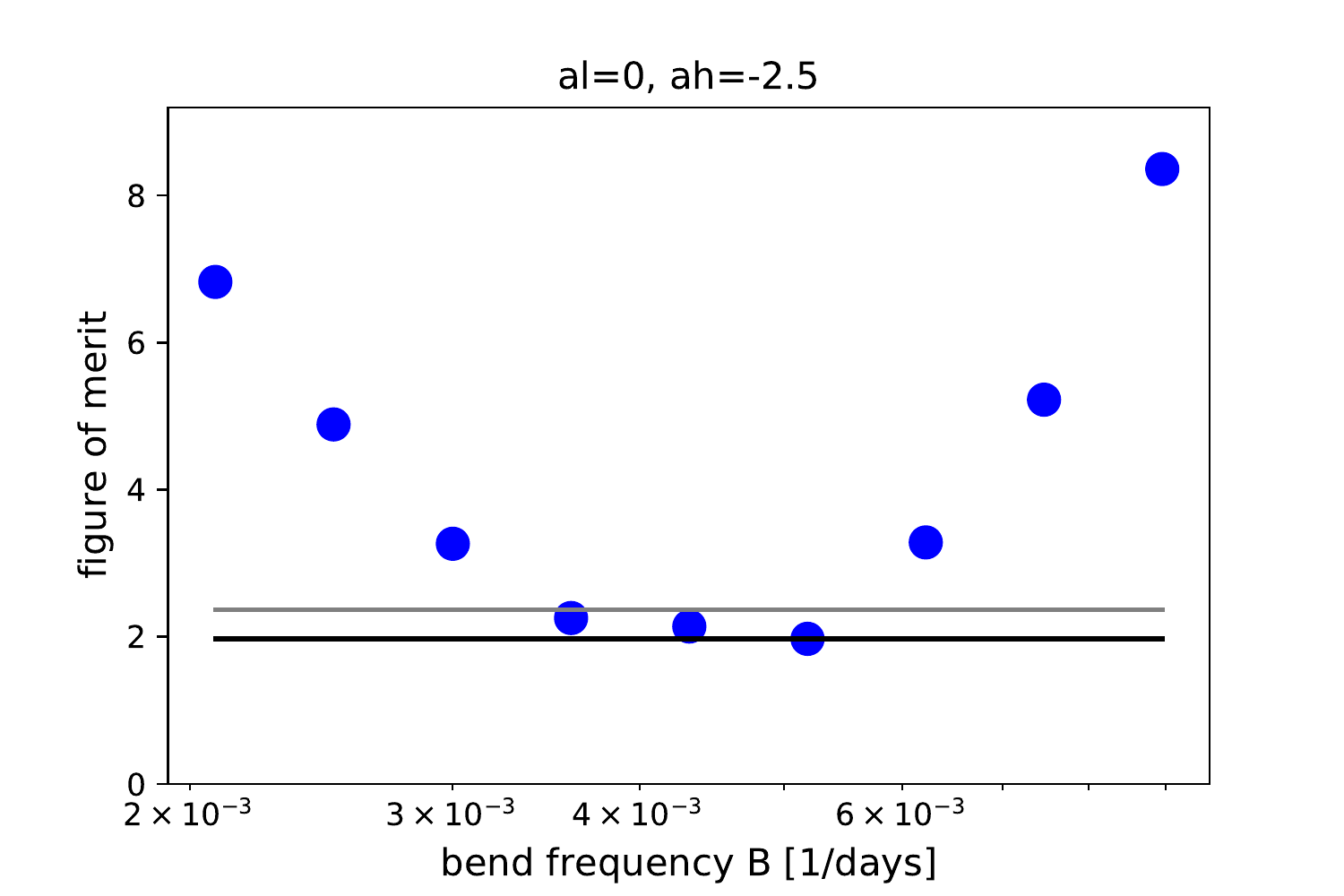}
    \includegraphics[width=0.33\textwidth]{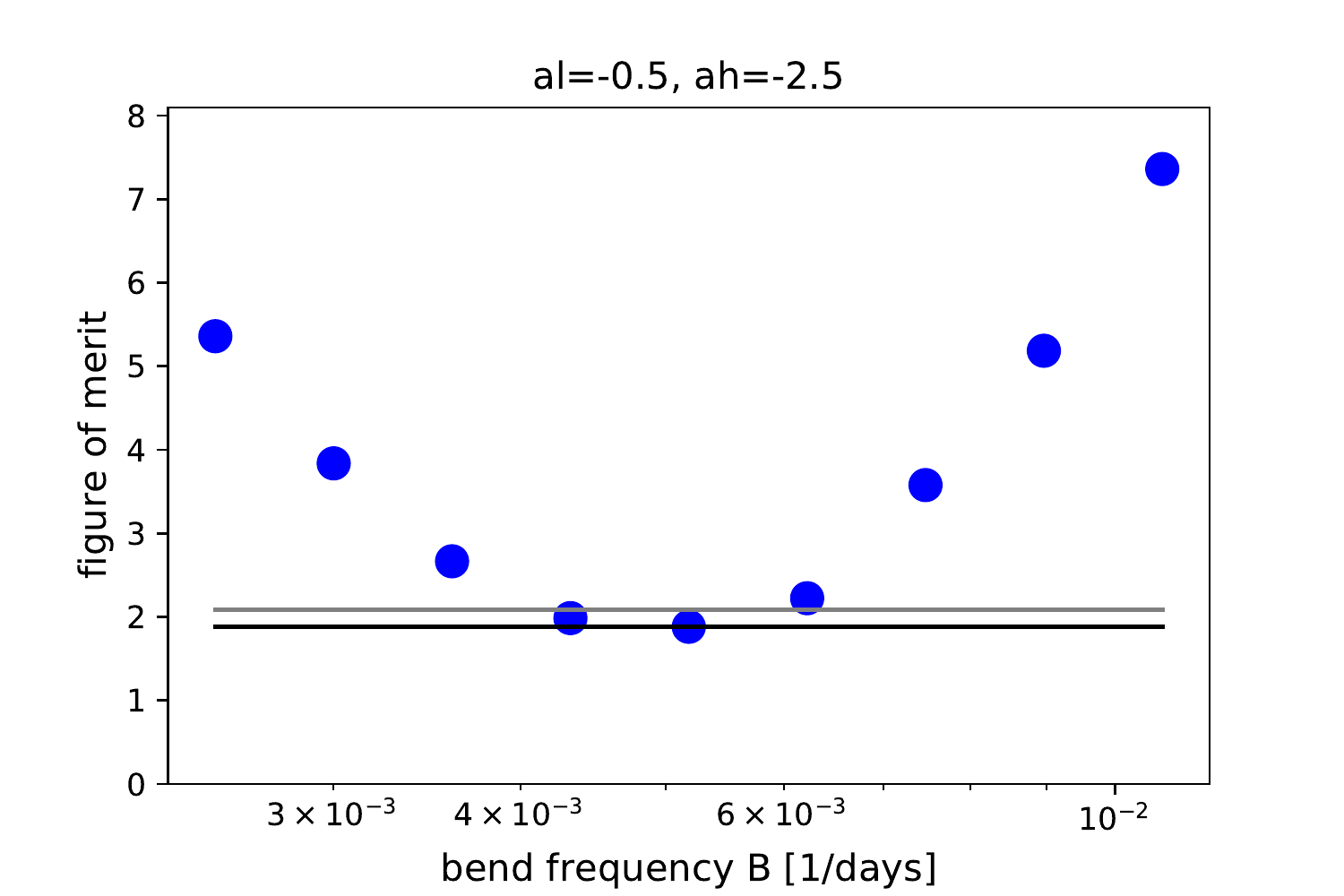}
    \includegraphics[width=0.33\textwidth]{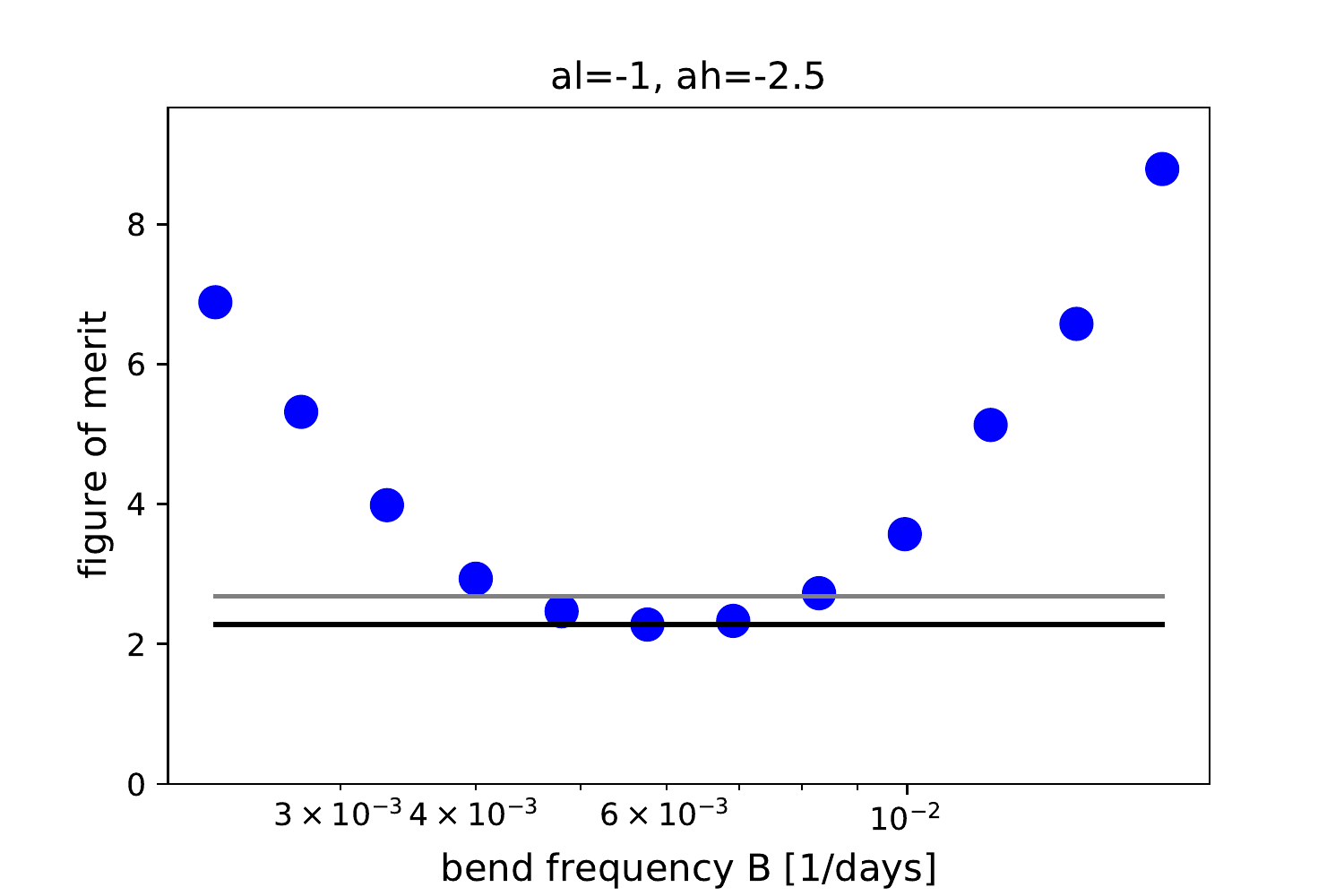}
    \includegraphics[width=0.33\textwidth]{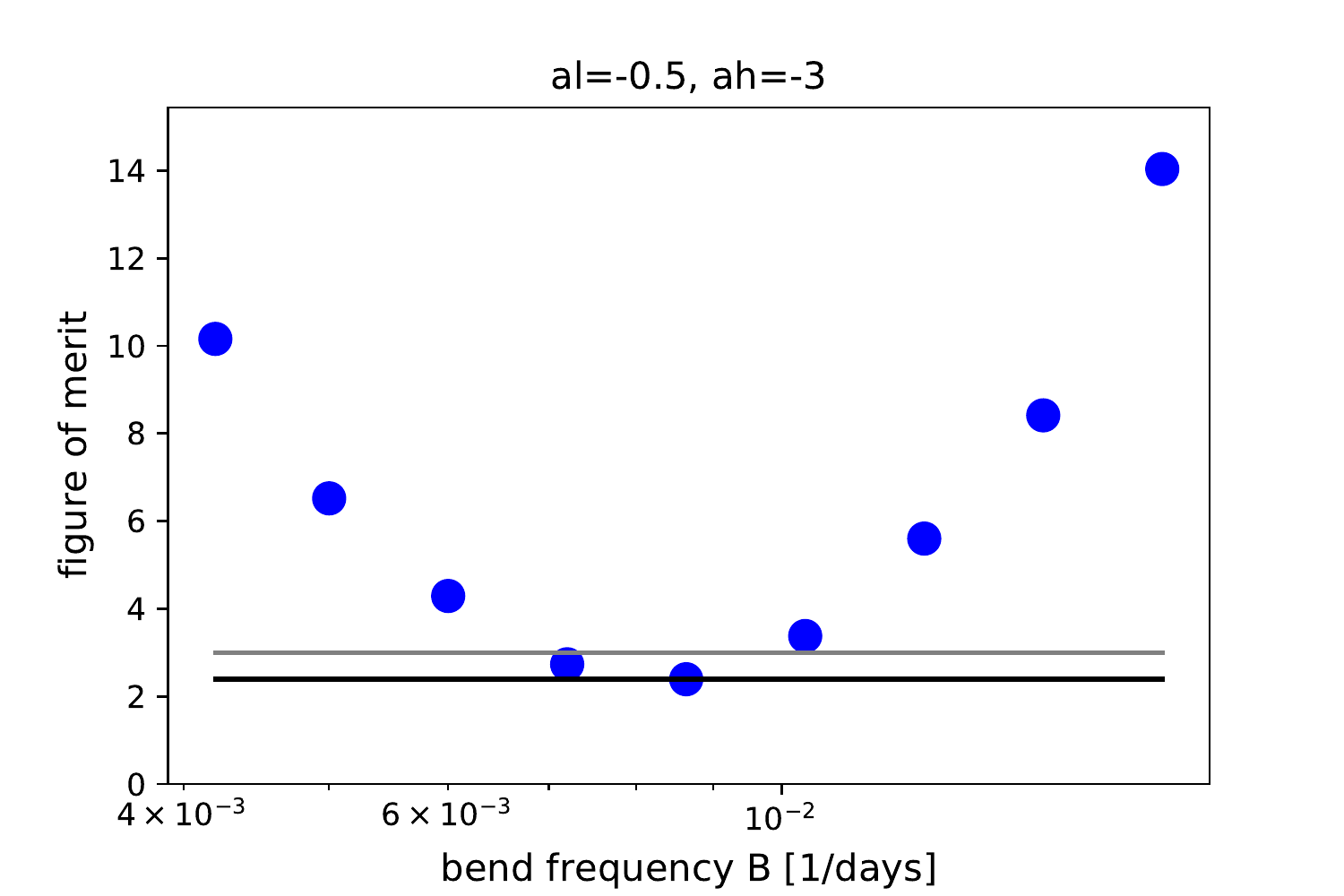}
    \includegraphics[width=0.33\textwidth]{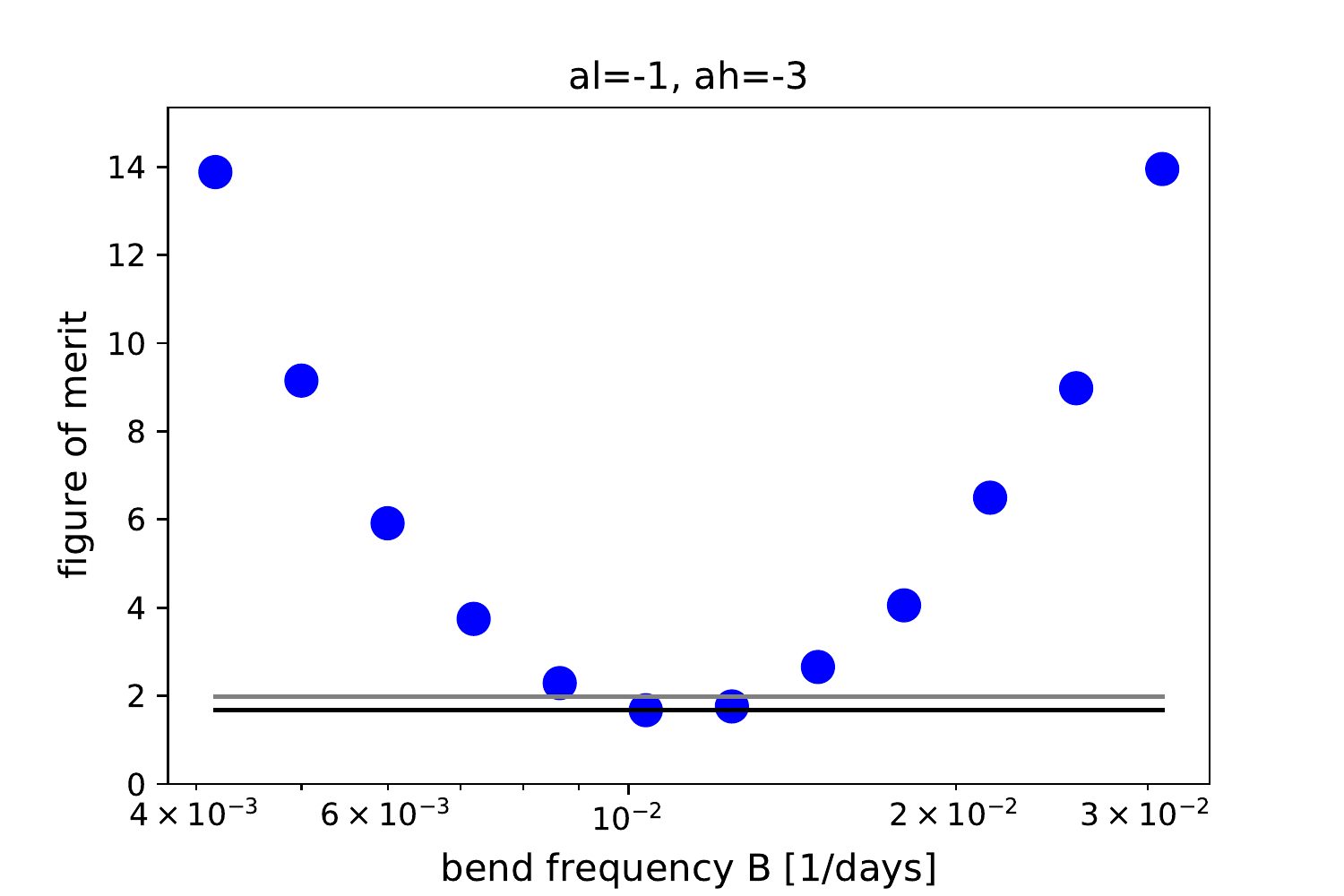}  
    \caption{Figure of merit values as a function of the bend frequency for $M=10^{8}M_\odot$ and \redd=0.1, $B$, for the better-fitting models with different power-law slopes ($\alpha_L$ and $\alpha_H$, as noted in the title of each panel).  Blue dots show the minimum $fom/N$ for each value of $B$ with all other parameters free, corresponding to the second iteration, with a finer grid in the parameters. The black horizontal line marks the lowest $fom/N$ achieved and the grey line marks the level of $min(fom)/N+\Delta$, from which error bounds are estimated. The value of $\Delta$ was obtained for each model as described in Sec. \ref{sec:errors}.}
    \label{fig:chi_B}
\end{figure*}

\begin{figure*}
    \centering
     \includegraphics[width=0.33\textwidth]{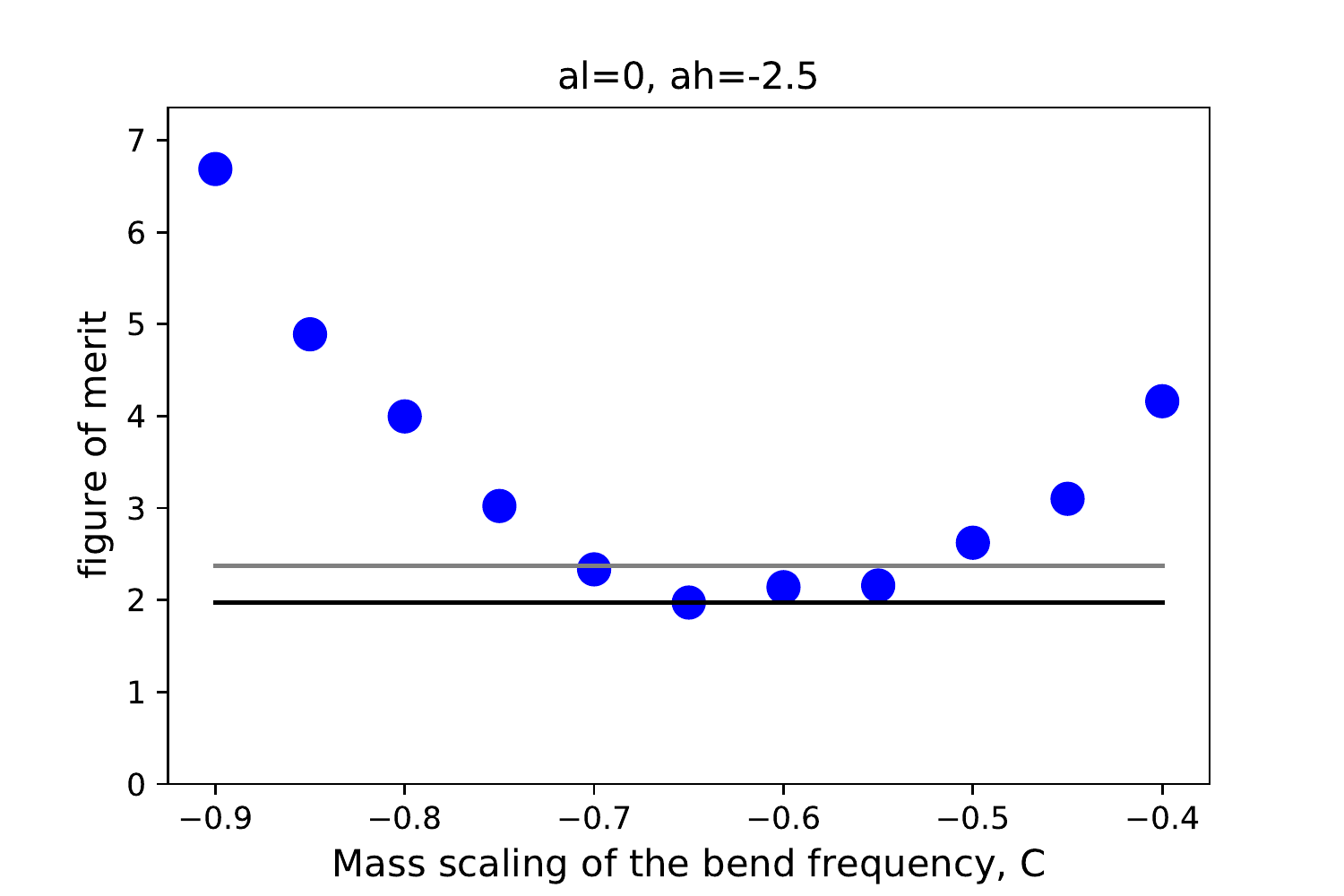}
    \includegraphics[width=0.33\textwidth]{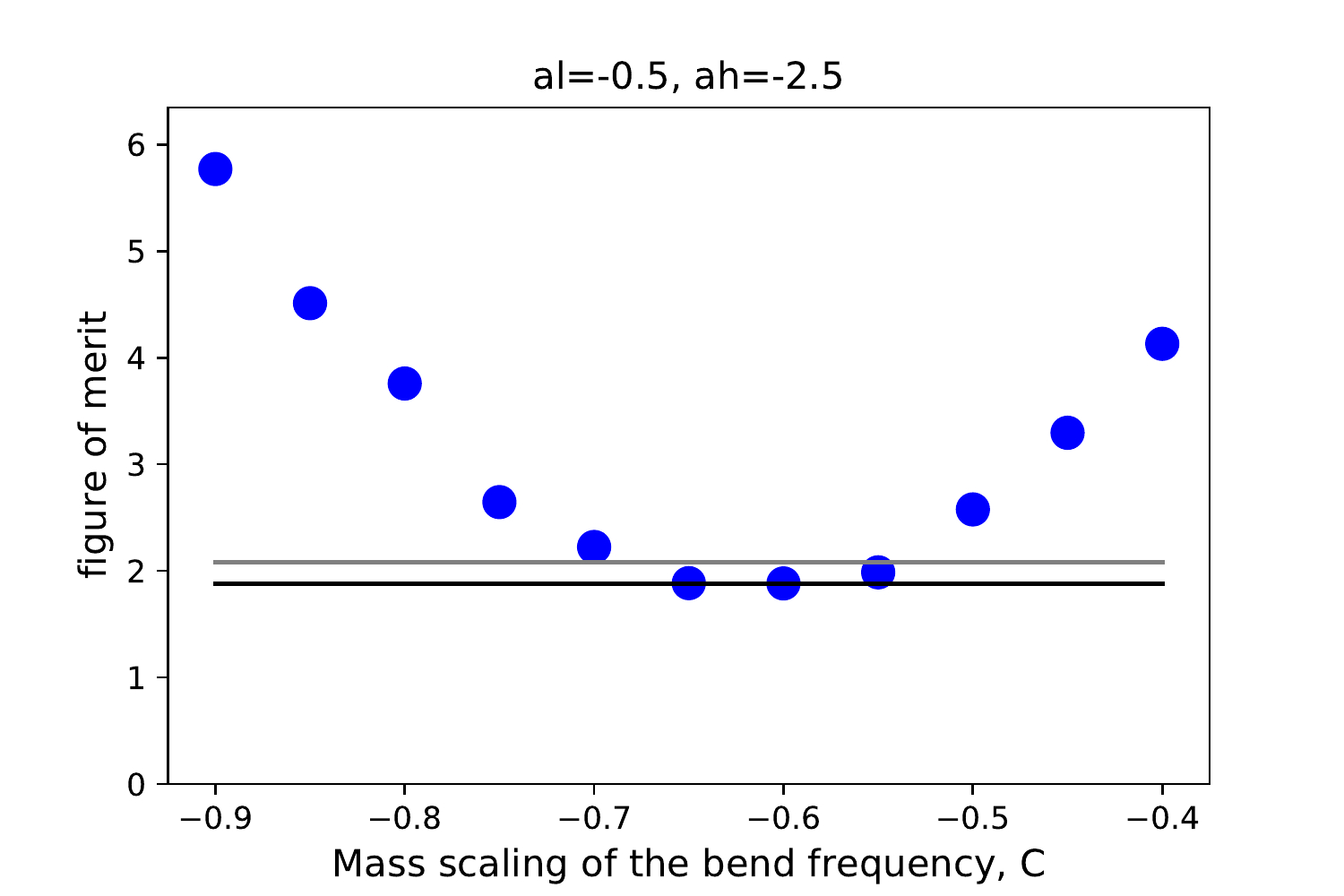}
    \includegraphics[width=0.33\textwidth]{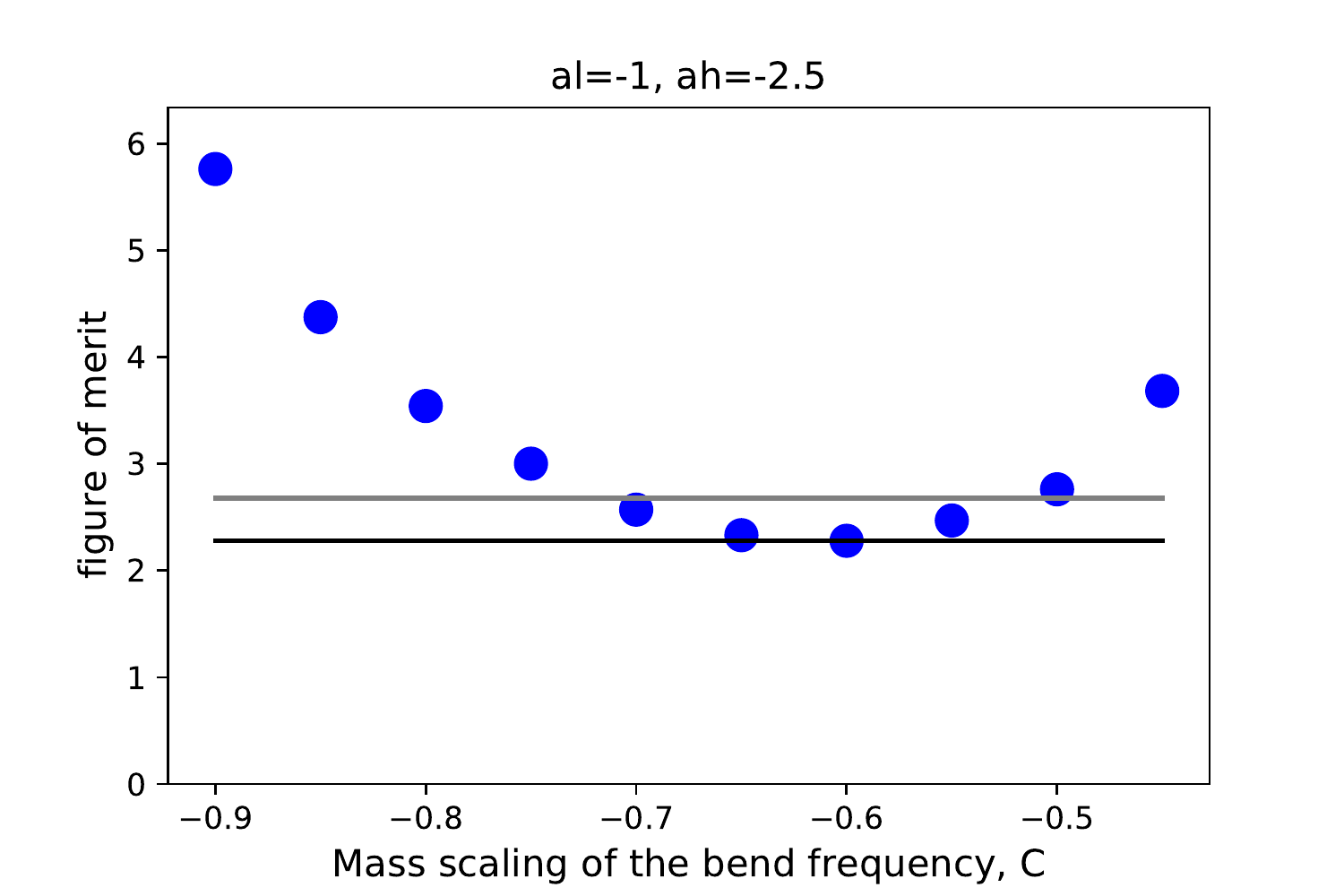}
    \includegraphics[width=0.33\textwidth]{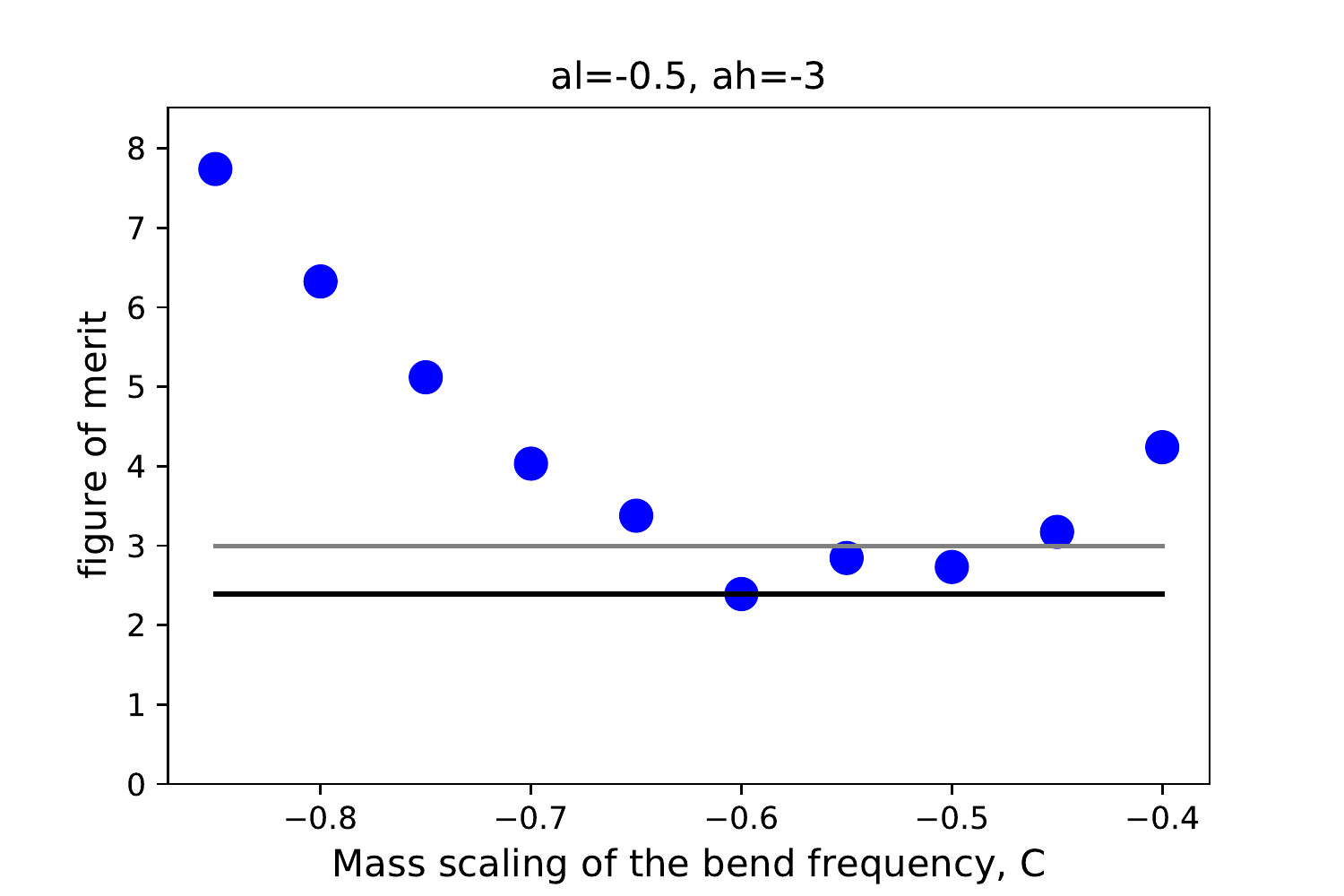}
    \includegraphics[width=0.33\textwidth]{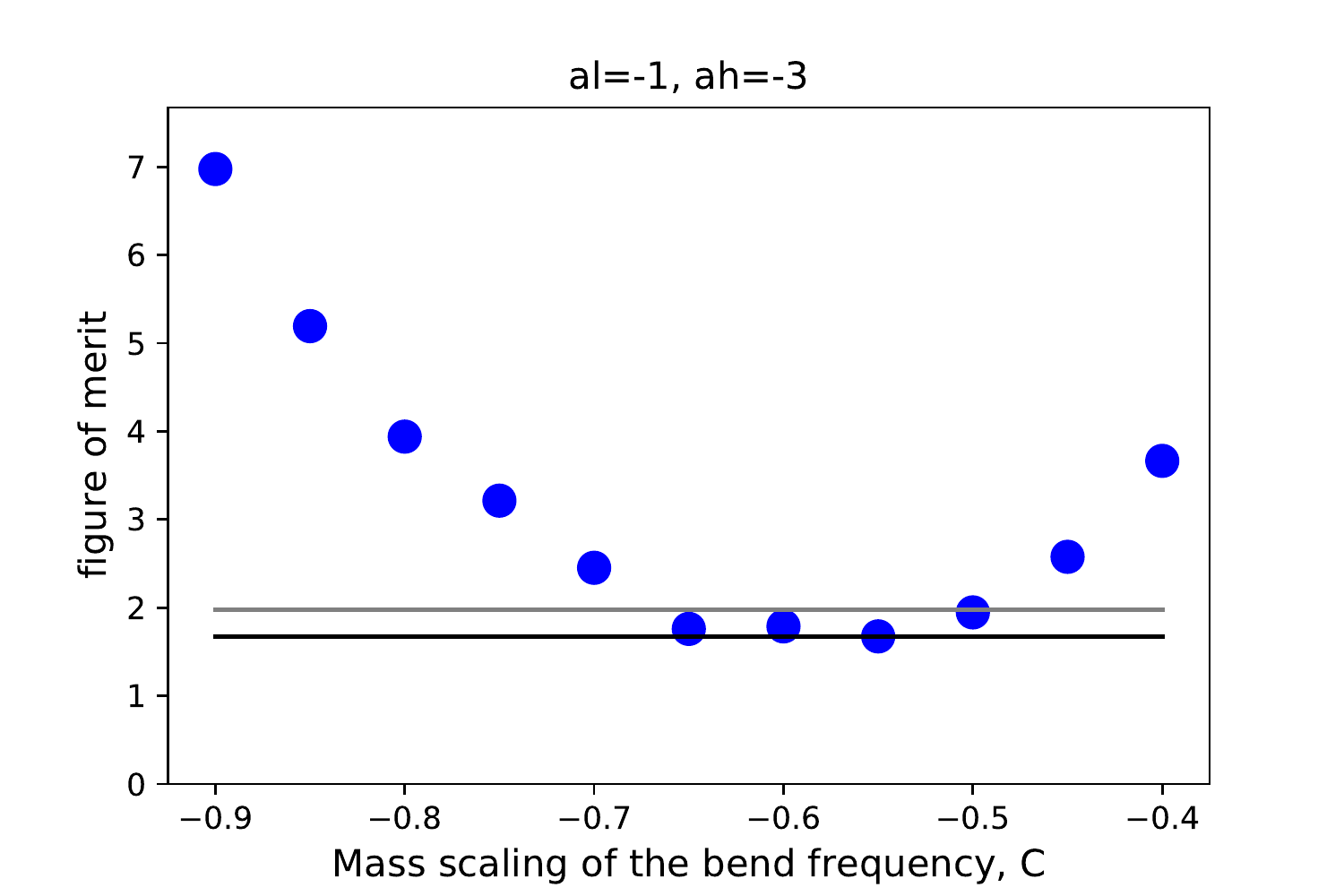}  
    \caption{As Fig.\ref{fig:chi_B} but for parameter $C$, which determines the dependence of the bend frequency on M, i.e $f_b\propto$M $^{C}$.}
    \label{fig:chi_C}
\end{figure*}

\begin{figure*}
    \centering
     \includegraphics[width=0.33\textwidth]{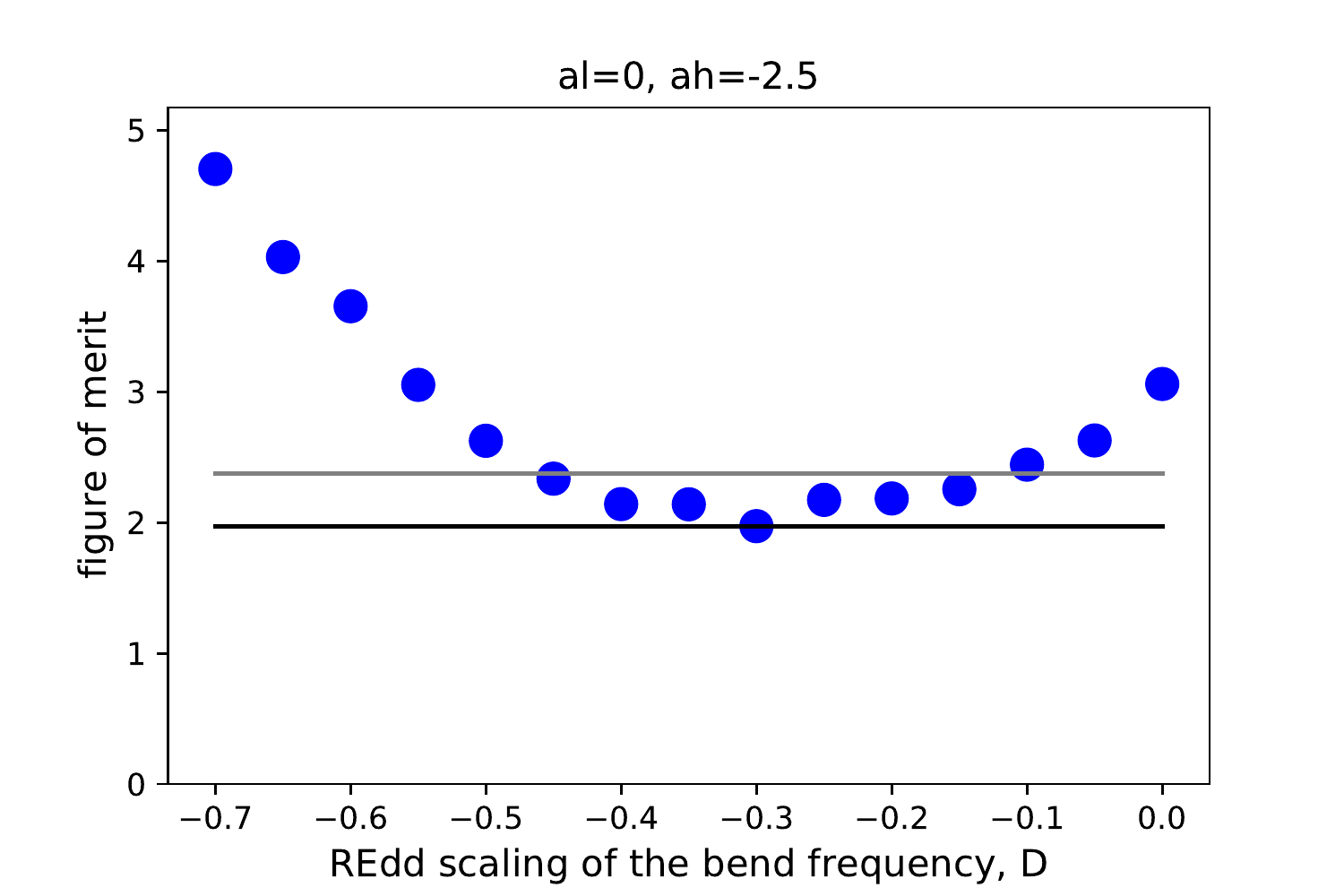}
    \includegraphics[width=0.33\textwidth]{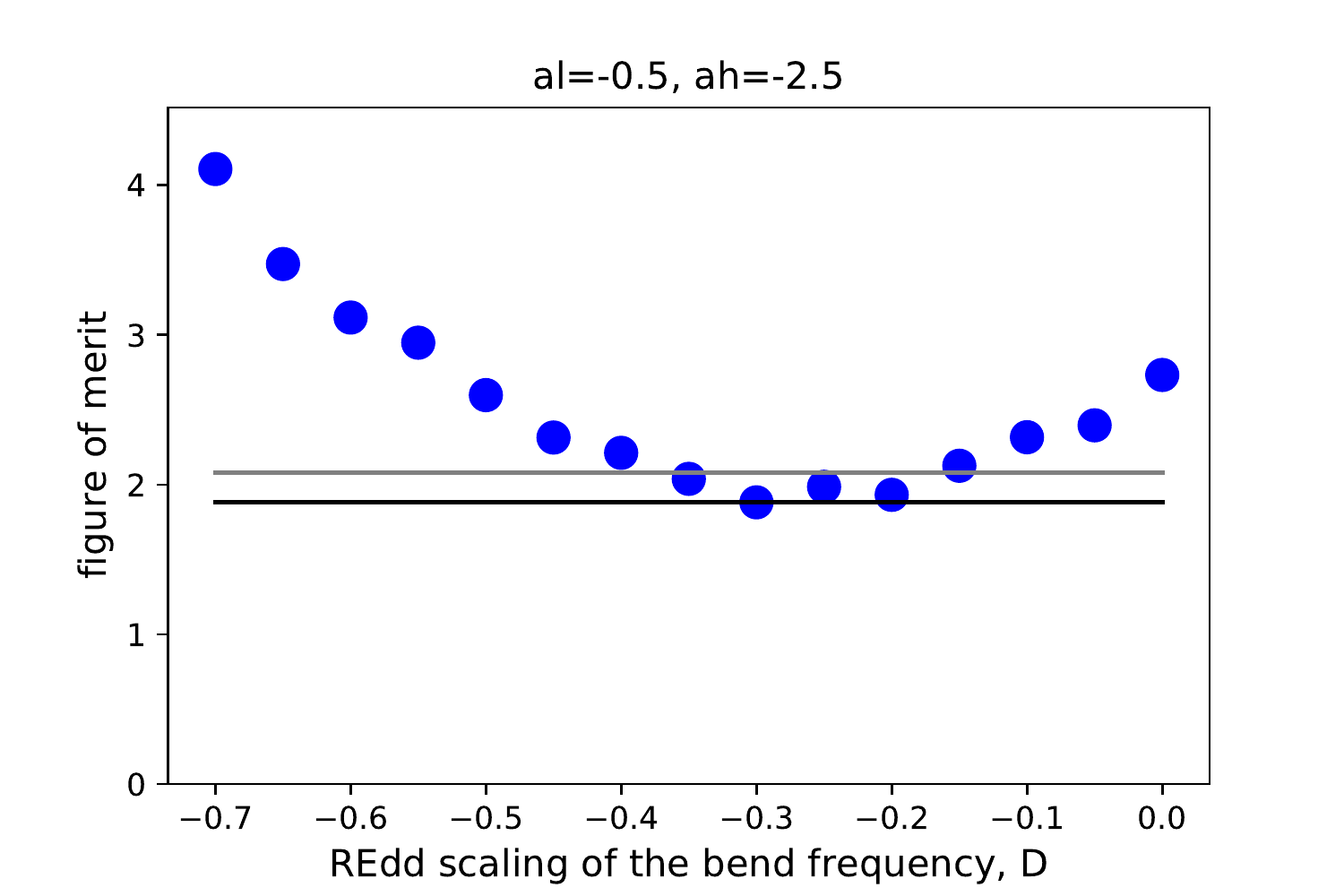}
    \includegraphics[width=0.33\textwidth]{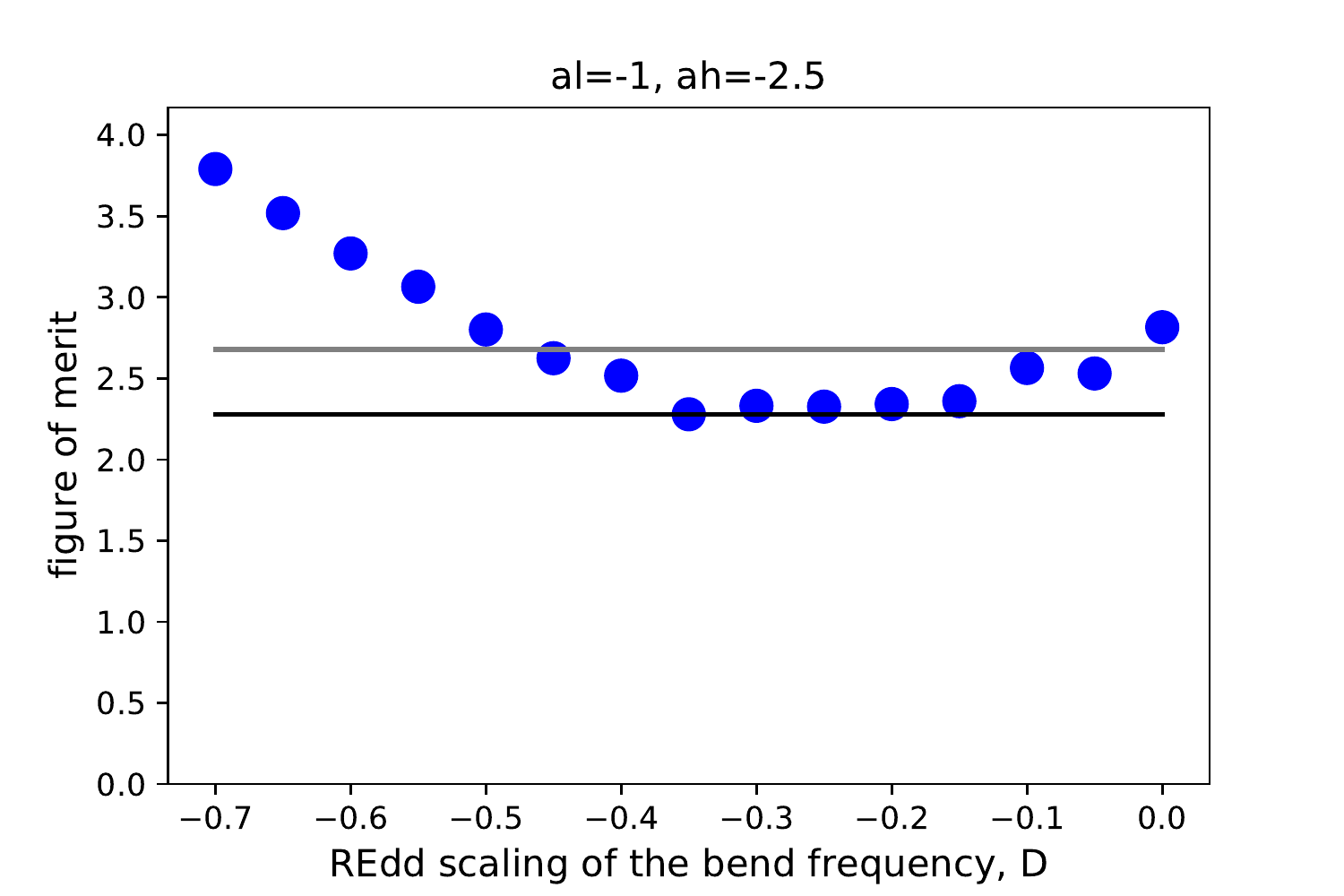}
    \includegraphics[width=0.33\textwidth]{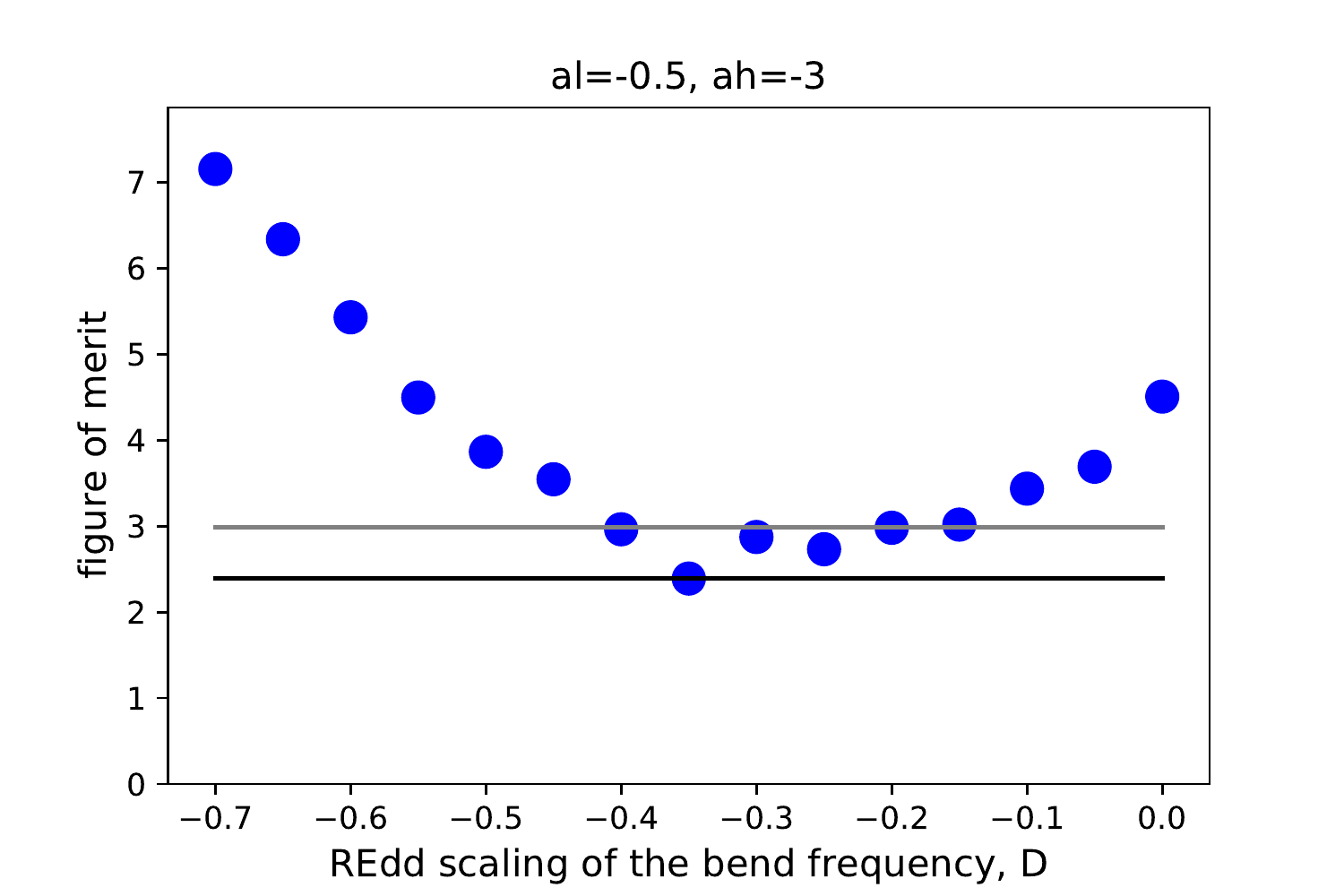}
    \includegraphics[width=0.33\textwidth]{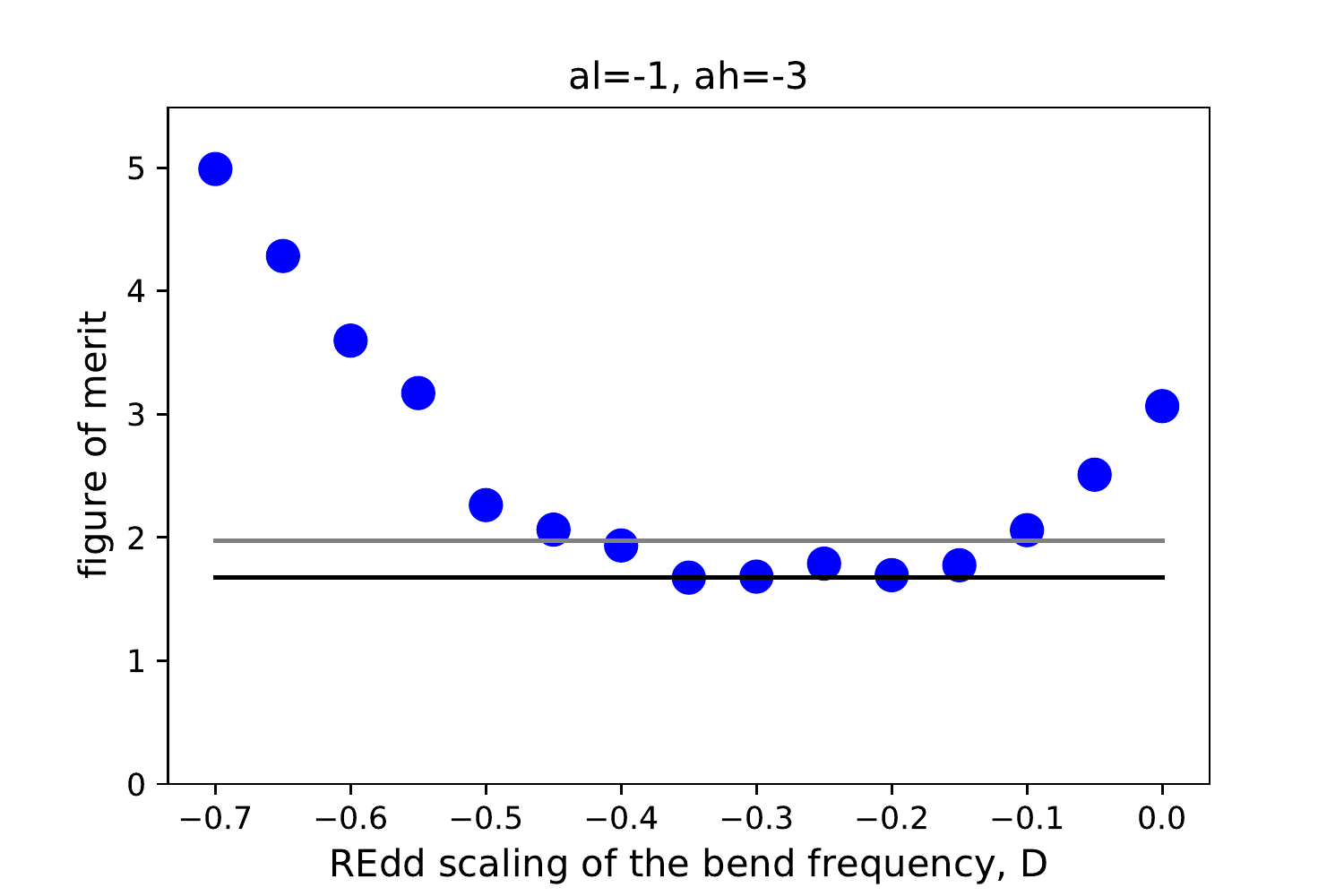}  
    \caption{As Fig.\ref{fig:chi_B} but for parameter $D$, which determines the dependence of the bend frequency on \redd , i.e $f_b\propto$\redd $^{D}$.}
    \label{fig:chi_D}
\end{figure*}

\begin{figure*}
    \centering
     \includegraphics[width=0.33\textwidth]{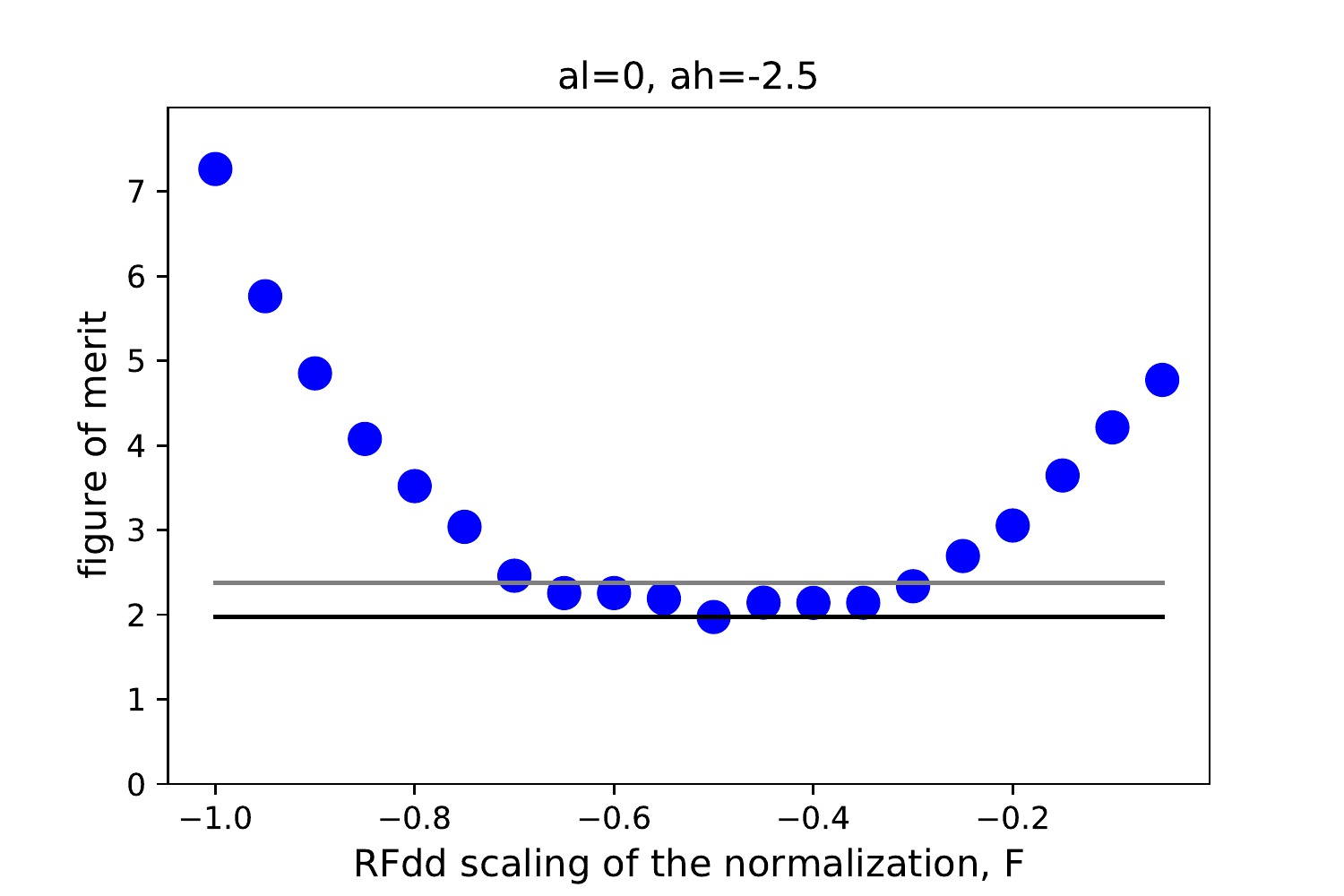}
    \includegraphics[width=0.33\textwidth]{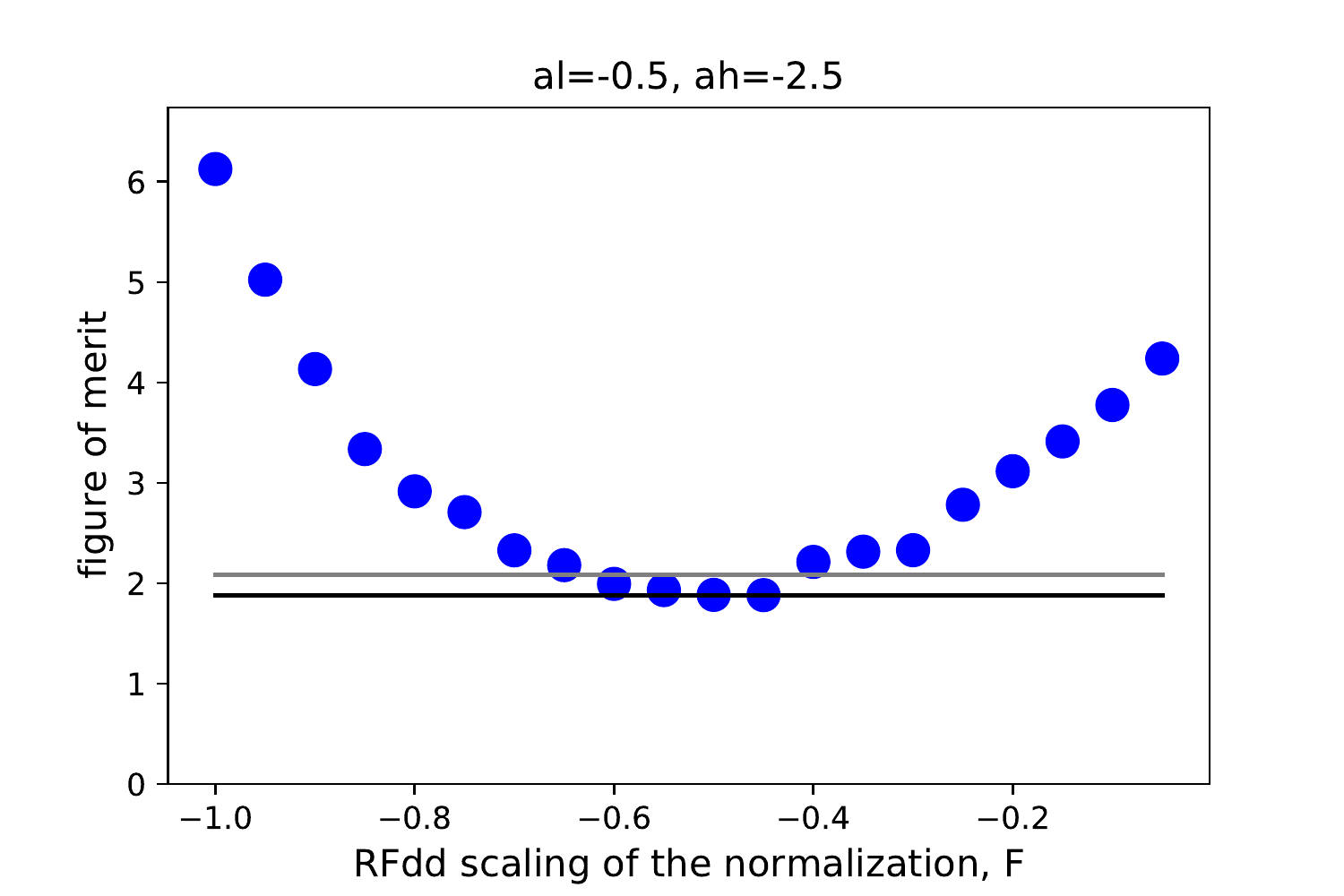}
    \includegraphics[width=0.33\textwidth]{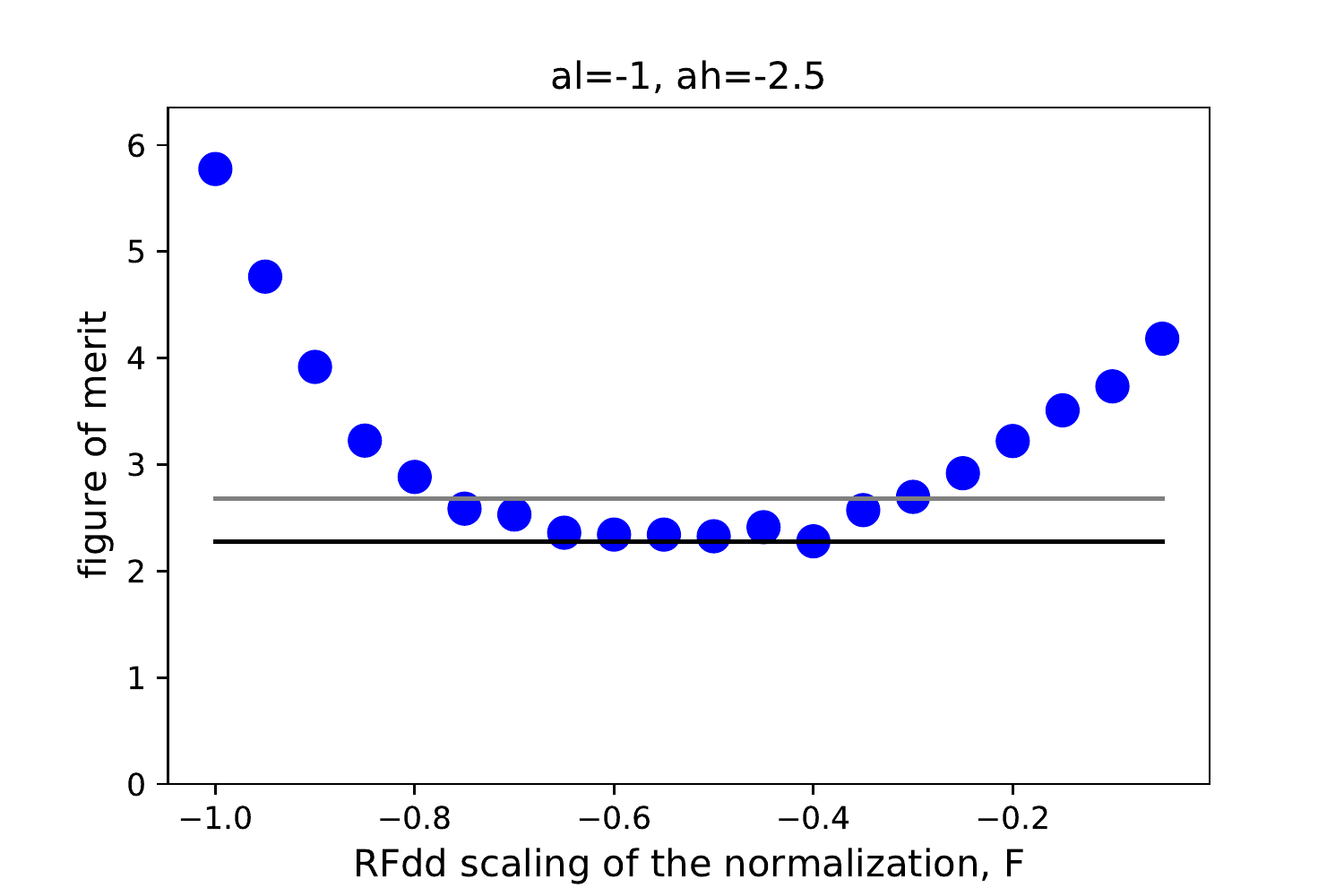}
    \includegraphics[width=0.33\textwidth]{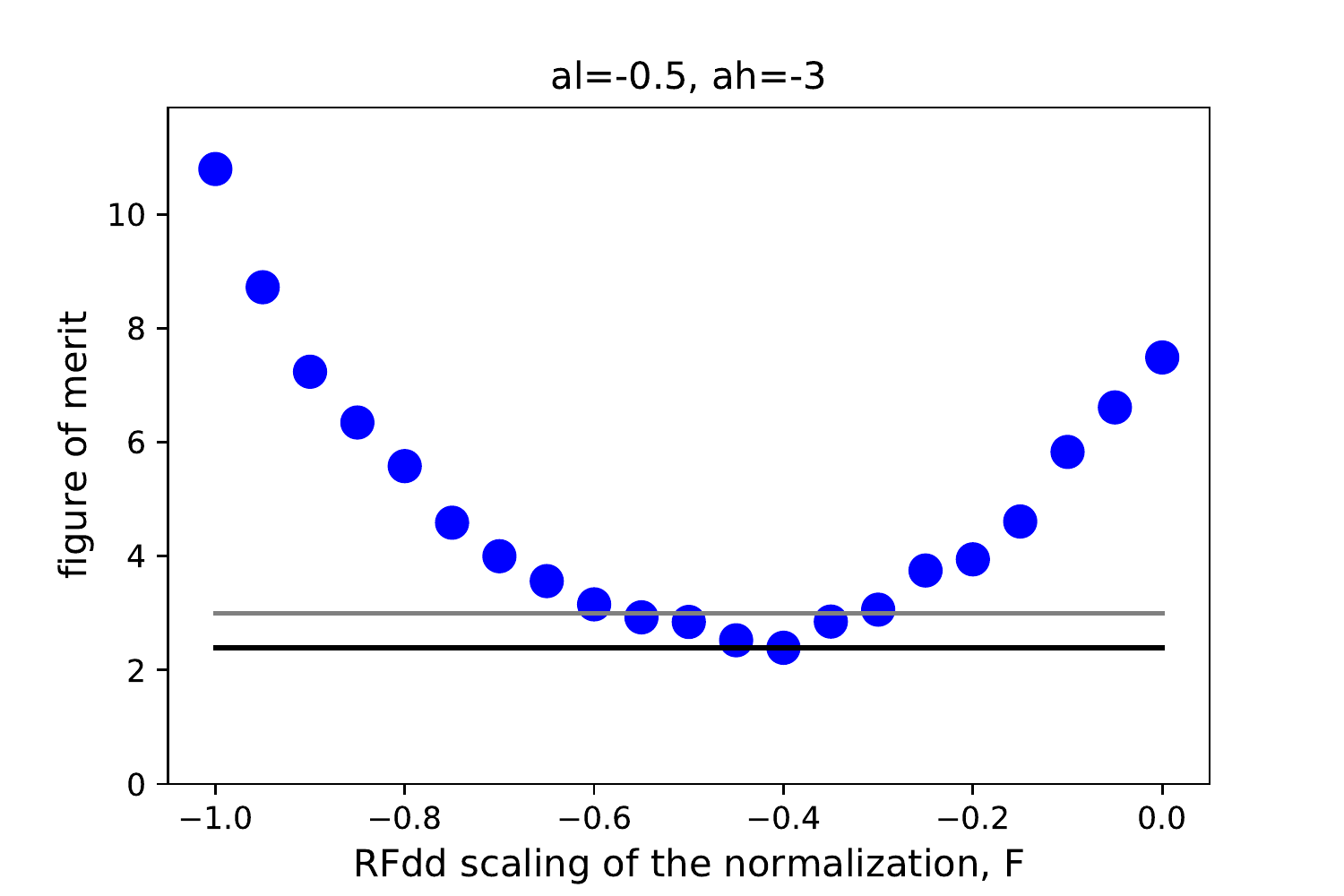}
    \includegraphics[width=0.33\textwidth]{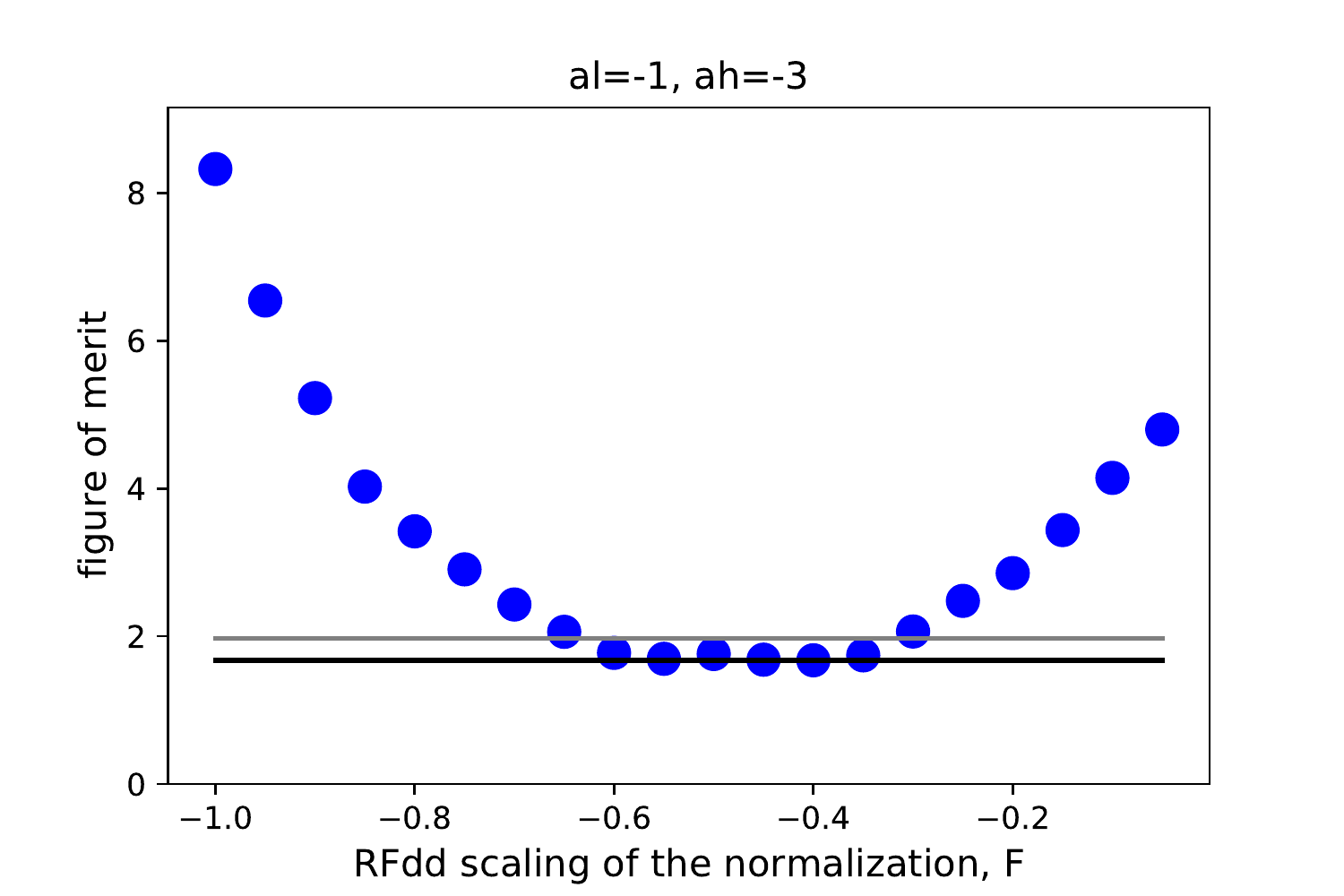}  
    \caption{As Fig.\ref{fig:chi_B} but for parameter $F$, which determines the dependence of the normalization of the power spectrum on \redd , i.e $A\propto$ \redd $^{F}$. 
    }
    \label{fig:chi_F}
\end{figure*}

\begin{figure*}
    \centering
     \includegraphics[width=0.45\textwidth]{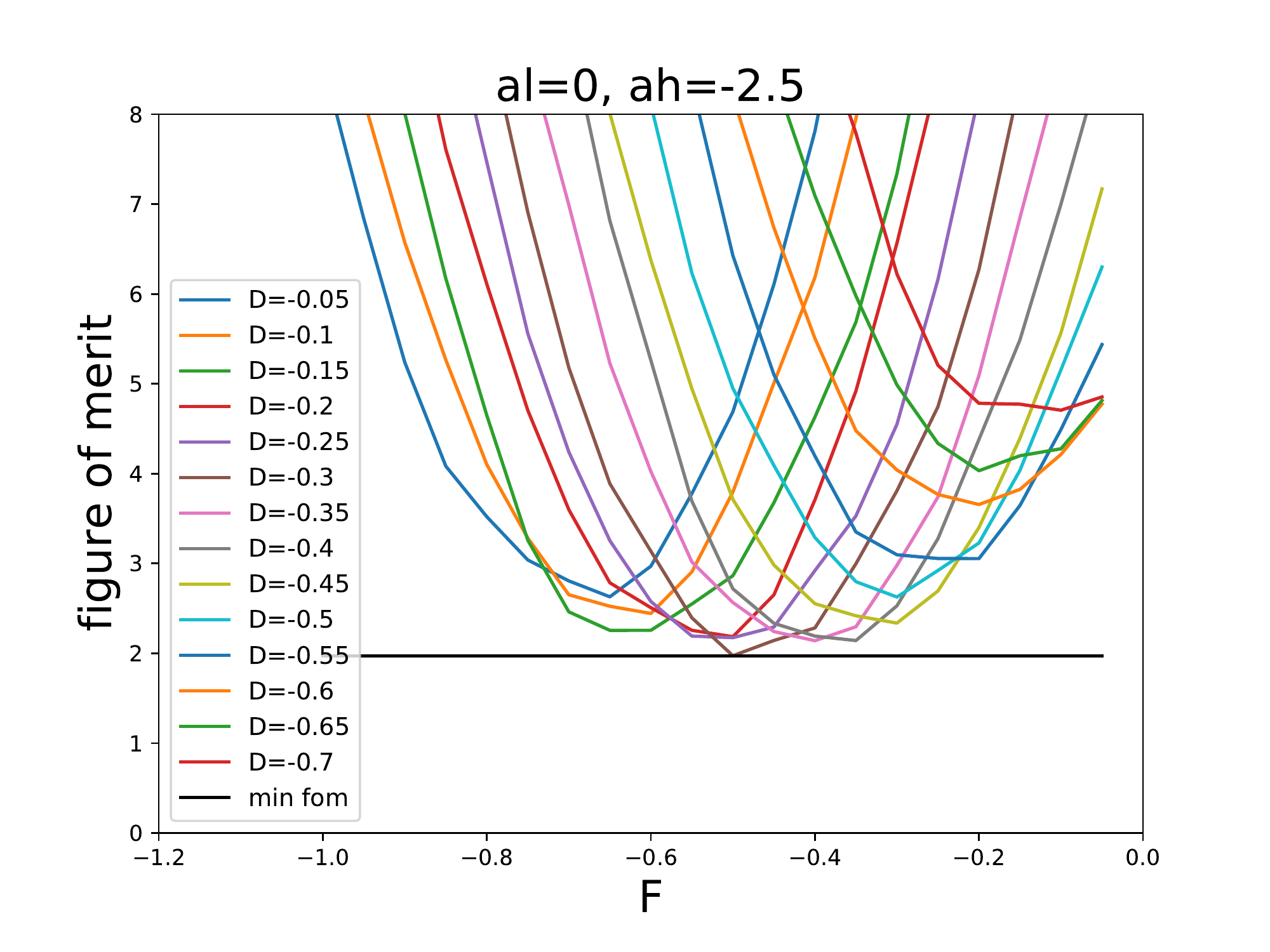}
    \includegraphics[width=0.45\textwidth]{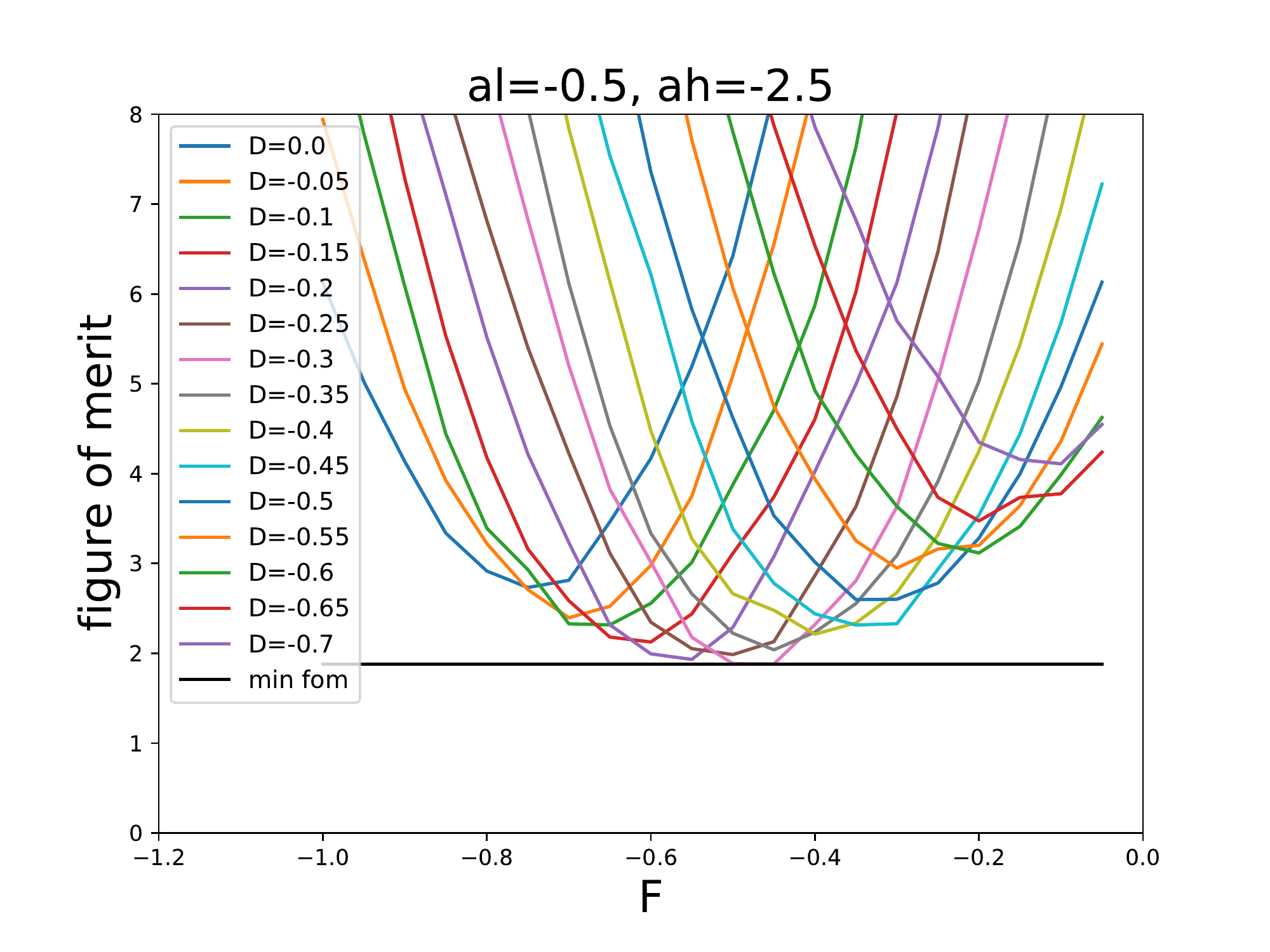}
    \includegraphics[width=0.45\textwidth]{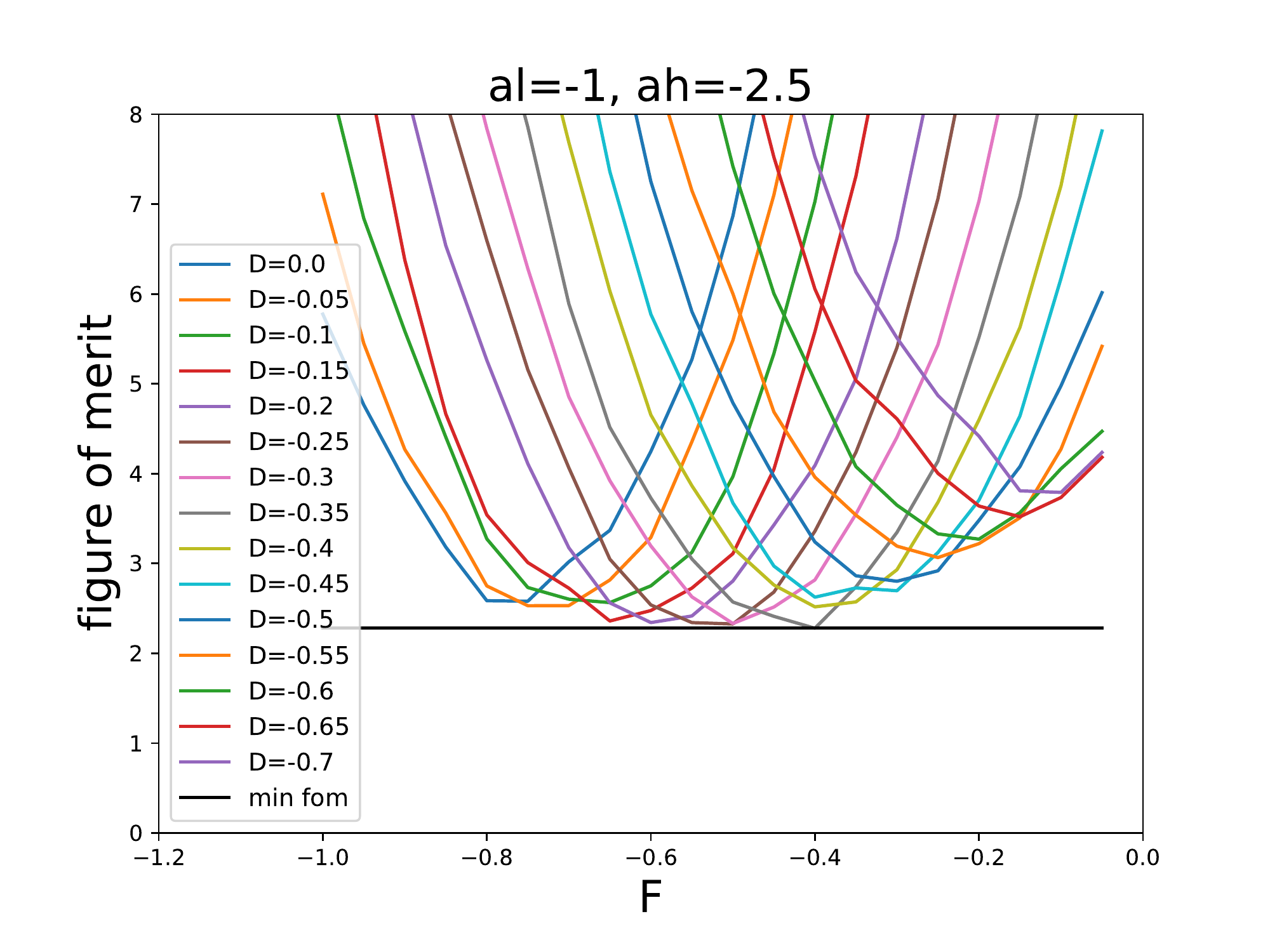}
    \includegraphics[width=0.45\textwidth]{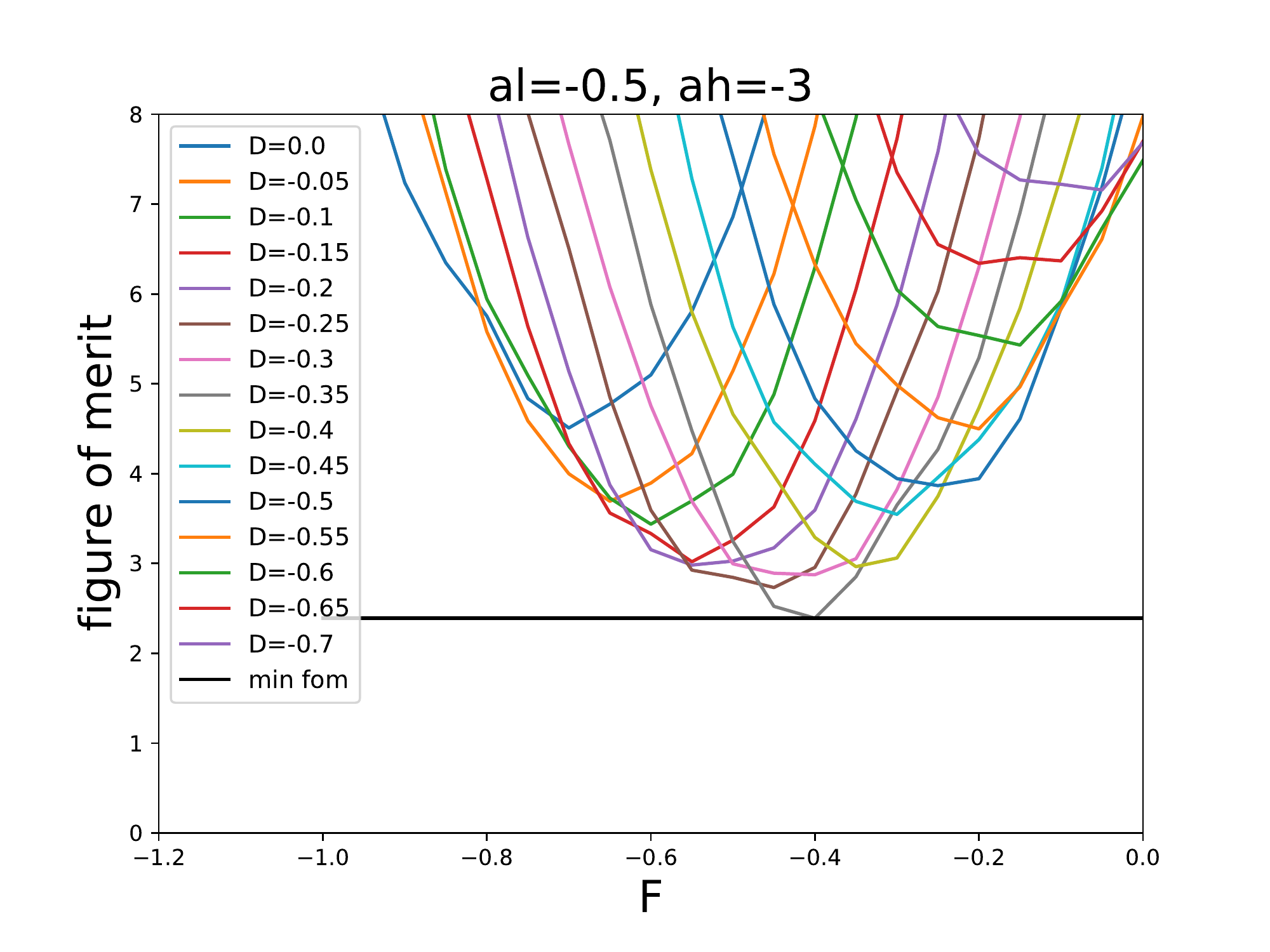}
    \includegraphics[width=0.45\textwidth]{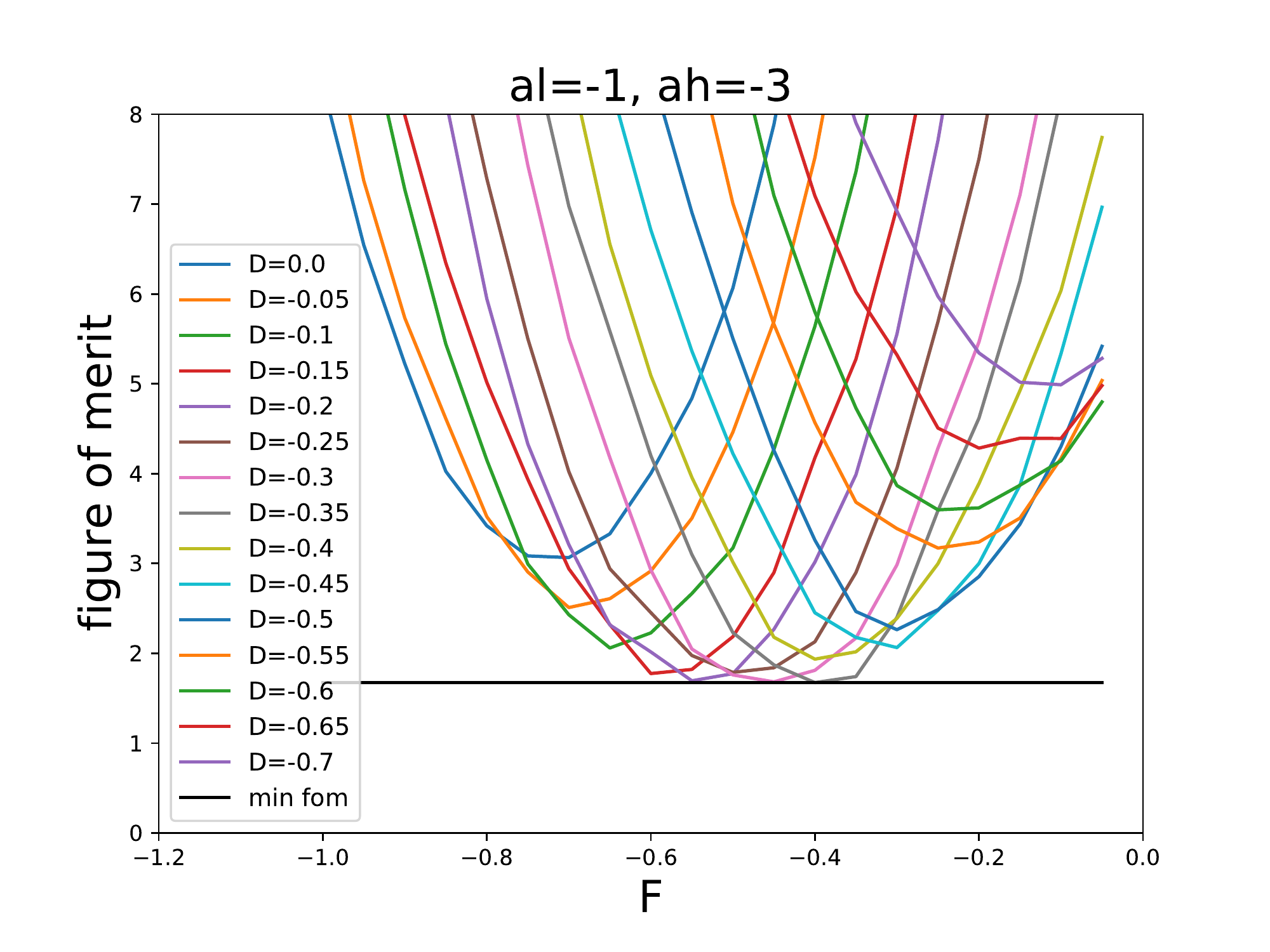}  
    \caption{Minimum $fom$ for given values of $D$, as a function of $F$. Both parameters relate to the dependence of the power spectrum on \redd\ and are therefore partially degenerate, which produces the flat profiles of $fom$ vs $D$. It is clear, however, that extreme values of $D$ produce overall worse fits, for any values of F. }
    \label{fig:chi_DF}
\end{figure*}

\end{document}